\documentclass[preprint,10pt]{elsarticle}

\usepackage{geometry}
\newgeometry{ hmargin={25mm,25mm}}

\usepackage{algorithm}
\usepackage[noend]{algpseudocode}
\usepackage{amsmath}
\usepackage{amssymb}
\usepackage{bm}
\usepackage{booktabs}
\usepackage{caption,subcaption}

\usepackage{multicol}
\usepackage[numbers]{natbib}
\usepackage{graphicx}

\DeclareMathOperator{\bp}{\bf p}

\DeclareMathOperator{\br}{\bf r}
\DeclareMathOperator{\bS}{\bf S}
\DeclareMathOperator{\bs}{\bf s}

\DeclareMathOperator{\bX}{\bf X}
\DeclareMathOperator{\bx}{\bf x}
\DeclareMathOperator{\bY}{\bf Y}
\DeclareMathOperator{\by}{\bf y}

\DeclareMathOperator{\bkappa}{\bm \kappa}

\DeclareMathOperator{\bmu}{\bm \mu}

\DeclareMathOperator{\bomega}{\bm \omega}

\DeclareMathOperator{\bsigma}{\bm \sigma}

\DeclareMathOperator{\btheta}{\bm \theta}

\newcommand{\toi}[2][t]{%
  \mathop{
    \mathrm{#2}^{( #1 )}
  }
}

\journal{arXiv}
\begin{document}

\begin{frontmatter}

\title{Global Seasonality Factor Investing}

\author{Andrew Butler, Adam Butler, Maciej Zawadzki}

\address{ReSolve Asset Management}

\begin{abstract}
  We study seasonality as a factor for predicting risk-adjusted returns across global futures markets. We document five seasonality anomalies that,
  in aggregate, are pervasive across bond, commodity and equity markets. We demonstrate that each seasonality factor produces
  excess returns that are not reasonably explained by established global time series carry, momentum and skewness factors. We apply
  recent innovations in statistical learning and fit nonlinear seasonality models at the individual market level, thereby allowing each
  feature-market pair to express its unique seasonal prediction pattern. We discuss candidate economic rationales for the existence and persistence
  of each seasonality anomaly. Walk-forward out-of-sample testing demonstrates that fast-moving seasonal patterns, such as day-of-week and day-of-month,
  produce strong gross performance but with extremely high turnover, resulting in negative performance net of transaction costs. In contrast,
  slower-moving seasonal patterns, such as day-of-quarter, day-of-half-year and day-of-year, produce positive gross performance and, with the
  exception of day-of-quarter, positive net performance. An equal-weight
  ensemble of seasonality models produces an out-of-sample Sharpe ratio that is competitive with, and uncorrelated to, established global time series
  factor portfolios. Empirical results demonstrate that adding seasonality to a global factor portfolio increases the out-of-sample Sharpe ratio by 
  approximately $30\%$ over the full evaluation period.
\end{abstract}


\end{frontmatter}


\section{Introduction} \label{sec:intro}
Factor investing has a long and established history in the financial literature. The seminal work of Fama and French 
\citep{Fama1988, Fama1992, Fama1993, Fama2015} expanded on the capital asset pricing model \citep{Treynor1962, Sharpe1963, Lintner1965}
and demonstrated that factors such as size, value and profitability can reasonably explain the cross-section of stock returns. In the last decade, 
several time series factors have demonstrated strong and pervasive predictive power for forecasting forward returns of global asset classes, 
including bonds, currencies, commodities and equities. For example, \citet{Moskowitz2012} document the time series momentum factor,
defined as the market's average return over a $12$-month period. The authors demonstrate that markets, in aggregate, exhibit a small positive
autocorrelation whereby a market's recent historical average return positively predicts its forward returns.  From a causal perspective, 
the time series momentum phenomenon can be explained by investor behavioural biases: investors tend to underreact to new information,
and then proceed to chase returns after a market trend has been established \citep{Hong1999, Asness2013, Baltas2012, Baltas2017, Hallerbach2017}.
Closely related to this work, \citet{MoskowitzNonlinear} extend the traditional linear time series model and present a nonlinear factor model
that is capable of capturing mean-reversion effects when market trends reach extreme levels.

Similarly, \citet{Koijen2018} document the `carry' factor, defined  as the return an investor would earn by holding an asset assuming no change in 
the underlying asset price. They demonstrate that in global futures markets,  the carry factor can be estimated as the difference between the spot 
price and futures price, which positively predicts the cross-section of market returns within each sector. More recently, \citet{Adam2023}
document a time series implementation of the global carry factor that produces positive excess returns across all major global asset classes. 
From a risk theory perspective, the carry factor can be explained by 
investors' demand for higher yields in order to compensate for exposure to tail risks, such as defaults, delistings or inflation risks 
\citep{Boons2020, Koijen2017}. Alternatively, from a behavioural perspective, investors may underreact to carry:
they expect the prices of high-yielding assets to fall by enough to offset the yield, as implied by uncovered 
interest parity, and those expected price declines do not fully materialize \citep{Froot1989}.

Of equal importance, \citet{Baltas2022} study the skewness factor, defined as the third statistical moment of market returns. They find that an asset's historical 
skewness is a strong negative predictor of global market returns. The economic explanation for the observed effect relates to investors'
willingness to tolerate lower returns for assets with lottery-like payouts and, conversely, to demand higher compensation for assets
with large left-tail risks. Taken together, the time series momentum, carry and skewness factors are some of the most widely studied factors 
in the quantitative finance literature \citep{Asness2015, Benuzzi2022,Hafez2020, He2015,Hong2015, Kim2016}.

In financial markets, seasonality is defined as the tendency for assets to display regular and predictable return patterns during specific
times of the year. Seasonal pricing anomalies have been widely studied in the financial literature and documented across a variety of asset classes 
\citep[see, for example,][]{Guo2022, Li2018, Long2020, Merc2023, Sorensen2002, Song2022}. Notably, in equity markets,  \citet{Thaler1987} 
studied the `January effect,' where stock prices tend to rise during the first month of the year. \citet{Bouman2002} documented
the `sell in May and go away' effect, whereby stock returns tend to be higher from November to April compared to the period from May to October.
\citet{Desai2012} and \citet{Heston2008} observe a persistent calendar month effect in the cross-section of stock returns. \citet{Keloharju2015} document a
similar calendar effect that is persistent in both stock and commodity returns. More recently, \citet{Back2013} document 
persistent seasonality in the volatility of commodity markets, which in turn affects the pricing of commodity options. \citet{Zaremba2017}
explore calendar year effects in fixed income and demonstrate that the anomalies observed in equity markets do not translate equally to
global bond markets.

This paper contributes to the established body of literature by consolidating many of the empirical findings mentioned above into a unified global
seasonality factor framework. We explore five seasonality anomalies: day-of-week, day-of-month, day-of-quarter, day-of-half-year and day-of-year,
and demonstrate the pervasiveness of seasonality effects across most global asset classes. The remainder of the paper is organized as follows. 
In Section \ref{sec:method} we follow recent work in statistical learning and present a framework for fitting nonlinear regression
models to forecast forward volatility-adjusted market returns based on seasonality features \citep{Butler2026, MoskowitzNonlinear}.
The framework fits prediction models to individual markets and therefore enables each feature-market pair to express its potentially unique
seasonality behaviour. In Section \ref{sec:results} we provide an empirical overview of the observed seasonality phenomena and
provide economic justification for their existence. Additionally, we present a comprehensive walk-forward analysis,
spanning more than 30 years of market data. Multi-factor regression analysis demonstrates that, in aggregate, each seasonality anomaly
produces gross excess returns that are not explained by the established global time series carry, momentum and skewness factors. From a 
practitioner perspective, we find that fast-moving seasonal patterns, such as day-of-week and day-of-month, produce strong gross performance 
but with a turnover that results in negative net performance. Lower-turnover
seasonal anomalies, such as day-of-quarter, day-of-half-year and day-of-year, produce positive gross performance and, 
with the exception of day-of-quarter, positive net performance. 
When observed in isolation, these seasonality effects produce excess returns that concentrate in specific asset classes, with portfolio net 
Sharpe ratios that are materially weaker than the Sharpe ratios of established global time series factors. However, an equal-weight ensemble of
seasonality models produces competitive and differentiated out-of-sample Sharpe ratios,
with returns that have near-zero correlation to the existing time series factors. We demonstrate that adding seasonality as an equal-weight
component to a global factor portfolio increases the Sharpe ratio over every out-of-sample evaluation period and improves the full-sample
Sharpe ratio by approximately $30\%$.

\section{Methodology} \label{sec:method}
We cast the market return forecast as an empirical conditional estimation problem \citep{MoskowitzNonlinear}. Following \citet{Butler2026},
we denote the price of the market, $k$, at time $t$, as $\toi{\bp_k}$, with market returns over the period given by:
${\toi{\br_k} =\toi{\bp_k}/\toi[t-1]{\bp_k} - 1}$. In order to control for the non-stationarity of market volatility, we let
$\bY  = \{\toi{\by}\}_{t=1}^T$ be the set of forward volatility-adjusted average market returns. Specifically:
\begin{equation}\label{eq:response}
\toi{\by_k} = \frac{1}{\sqrt{m}}\sum_{i=0}^{m-1} \toi[t+i+l]{\br_k}/\toi[t+i]{\hat{\bsigma}_k},
\end{equation}
with latency $l=2$, executing the trade signal at the closing price the day after the feature calculation, and forecast horizon $m$. The market
volatility, $\toi{\hat{\bsigma}_k}$, is estimated using a $40$-day exponential moving average. Volatility-adjusted market returns are then decomposed 
additively as:
\begin{equation}\label{eq:response_demean}
\toi{\by_k} = \bmu_k + \toi{\bs_k} + \toi{\epsilon_k},
\end{equation}
where $\bmu_k$ is the estimated market drift, $\toi{\bs_k}$ is the market's time-varying seasonality and $ \toi{\epsilon_k}$ denotes the irreducible
error. We estimate the market drift, $\bmu_k$, as the sample mean over the available training dataset. Therefore, the final seasonality
response variables, $\bS  = \{\toi{\bs}\}_{t=1}^T$, are standardized to have approximately zero mean and unit variance.  

For each seasonality anomaly, we let $\bX =  \{\toi{\bx}\}_{t=1}^T$ denote the associated feature variable. Feature variables are encoded 
as integer values, representing the integer day within their sampling period. Specifically:
\begin{itemize}
    \item {\underline{Day-of-week}}: Day integers (0--4) within a business week.
    \item {\underline{Day-of-month}}: Day integers (0--30) within a month.
    \item {\underline{Day-of-quarter}}: Day integers (0--91) within a quarter.
    \item {\underline{Day-of-half-year}}: Day integers (0--182) within a half-year.
    \item {\underline{Day-of-year}}: Day integers (0--365) within a year.
\end{itemize}

Given a training dataset ${D = \{(\toi{\bx}, \toi{\bs}) \}_{t=1}^T}$, we estimate a 
univariate prediction model $f \colon \mathbb{R} \times \mathbb{R}^{d_\theta } \to \mathbb{R}$ such that: 
\begin{equation}\label{eq:expectation}
\toi{\hat{\bs}_k} =  f(\toi{\bx_k}, \btheta) = \mathbb{E}_D[\bS \mid \bX = \toi{\bx_k}].
\end{equation}
Under the assumption that seasonality effects are likely to be nonlinear, we apply a non-parametric Nadaraya--Watson kernel regression
\citep{Nad1964, Watson1964}, which produces locally weighted conditional expected values in the neighbourhood of each feature value $\toi{\bx_k}$.
Below is a brief description of the model, as originally described in  \citet{Butler2026}.

\subsection{Kernel regression}
The Nadaraya--Watson kernel regression \citep{Nad1964, Watson1964} is a non-parametric method that models the response function, $\toi{\bs_k}$, as a 
convex combination of response values, $\{  \toi{\bs_k} \}_{t=1}^T$, in the neighbourhood of $\toi{\bx_k}$. We define a non-negative weight in the neighbourhood of
$\toi{\bx_k}$ as:
\begin{equation} \label{eq:kernel_weight}
\bomega(\toi[j]{\bx_k} - \toi{\bx_k}) = \frac{\bkappa_b( \toi[j]{\bx_k} - \toi{\bx_k}) } { \sum_{i=1}^T \bkappa_b( \toi[i]{\bx_k} - \toi{\bx_k}) },
\end{equation}
where $\bkappa_b \colon \mathbb{R}  \to \mathbb{R}$  is a kernel with bandwidth parameter $b > 0$. The final regression estimate is therefore:
\begin{equation} \label{eq:kernel_est}
\toi{\hat{\bs}_k} = \sum_{j=1}^T \bomega(\toi[j]{\bx_k} - \toi{\bx_k}) \toi[j]{\bs_k}.
\end{equation}

The kernel function, $\bkappa_b( \toi[j]{\bx_k} - \toi{\bx_k})$, assigns a non-negative weight to each observation, $\{  \toi{\bx_k} \}_{t=1}^T$, that decreases with
the distance to the observation $\toi{\bx_k}$, while the bandwidth controls the span of the local neighbourhood. Common choices of kernel function include
uniform,  $\bkappa_b(\bx) = \mathbb{I}(|| \frac{\bx}{b} || \leq 1)$, Epanechnikov, $\bkappa_b(\bx) = (1 - ||\frac{\bx}{ b}||^2) \mathbb{I}(|| \frac{\bx}{ b} || \leq 1)$ 
and Gaussian, $\bkappa_b(\bx) = \frac{1}{\sqrt{2\pi}} e^{ -0.5 ||\frac{\bx}{b}||^2}$. The choice of kernel function and bandwidth is important as it determines the
degree of nonlinearity in the model. A small bandwidth will result in a model that closely fits the data, potentially capturing fine-grained patterns 
but also being more susceptible to noise.  A large bandwidth, on the other hand, will produce a smoother model that may overlook subtle patterns.
In all experiments below, kernel regression models use a Gaussian kernel with a bandwidth of $0.50$.

\section{Experimental Results} \label{sec:results}
We consider an asset universe of $50$ global futures markets, described in Table \ref{table:universe}.
 Daily price data is provided by Commodity Systems Inc. and is available as early as January $1950$ and extends through June $2026$. 
 Each futures market is rolled on a calendar basis in order to remain invested in the nearest, most liquid available contract. 
We evaluate the strength and persistence of the seasonality anomalies through the following experimental design: for each market, feature and
forecast horizon, the seasonality model is fit using all available data prior to January $1$, $1995$. A market enters the portfolio universe after
accumulating at least $10$ years of training data. For robustness, we consider a wide range of 
forecast horizons: ${m \in \{1, 5, 21, 42, 63, 84, 105, 126\}}$, corresponding to daily, weekly and 1--6 months. Forecast horizons are paired with
features such that the forecast length is strictly less than the respective feature periodicity. For example, the day-of-week feature is paired
with daily forecast horizons, whereas the day-of-month feature is paired with daily and weekly forecast horizons, and so on. Out-of-sample performance
is evaluated by walk-forward training and testing: each model is refit every year using all available data, and model predictions are applied to the
following year's out-of-sample segment. For each forecast horizon, we apply an embargo period of $m+1$ days to ensure no overlap between training and test data.
The testing period is between January $1$, $1995$ and June $30$, $2026$. For each feature, portfolios are formed by averaging the predictions across 
each forecast horizon. Individual market positions are formed  by smoothing the forecast by an $m$-day moving average and then dividing the final market prediction 
by the $40$-day market volatility estimate. Portfolios are rebalanced on a daily basis assuming trade execution at the next-day market closing price.
Final portfolio results are scaled to a volatility of $20\%$ and are net of slippage and  commission transaction costs, 
described in more detail in Table \ref{table:universe}.

For each feature we perform a multi-factor regression analysis in order to investigate the relationship between the seasonality
factor and well-established time series factors: carry, momentum and skewness (skew). Regression analysis is performed at the sector and total 
portfolio level, where the seasonality portfolio (or sector) returns, $R_{\text{Season}}^{(t)}$, are modelled as:
\begin{equation}\label{eq:regression}
R_{\text{Season}}^{(t)} = \alpha + \beta_{\text{Carry}} \cdot R_{\text{Carry}}^{(t)} + \beta_{\text{Momentum}} \cdot R_{\text{Momentum}}^{(t)}
 + \beta_{\text{Skew}} \cdot R_{\text{Skew}}^{(t)} + \toi{\epsilon}.
\end{equation}
The other factor portfolios are rebalanced daily and are formed by inverse-volatility weighting. Following \citet{Adam2023},
the time series carry estimate of each market is evaluated by computing the pairwise average percent difference in price between each pair of futures
contracts over the 1- to 12-month term structure. The time series momentum estimate for each market is computed  as the volatility-adjusted average 
return observed over a $252$-day lookback. Similarly, the skewness feature for each market is estimated using the empirical 
skewness value observed over a $252$-day lookback. We begin our analysis by first considering each seasonality feature individually, as described in 
Section \ref{sec:method}. We then consider a seasonality ensemble factor portfolio, formed by averaging across each feature and forecast horizon pair.
We conclude by measuring the value-add of including seasonality as an equal-weight component in a global factor portfolio.


\subsection{Seasonality Anomaly: Day-of-Week} \label{sec:day_of_week}
\begin{figure}[H]
  \centering
  \begin{subfigure}[b]{0.45\linewidth}
    \includegraphics[width=\linewidth ]{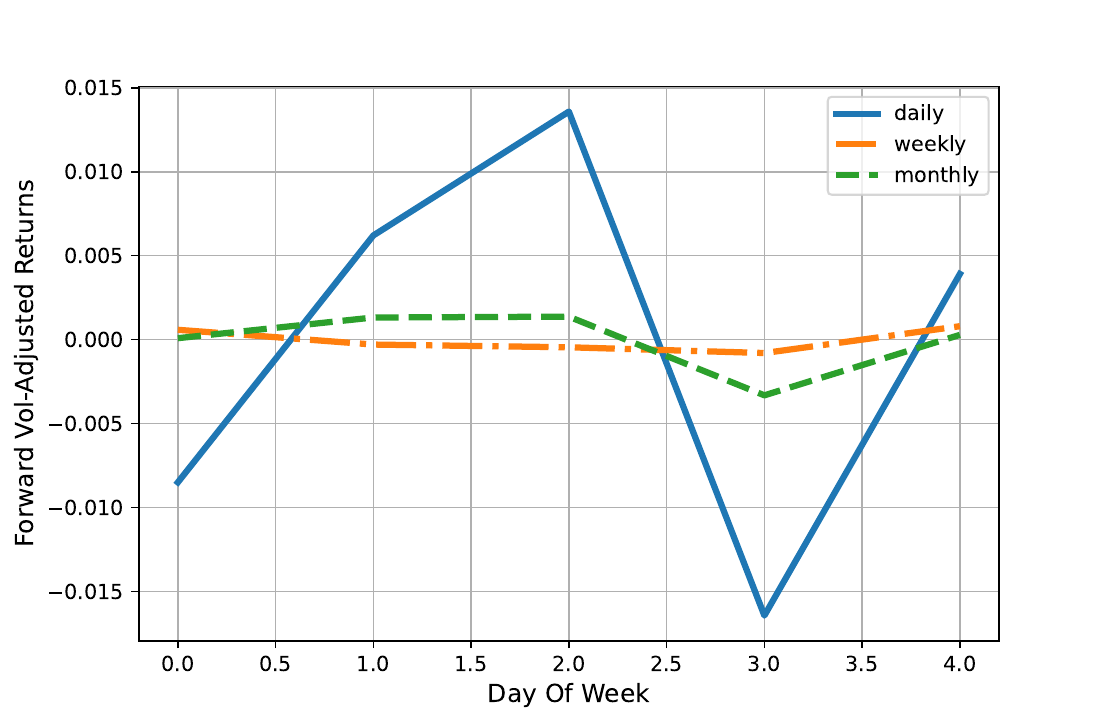}
    \caption{Bond Sector}
  \end{subfigure}
  \begin{subfigure}[b]{0.45\linewidth}
    \includegraphics[width=\linewidth ]{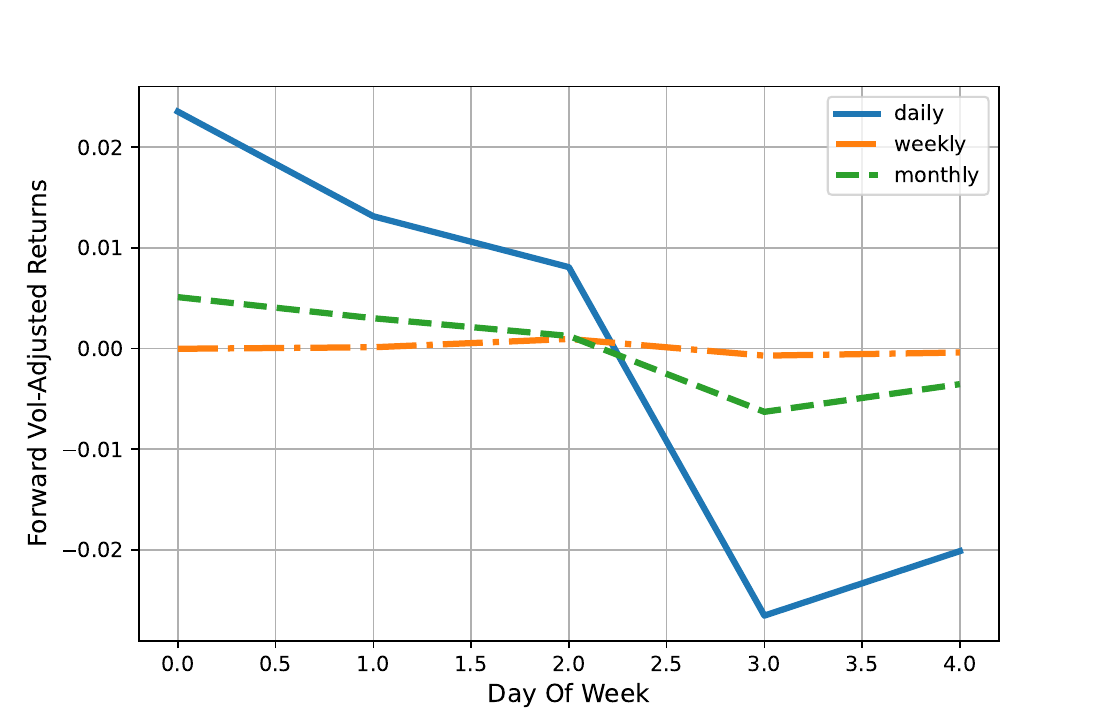}
    \caption{Commodity Sector}
  \end{subfigure}
  \begin{subfigure}[b]{0.45\linewidth}
    \includegraphics[width=\linewidth ]{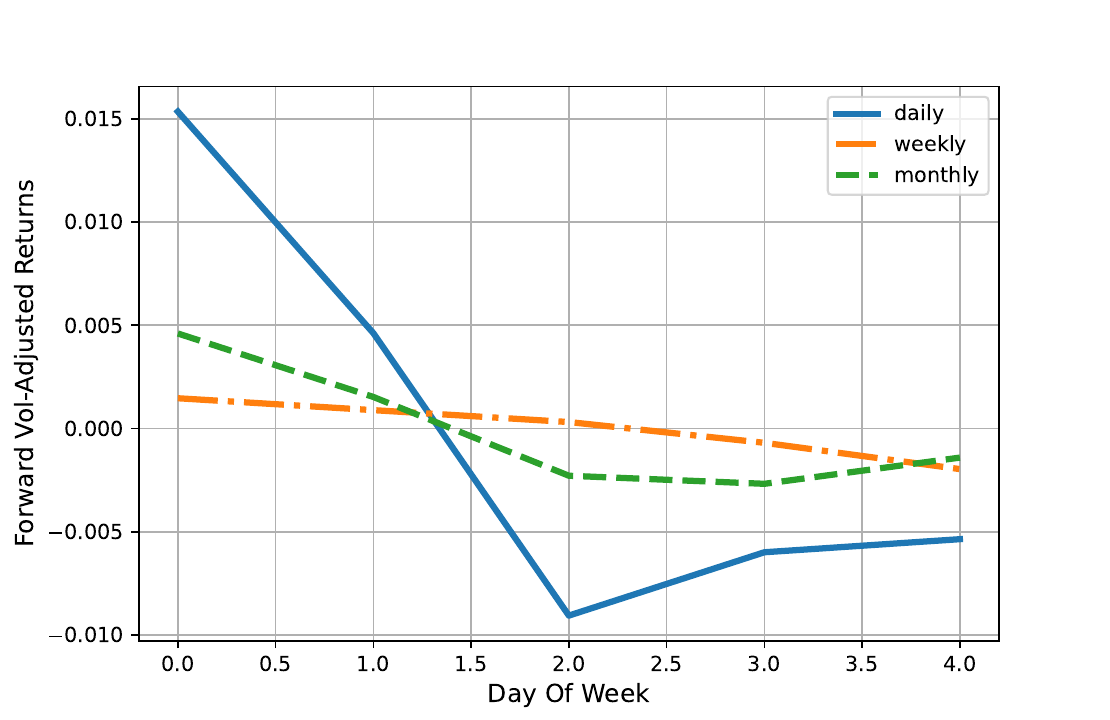}
    \caption{Currency Sector}
  \end{subfigure}
  \begin{subfigure}[b]{0.45\linewidth}
    \includegraphics[width=\linewidth ]{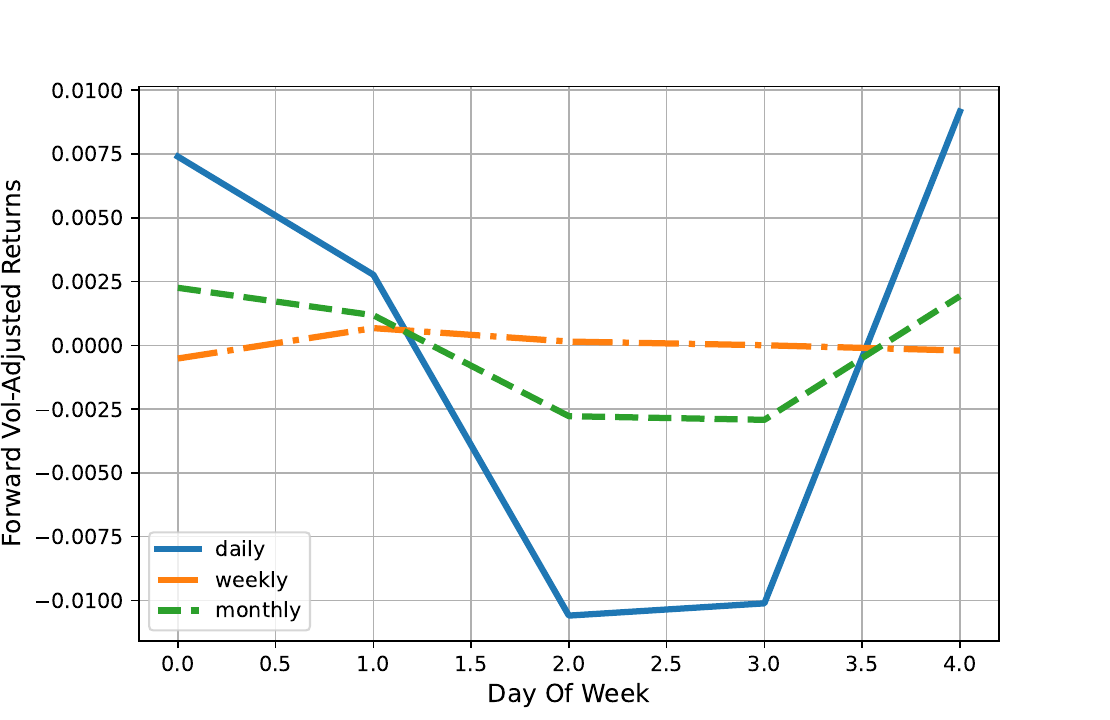}
    \caption{Equity Sector}
  \end{subfigure}
  
  \caption{Estimated forward seasonality response as a function of the day-of-week seasonality factor. 
  Models are fit using all available training data from January $1950$ through June $2026$. }
  \label{fig:day-of-week}
\end{figure}

Figure \ref{fig:day-of-week} reports the estimated forward response as a function of the day-of-week seasonality feature. For brevity, the response 
functions are pooled and fit across bond, commodity, currency and equity sectors. For completeness, a full catalog of all individual market models
is presented in \ref{sec:appendix}.  First, as expected, we note that the day-of-week feature produces near-zero
response predictions for any forecast horizon greater than one day, and thus provides justification for rejecting feature and forecast horizon pairs 
where the forecast length is greater than the feature periodicity. Recall that the feature values (0--4) represent Monday--Friday and 
that trade execution is performed at the close of the day after signal generation. We observe a common seasonality pattern shared across each
sector, namely, that the forecast on Thursday is negative. In other words, there is a consistent end-of-week phenomenon that seeks to go short at the 
close on Friday, and therefore the portfolio is positioned net-short at the open on Monday. Moreover, all markets seek to quickly reverse 
the short signal by going long at the close on either Monday or Tuesday and remaining long until mid-week. We hypothesize that this pervasive
day-of-week phenomenon may plausibly reflect a weekend effect, whereby investors face two days of unhedgeable information accumulation, which is then
incorporated into market prices on Monday. New information results in uncertainty, which in turn systematically drives prices downward on average. The potential
for investors to overreact to market uncertainty, combined with their ability to hedge and re-position as markets reopen after the weekend,
is reflected in the reversal of return expectations in the days following the expected Monday decline in prices.

\begin{table}[H]
\centering
\small

\begin{tabular}{llllll}
\toprule
 & Alpha & Carry & Momentum & Skew & R2 \\
\midrule
Agriculture & 0.0508** (0.0203)  & 0.0357*** (0.0137)  & -0.0257* (0.0136)  & 0.0274** (0.0112)  & 0.0018 \\
Bond & -0.0004 (0.009)  & 0.0882*** (0.0133)  & -0.0279** (0.0123)  & 0.0443*** (0.0124)  & 0.0113 \\
Currency & 0.0055 (0.0075)  & 0.0349*** (0.0132)  & -0.0007 (0.0121)  & -0.0257** (0.0123)  & 0.0011 \\
Energy & 0.0377*** (0.0146)  & -0.0181 (0.0163)  & 0.0649*** (0.016)  & 0.013 (0.0115)  & 0.0032 \\
Equity & -0.0048 (0.0075)  & 0.0004 (0.0114)  & -0.0602*** (0.0115)  & -0.0425*** (0.0116)  & 0.0067 \\
Metal & 0.0036 (0.0067)  & 0.0188* (0.0111)  & 0.0306*** (0.0115)  & 0.0575*** (0.0116)  & 0.0053 \\
Portfolio & 0.0932*** (0.035)  & 0.0159 (0.0123)  & 0.0174 (0.0125)  & 0.0172 (0.0114)  & 0.0013 \\
\bottomrule
\end{tabular}

\caption{Multi-factor regression analysis for the day-of-week seasonality factor regressed on time series 
carry, momentum and skewness factors. The intercept term is annualized. Statistical significance
at the $0.10$, $0.05$ and $0.01$ level is denoted by $*$, $**$ and $***$, respectively. 
The analysis period is January $1995$ through June $2026$. }
\label{table:day_of_week}
\end{table}

Table \ref{table:day_of_week} reports the results of the multi-factor regression analysis, evaluated at the individual sector and the total portfolio
level. We observe no strong statistical dependency on the other time series factors, with the majority of $R^2$ values reported as less than $1\%$. The portfolio 
annualized alpha is $9.32\%$, which is statistically significant at the $1\%$ level. However, the majority of the alpha is derived from 
agriculture $(5.08\%)$ and energy $(3.77\%)$ markets, whereas all other sectors display statistically insignificant excess returns.

\begin{figure}[ht]
  \centering
  \begin{subfigure}[b]{0.49\linewidth}
    \includegraphics[trim={0.5cm 1.cm 1cm 1.cm}, clip, width=\linewidth ]{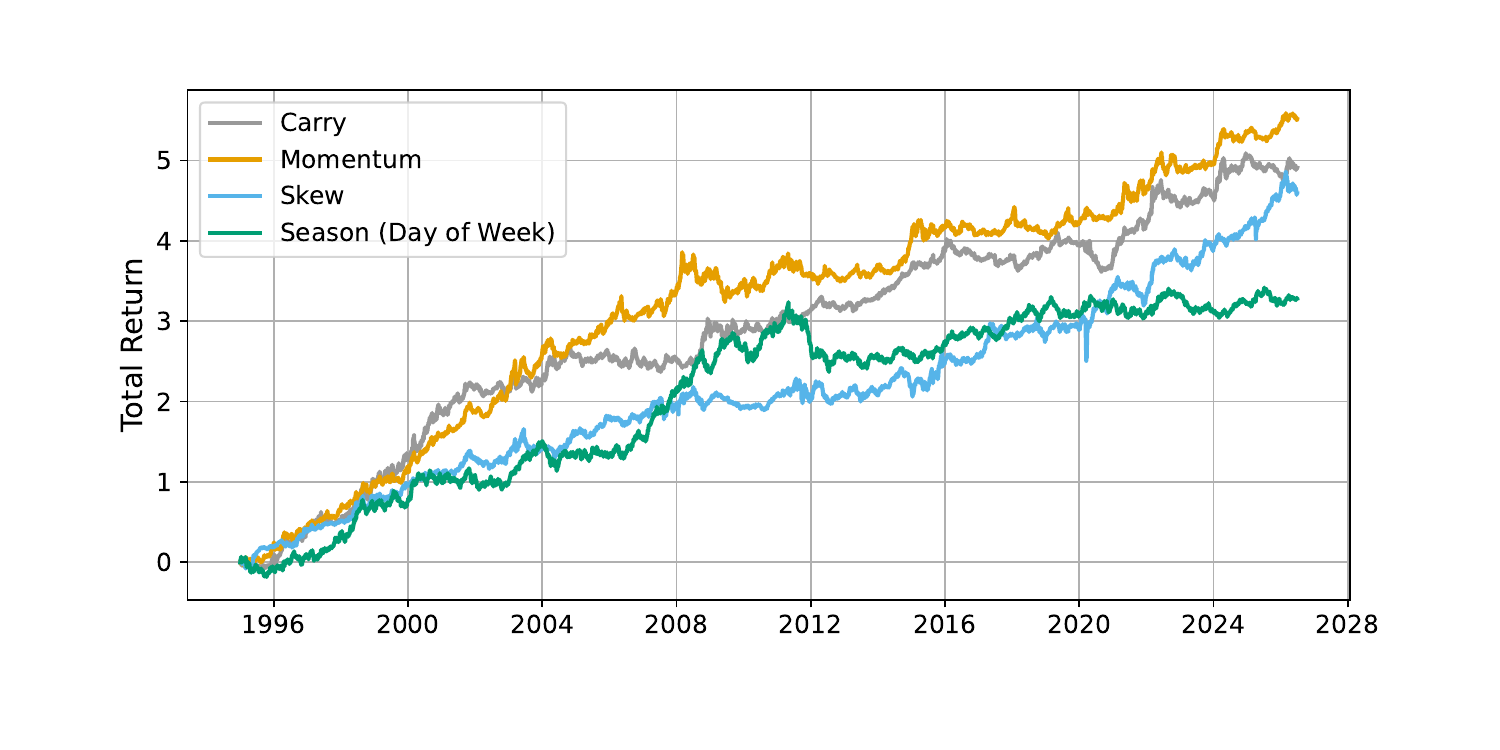}
    \caption{Gross out-of-sample total return.}
    \label{fig:day_of_week_eq_gross}
  \end{subfigure}
  \begin{subfigure}[b]{0.49\linewidth}
    \includegraphics[width=\linewidth ]{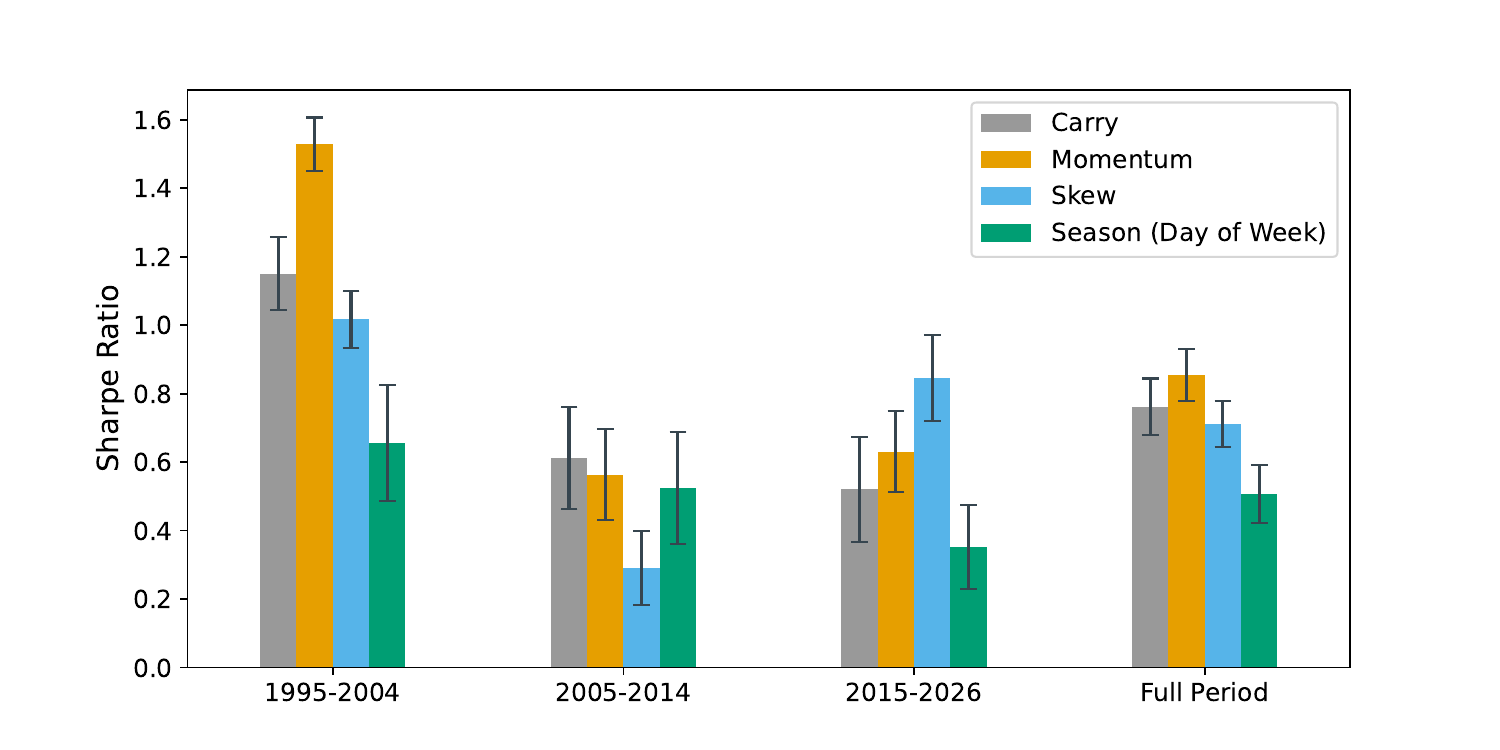}
    \caption{Gross out-of-sample Sharpe ratio.}
    \label{fig:day_of_week_sharpe_gross}
  \end{subfigure}

  \caption{Gross out-of-sample total return and Sharpe ratio for models fit using the day-of-week seasonality factor.
  The out-of-sample period is January $1995$ through June $2026$.}
  \label{fig:day_of_week_gross}
\end{figure}

\begin{figure}[ht]
  \centering
  \begin{subfigure}[b]{0.49\linewidth}
    \includegraphics[trim={0.5cm 1.cm 1cm 1.cm}, clip, width=\linewidth ]{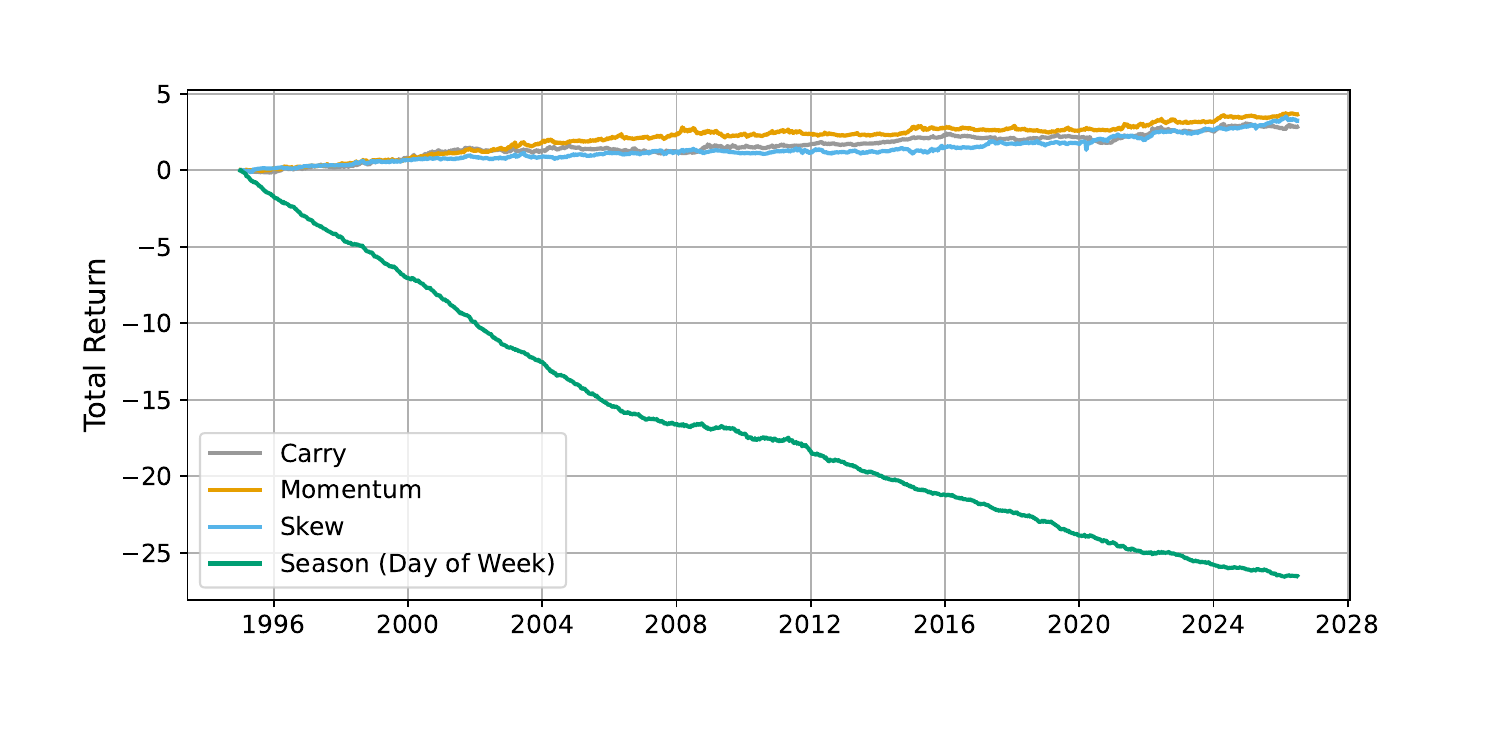}
    \caption{Net out-of-sample total return.}
    \label{fig:day_of_week_eq_net}
  \end{subfigure}
  \begin{subfigure}[b]{0.49\linewidth}
    \includegraphics[width=\linewidth ]{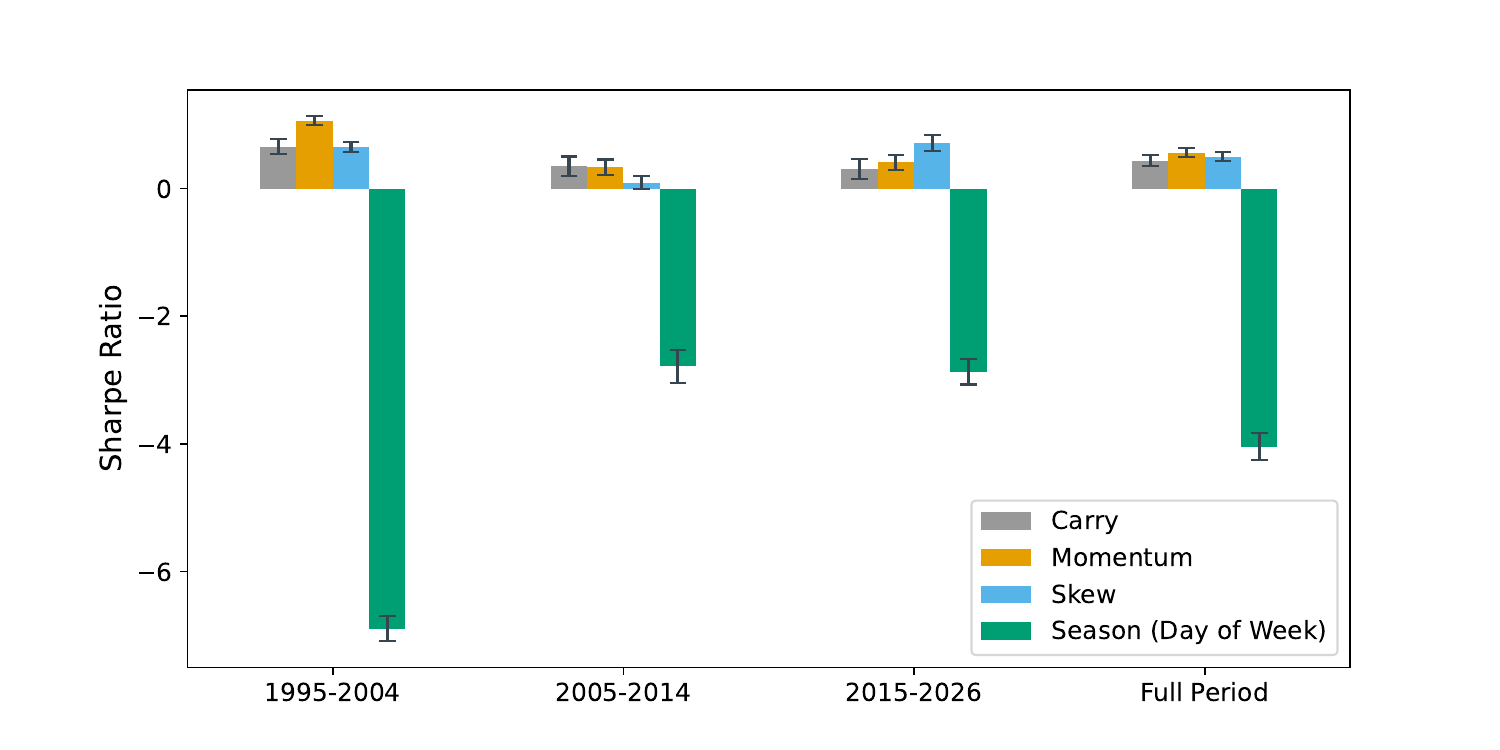}
    \caption{Net out-of-sample Sharpe ratio.}
    \label{fig:day_of_week_sharpe_net}
  \end{subfigure}

  \caption{Net out-of-sample total return and Sharpe ratio for models fit using the day-of-week seasonality factor.
  The out-of-sample period is January $1995$ through June $2026$.}
  \label{fig:day_of_week_net}
\end{figure}

Figure \ref{fig:day_of_week_gross} reports the gross out-of-sample total return and Sharpe ratio for the day-of-week seasonality factor portfolio. 
The day-of-week seasonality factor displays a gross Sharpe ratio of approximately $0.50$ over the full evaluation period,
which is significantly positive but is generally weaker than the gross Sharpe ratios of the other factor portfolios. In contrast, Figure \ref{fig:day_of_week_net}
reports a strong negative net out-of-sample total return, with a Sharpe ratio of approximately $-4.0$ over the full period. The empirical results clearly highlight that while
the day-of-week effect is significant (particularly in commodities), the turnover of the strategy is exorbitant, and therefore a naive application of the
factor strategy results in substantial transaction costs, above and beyond the expected factor alpha. 


\subsection{Seasonality Anomaly: Day-of-Month} \label{sec:day_of_month}
\begin{figure}[H]
  \centering
  \begin{subfigure}[b]{0.43\linewidth}
    \includegraphics[width=\linewidth ]{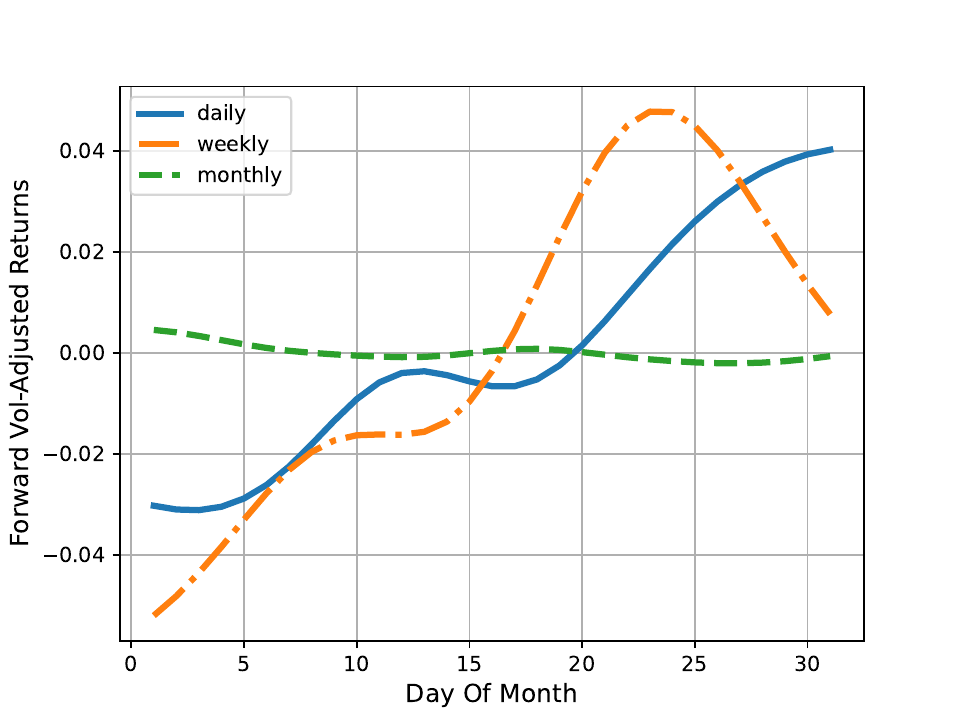}
    \caption{Bond Sector}
  \end{subfigure}
  \begin{subfigure}[b]{0.43\linewidth}
    \includegraphics[width=\linewidth ]{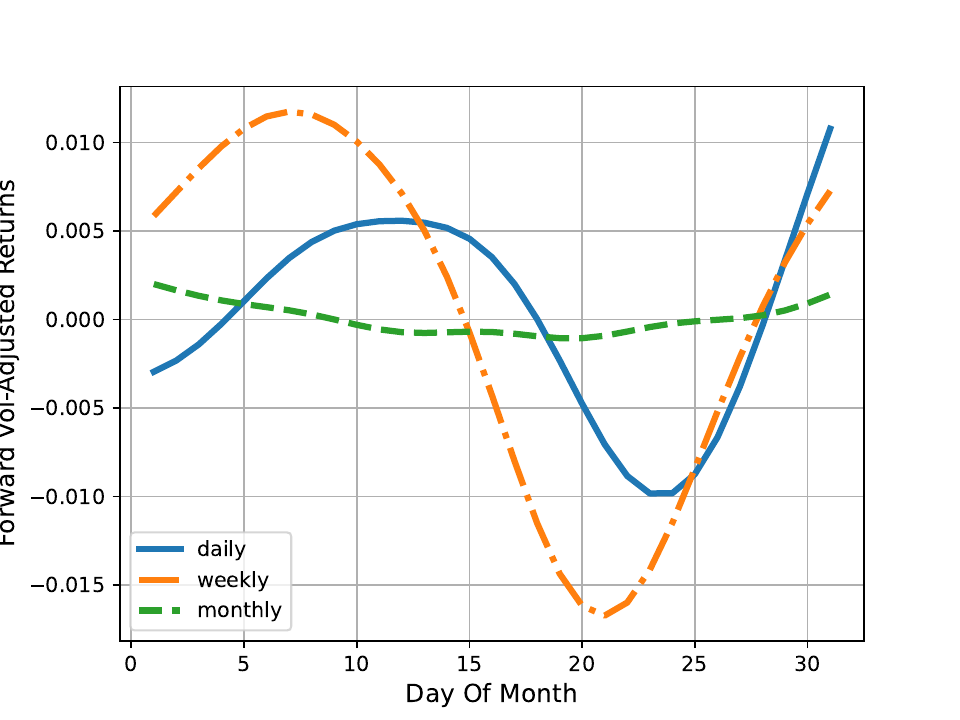}
    \caption{Commodity Sector}
  \end{subfigure}
  \begin{subfigure}[b]{0.43\linewidth}
    \includegraphics[width=\linewidth ]{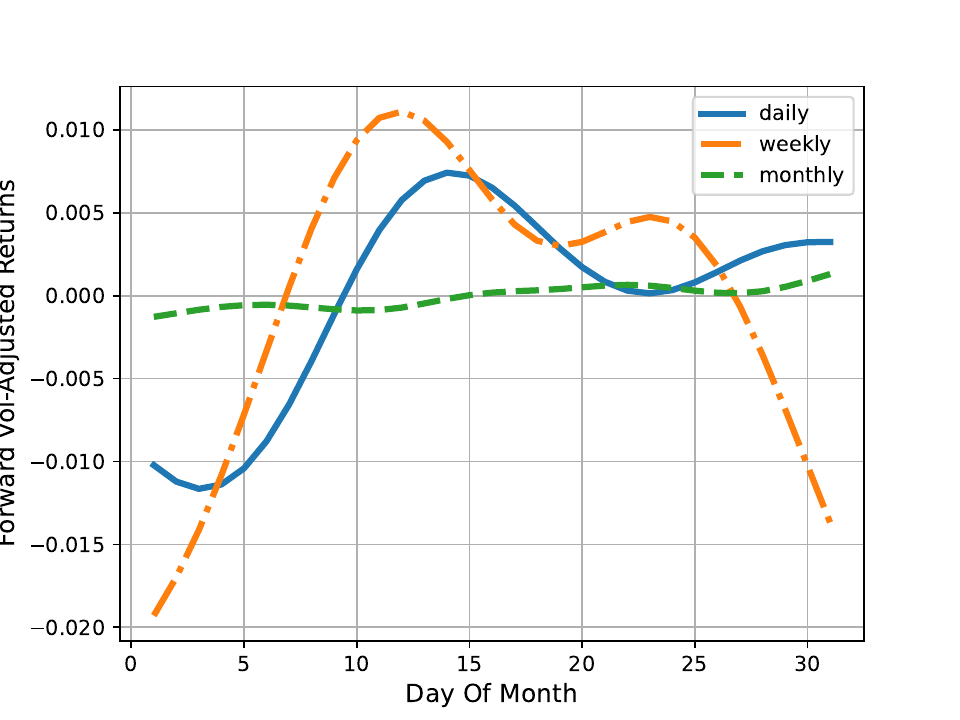}
    \caption{Currency Sector}
  \end{subfigure}
  \begin{subfigure}[b]{0.43\linewidth}
    \includegraphics[width=\linewidth ]{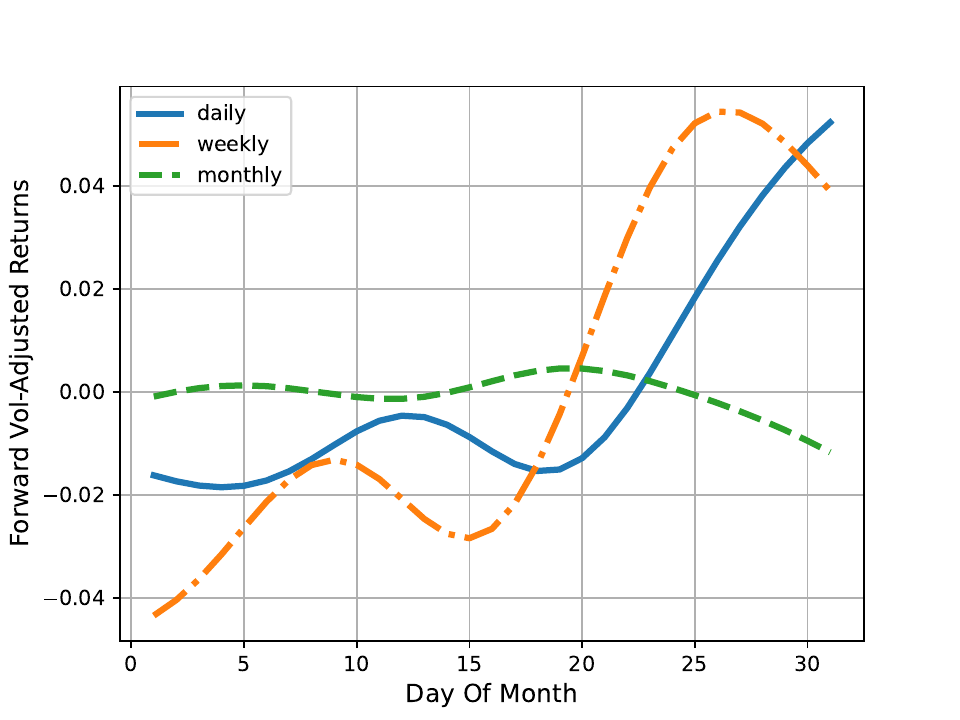}
    \caption{Equity Sector}
  \end{subfigure}
  
  \caption{Estimated forward seasonality response as a function of the day-of-month seasonality factor. 
  Models are fit using all available training data from January $1950$ through June $2026$. }
  \label{fig:day-of-month}
\end{figure}

Figure \ref{fig:day-of-month} reports the estimated forward response as a function of the day-of-month seasonality feature. With the exception 
of currency markets, all other sectors report a common end-of-month effect: markets display positive returns in the days leading up to the end of the month,
typically followed by negative returns in equity and bond markets at the beginning of each month. A potential causal explanation for the observed 
effect may relate to institution-driven policy rebalancing, in combination with routine, predictable end-of-month cash flows, that create positive demand
for assets that have underperformed and excess supply for assets that have outperformed over the month. Moreover, commodity markets display negative
expected returns around the third week of the month, potentially related to predictable commodity flows whereby commercial hedgers adjust and roll positions
coincident with commercial production flows.

Table \ref{table:day_of_month} reports the results of the multi-factor regression analysis. As before, we observe no strong statistical 
dependency on the other time series factors, with the majority of factor loadings being  less than $0.05$ in magnitude and all
$R^2$ values reported as less than $1\%$. Moreover, we report a statistically significant 
portfolio alpha of $15.38\%$, driven largely by the significant alpha in bond $(3.47\%)$ and equity $(6.52\%)$ sectors. All other sectors report
alpha values ranging from $0.71\%$ to $1.44\%$, which cumulatively account for approximately one third of the total portfolio alpha. 

\begin{figure}[ht]
  \centering
  \begin{subfigure}[b]{0.49\linewidth}
    \includegraphics[trim={0.5cm 1.cm 1cm 1.cm}, clip, width=\linewidth ]{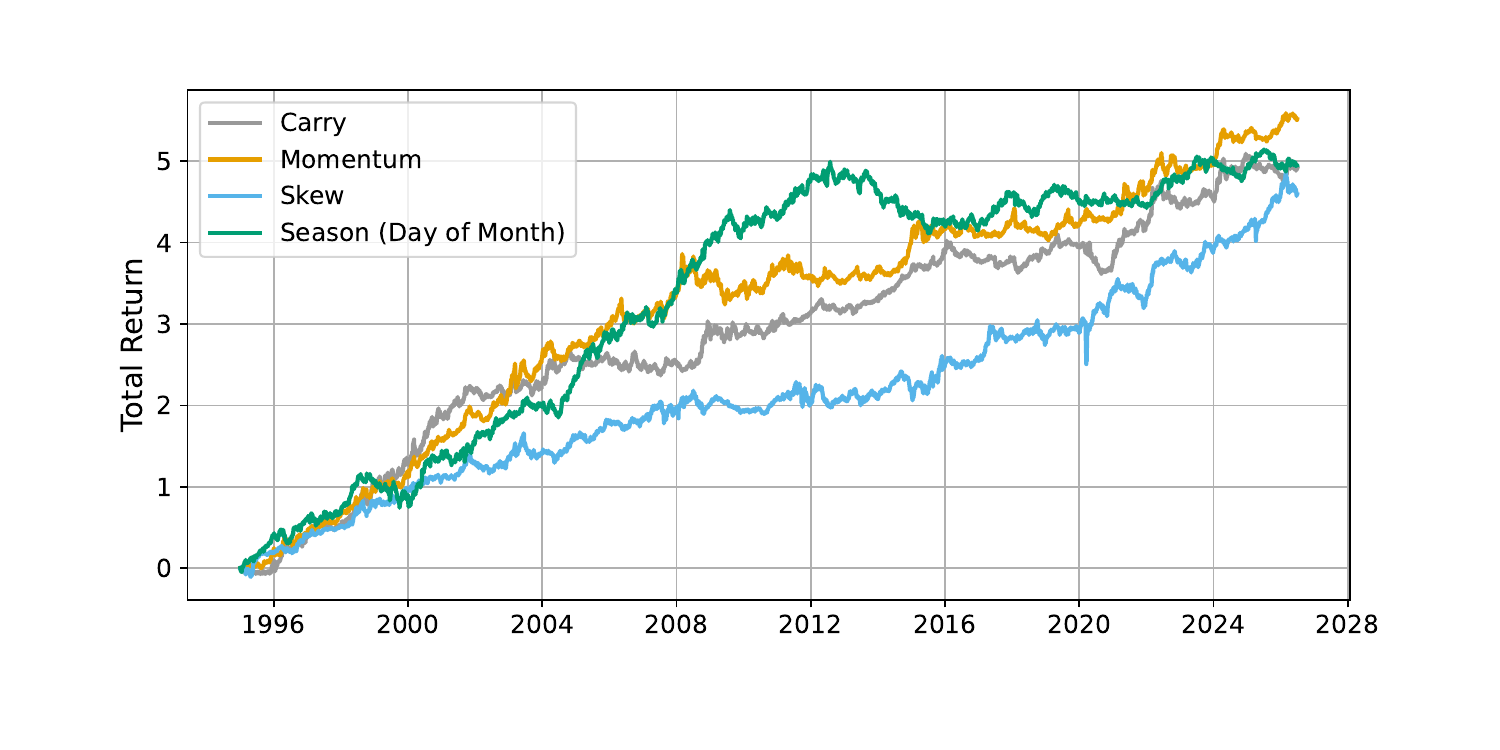}
    \caption{Gross out-of-sample total return.}
    \label{fig:day_of_month_eq_gross}
  \end{subfigure}
  \begin{subfigure}[b]{0.49\linewidth}
    \includegraphics[width=\linewidth ]{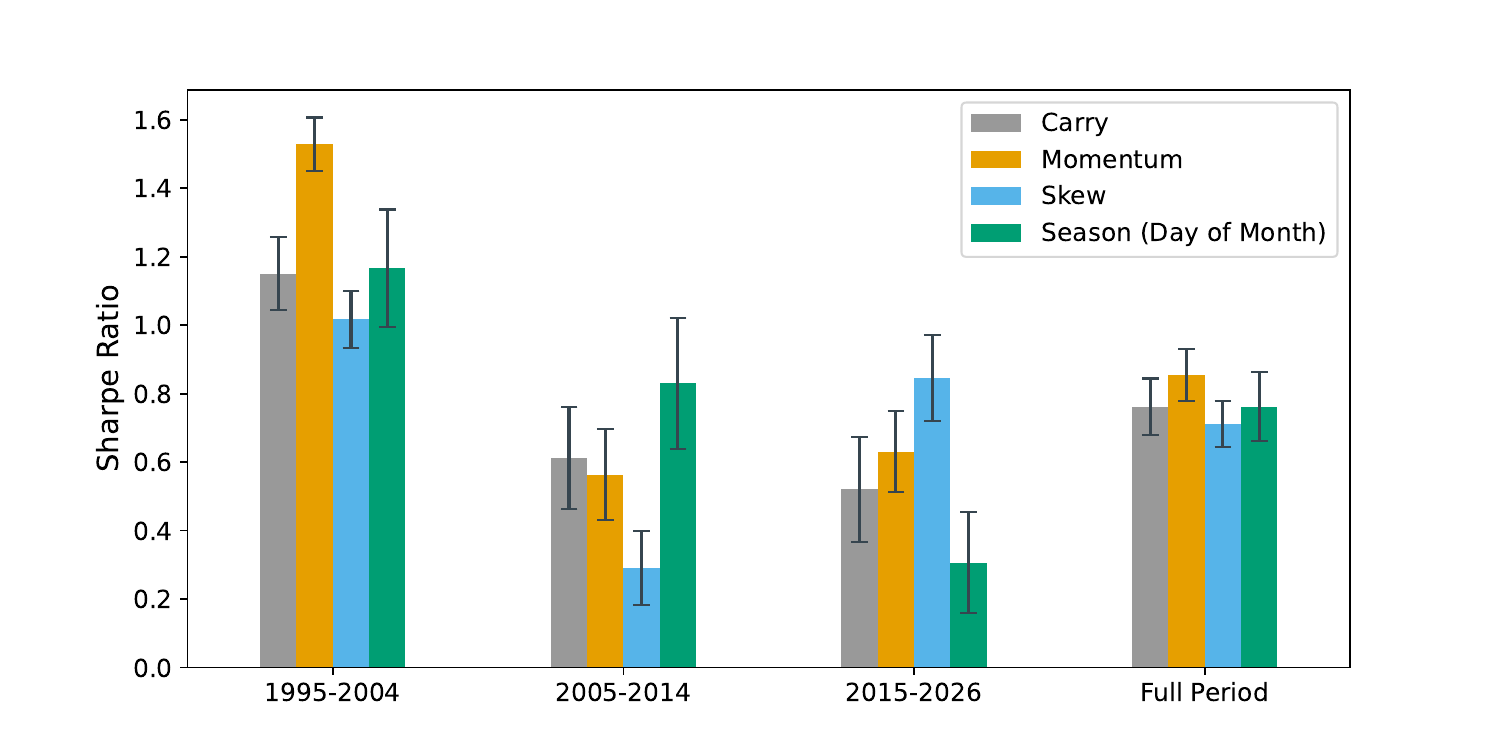}
    \caption{Gross out-of-sample Sharpe ratio.}
    \label{fig:day_of_month_sharpe_gross}
  \end{subfigure}

  \caption{Gross out-of-sample total return and Sharpe ratio for models fit using the day-of-month seasonality factor.
  The out-of-sample period is January $1995$ through June $2026$.}
  \label{fig:day_of_month_gross}
\end{figure}

\begin{figure}[ht]
  \centering
  \begin{subfigure}[b]{0.49\linewidth}
    \includegraphics[trim={0.5cm 1.cm 1cm 1.cm}, clip, width=\linewidth ]{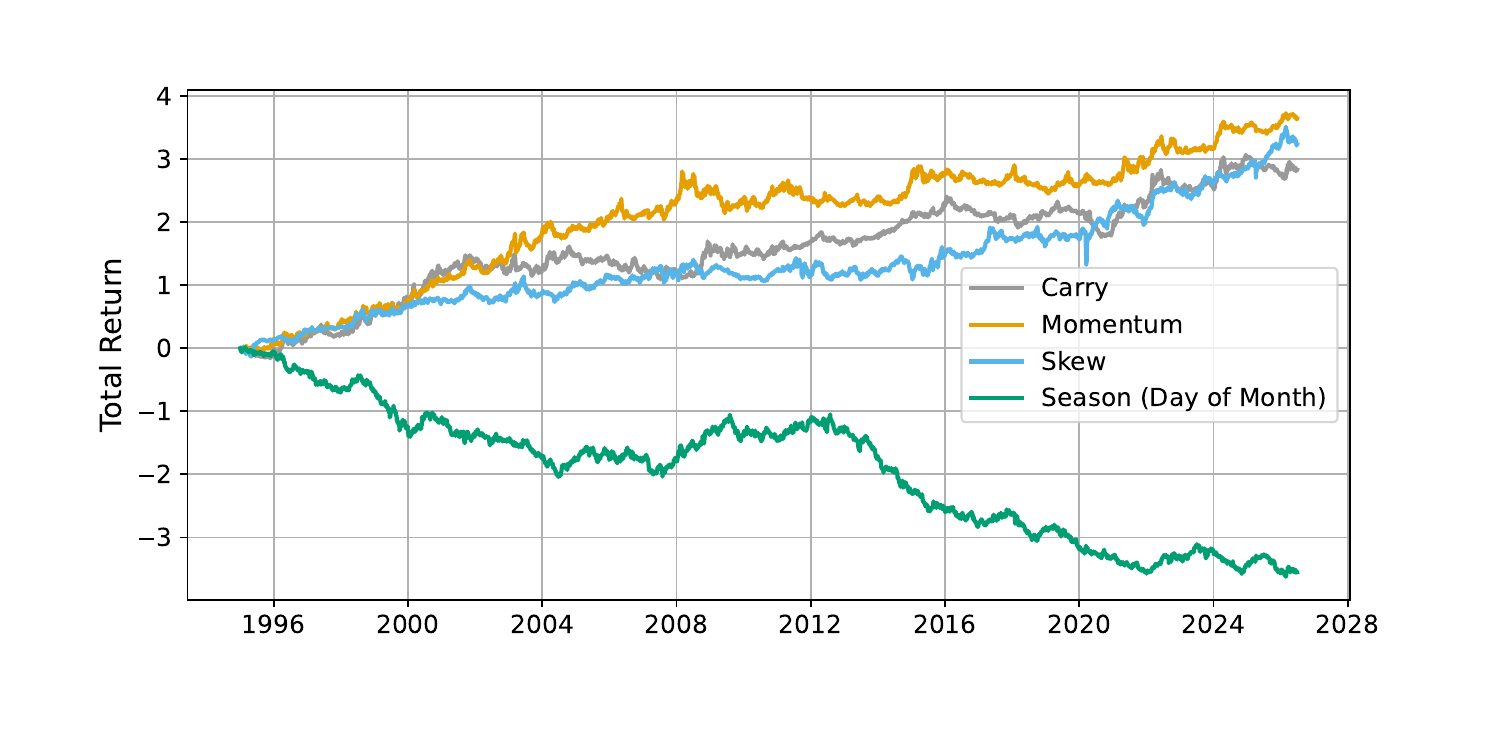}
    \caption{Net out-of-sample total return.}
    \label{fig:day_of_month_eq_net}
  \end{subfigure}
  \begin{subfigure}[b]{0.49\linewidth}
    \includegraphics[width=\linewidth ]{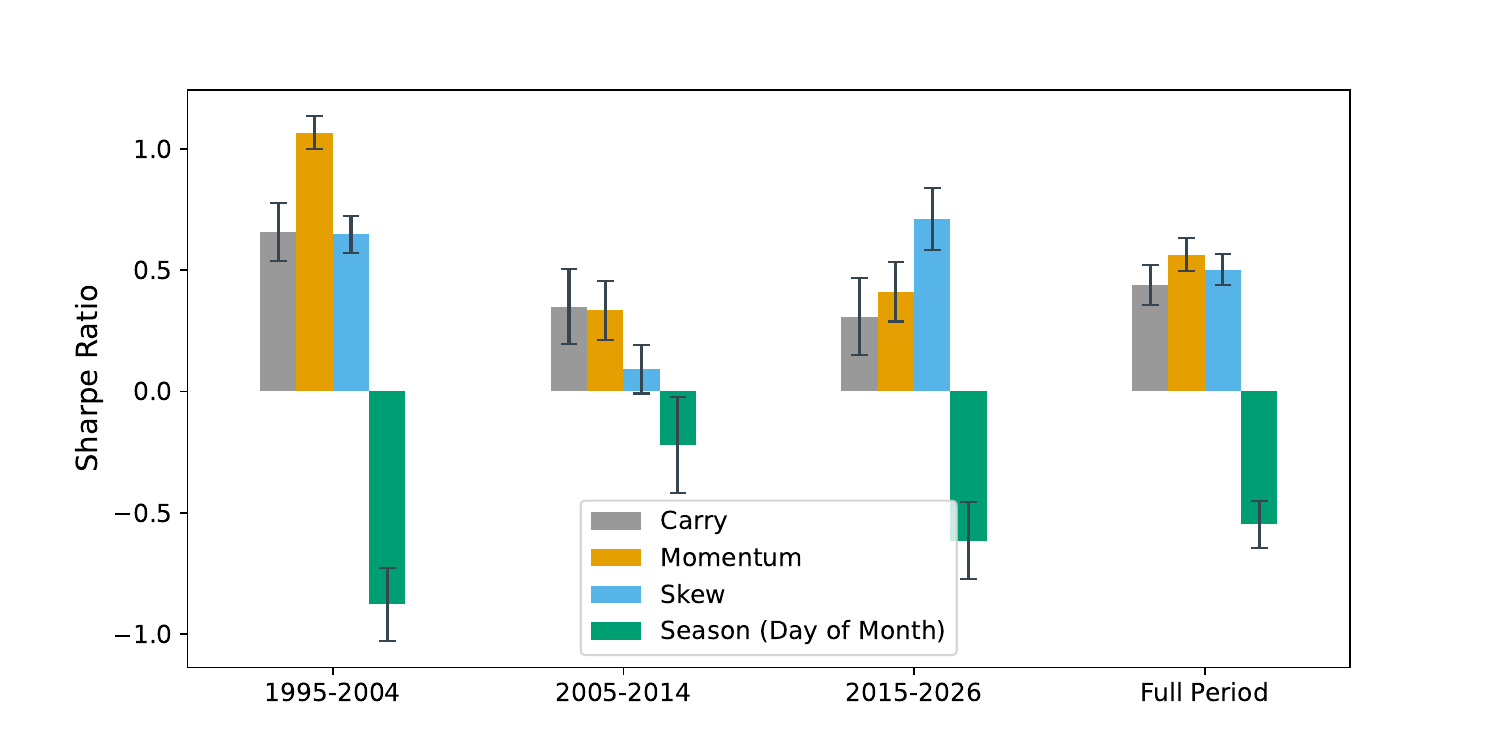}
    \caption{Net out-of-sample Sharpe ratio.}
    \label{fig:day_of_month_sharpe_net}
  \end{subfigure}

  \caption{Net out-of-sample total return and Sharpe ratio for models fit using the day-of-month seasonality factor.
  The out-of-sample period is January $1995$ through June $2026$.}
  \label{fig:day_of_month_net}
\end{figure}

Figure \ref{fig:day_of_month_gross} reports the gross out-of-sample total return and Sharpe ratio for the day-of-month seasonality factor portfolio. 
We observe a gross Sharpe ratio of approximately $0.75$ over the full evaluation period,
which is competitive with the gross Sharpe ratios of the other factor portfolios. However, as before, Figure \ref{fig:day_of_month_net} reports a strong negative
net total return and net Sharpe ratio of approximately $-0.50$, again highlighting the difficulty of capturing the seasonal factor returns in 
the presence of transaction cost frictions.

\begin{table}[H]
\centering
\small

\begin{tabular}{llllll}
\toprule
 & Alpha & Carry & Momentum & Skew & R2 \\
\midrule
Agriculture & 0.0107 (0.0119)  & -0.032** (0.0137)  & 0.0249* (0.0136)  & 0.0293*** (0.0112)  & 0.0014 \\
Bond & 0.0347* (0.0185)  & 0.0778*** (0.0133)  & 0.0194 (0.0123)  & -0.0077 (0.0125)  & 0.0072 \\
Currency & 0.0144 (0.0092)  & -0.008 (0.0131)  & -0.0559*** (0.0121)  & 0.055*** (0.0123)  & 0.0048 \\
Energy & 0.0126 (0.0095)  & -0.0054 (0.0163)  & -0.0124 (0.016)  & 0.0503*** (0.0115)  & 0.0024 \\
Equity & 0.0652** (0.0268)  & 0.0224** (0.0114)  & 0.0364*** (0.0115)  & -0.0566*** (0.0116)  & 0.0037 \\
Metal & 0.0071 (0.0065)  & 0.0103 (0.0111)  & 0.0172 (0.0115)  & -0.0329*** (0.0116)  & 0.0012 \\
Portfolio & 0.1538*** (0.0351)  & 0.0103 (0.0123)  & 0.0036 (0.0125)  & -0.0262** (0.0114)  & 0.0007 \\
\bottomrule
\end{tabular}

\caption{Multi-factor regression analysis for the day-of-month seasonality factor regressed on time series 
carry, momentum and skewness factors. The intercept term is annualized. Statistical significance
at the $0.10$, $0.05$ and $0.01$ level is denoted by $*$, $**$ and $***$, respectively. 
The analysis period is January $1995$ through June $2026$. }
\label{table:day_of_month}
\end{table}

\subsection{Seasonality Anomaly: Day-of-Quarter} \label{sec:day_of_quarter}
\begin{figure}[H]
  \centering
  \begin{subfigure}[b]{0.43\linewidth}
    \includegraphics[width=\linewidth ]{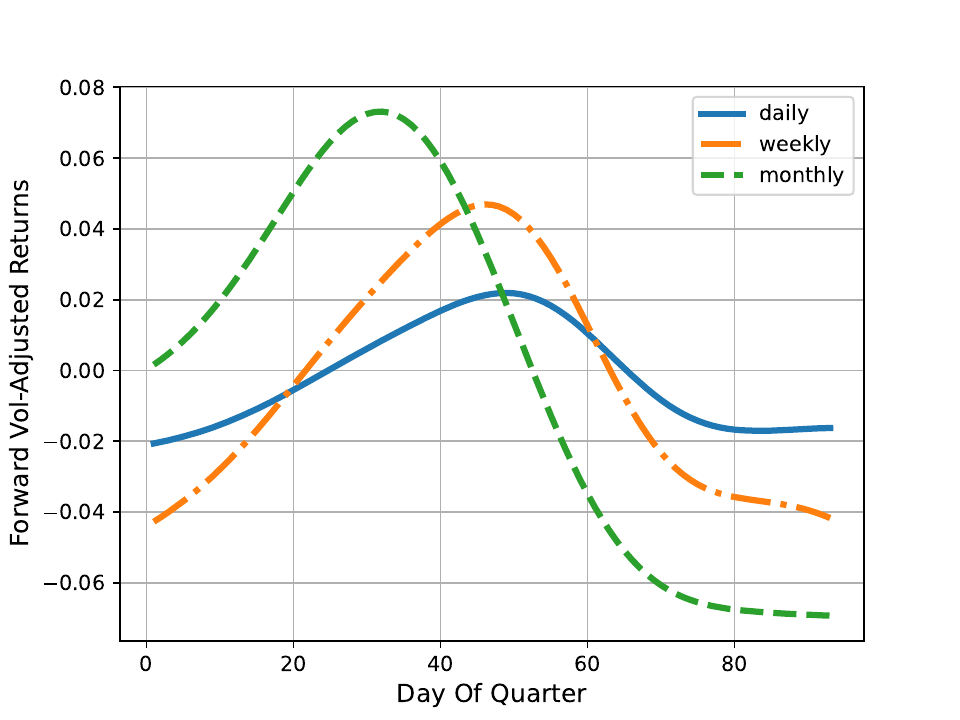}
    \caption{Bond Sector}
  \end{subfigure}
  \begin{subfigure}[b]{0.43\linewidth}
    \includegraphics[width=\linewidth ]{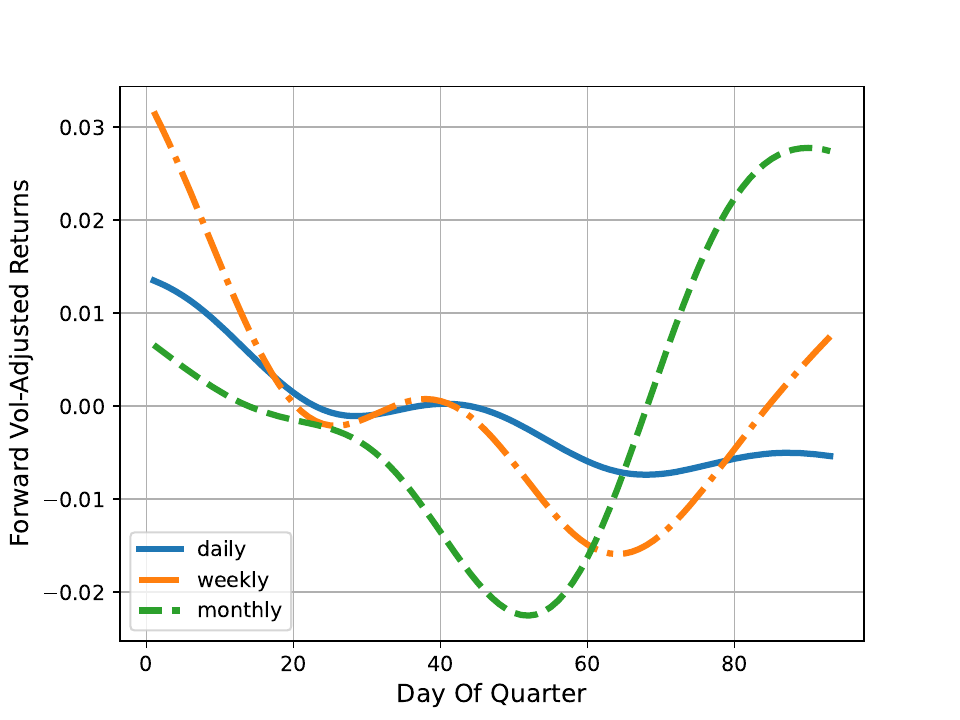}
    \caption{Commodity Sector}
  \end{subfigure}
  \begin{subfigure}[b]{0.43\linewidth}
    \includegraphics[width=\linewidth ]{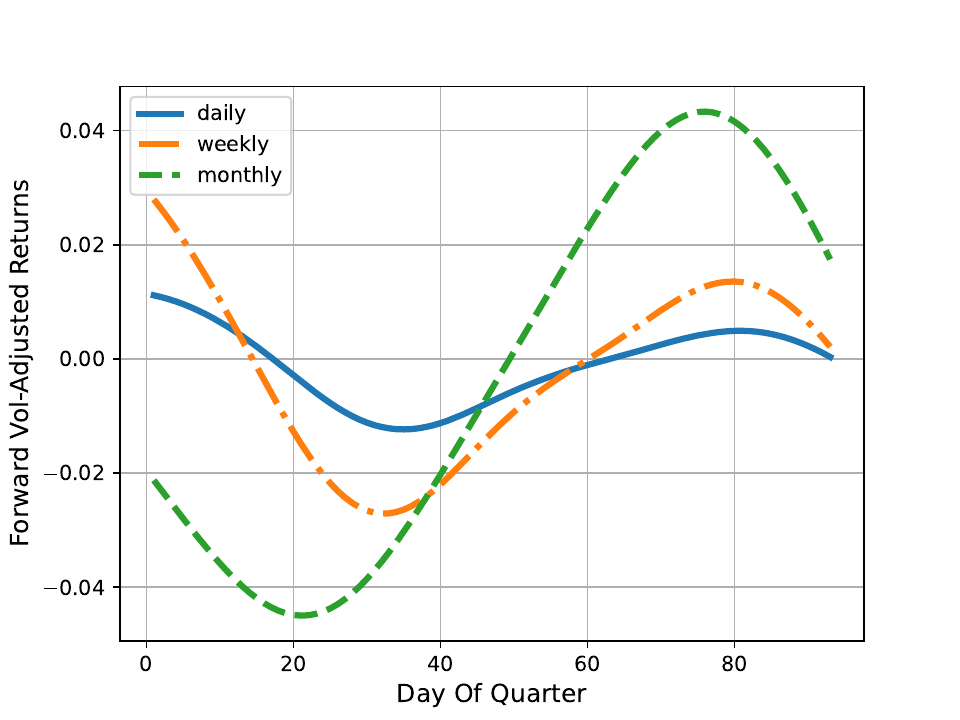}
    \caption{Currency Sector}
  \end{subfigure}
  \begin{subfigure}[b]{0.43\linewidth}
    \includegraphics[width=\linewidth ]{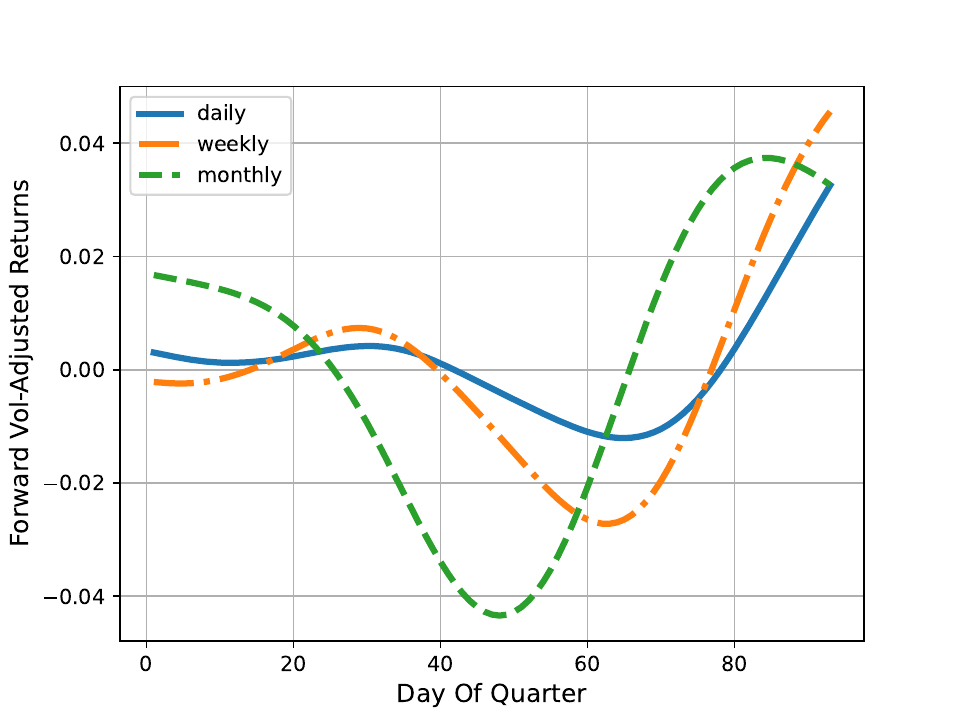}
    \caption{Equity Sector}
  \end{subfigure}
  
  \caption{Estimated forward seasonality response as a function of the day-of-quarter seasonality factor. 
  Models are fit using all available training data from January $1950$ through June $2026$. }
  \label{fig:day-of-quarter}
\end{figure}

Figure \ref{fig:day-of-quarter} reports the estimated forward response as a function of the day-of-quarter seasonality feature. In comparison to the 
day-of-week and day-of-month effects, the day-of-quarter anomaly displays more heterogeneous seasonality effects. For example, bond markets typically exhibit persistent negative
returns at quarter end, whereas equity markets display positive returns. In contrast, both commodity and currency markets exhibit positive returns 
at the beginning of the quarter, followed by intermediate weakness observed in the weeks thereafter. Table \ref{table:day_of_quarter} reports the results of the
multi-factor regression analysis for the day-of-quarter factor portfolio. Again, we observe no strong statistical 
dependency on the other time series factors, with the majority of  $R^2$ values reported as less than $1\%$. We do observe a modest (approximately $0.10$)
factor loading on the carry factor that is statistically significant across bond, currency and metal sectors. Nonetheless, we report a statistically significant
portfolio alpha of $10.87\%$, which is relatively evenly distributed across sectors, with alpha values ranging from $0.78\%$ to $2.32\%$.

\begin{figure}[ht]
  \centering
  \begin{subfigure}[b]{0.49\linewidth}
    \includegraphics[trim={0.5cm 1.cm 1cm 1.cm}, clip, width=\linewidth ]{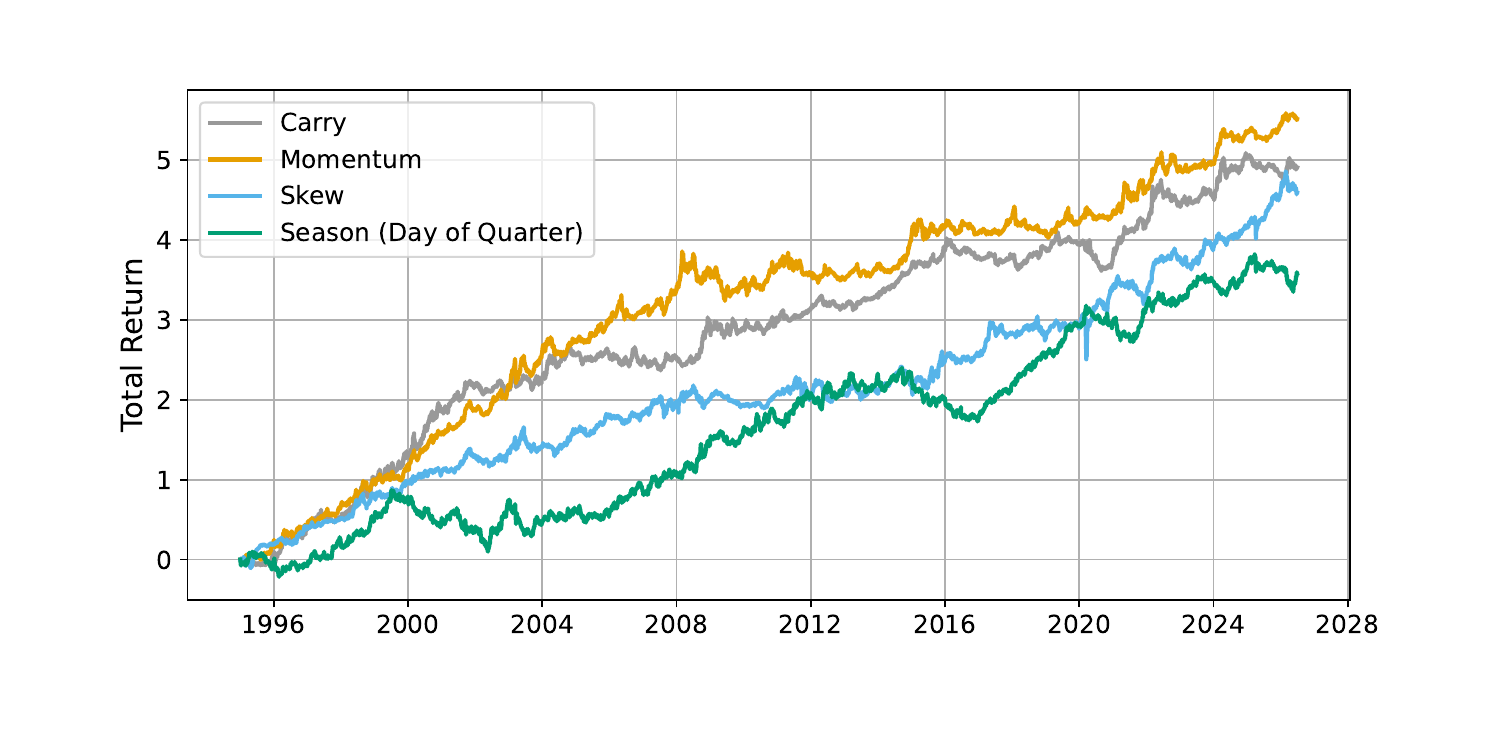}
    \caption{Gross out-of-sample total return.}
    \label{fig:day_of_quarter_eq_gross}
  \end{subfigure}
  \begin{subfigure}[b]{0.49\linewidth}
    \includegraphics[width=\linewidth ]{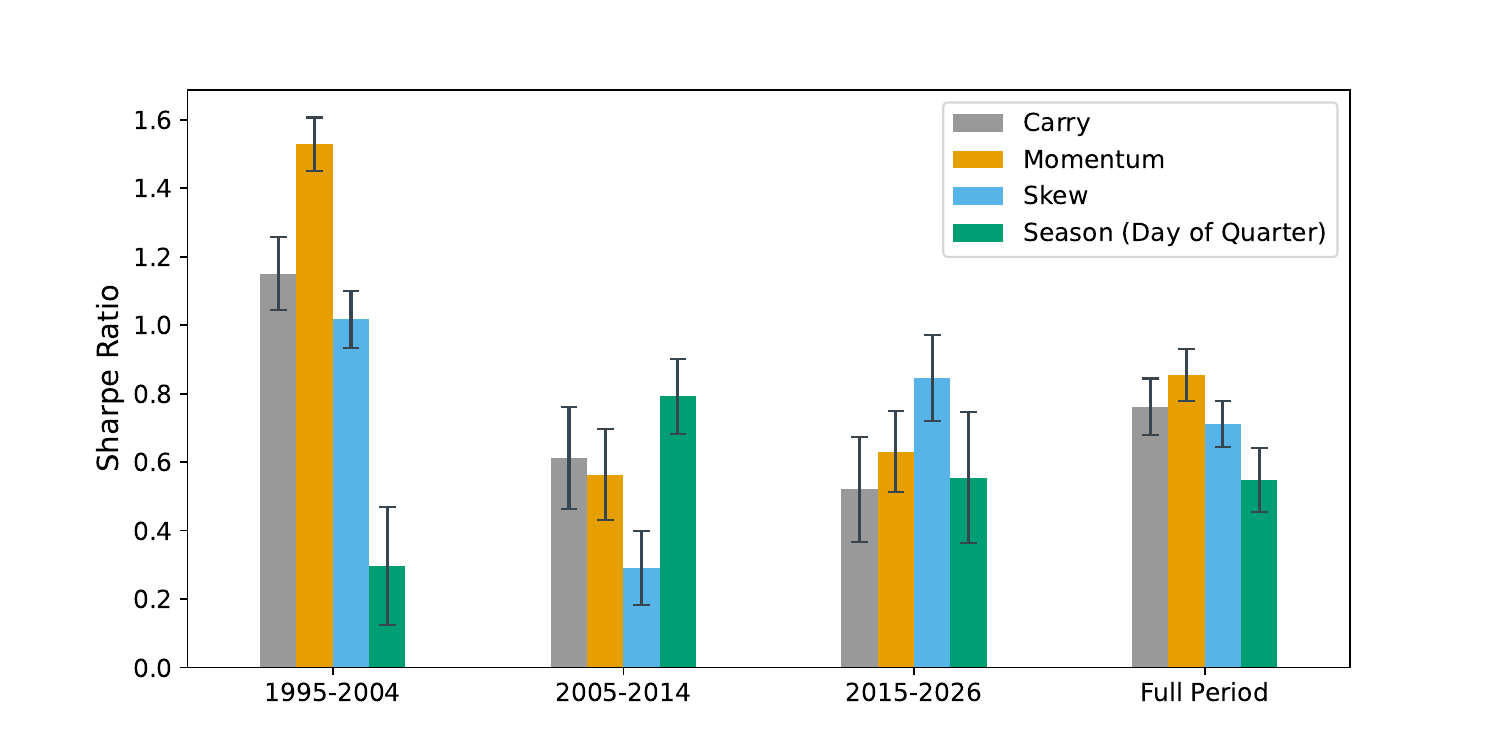}
    \caption{Gross out-of-sample Sharpe ratio.}
    \label{fig:day_of_quarter_sharpe_gross}
  \end{subfigure}

  \caption{Gross out-of-sample total return and Sharpe ratio for models fit using the day-of-quarter seasonality factor.
  The out-of-sample period is January $1995$ through June $2026$.}
  \label{fig:day_of_quarter_gross}
\end{figure}

\begin{figure}[h]
  \centering
  \begin{subfigure}[b]{0.49\linewidth}
    \includegraphics[trim={0.5cm 1.cm 1cm 1.cm}, clip, width=\linewidth ]{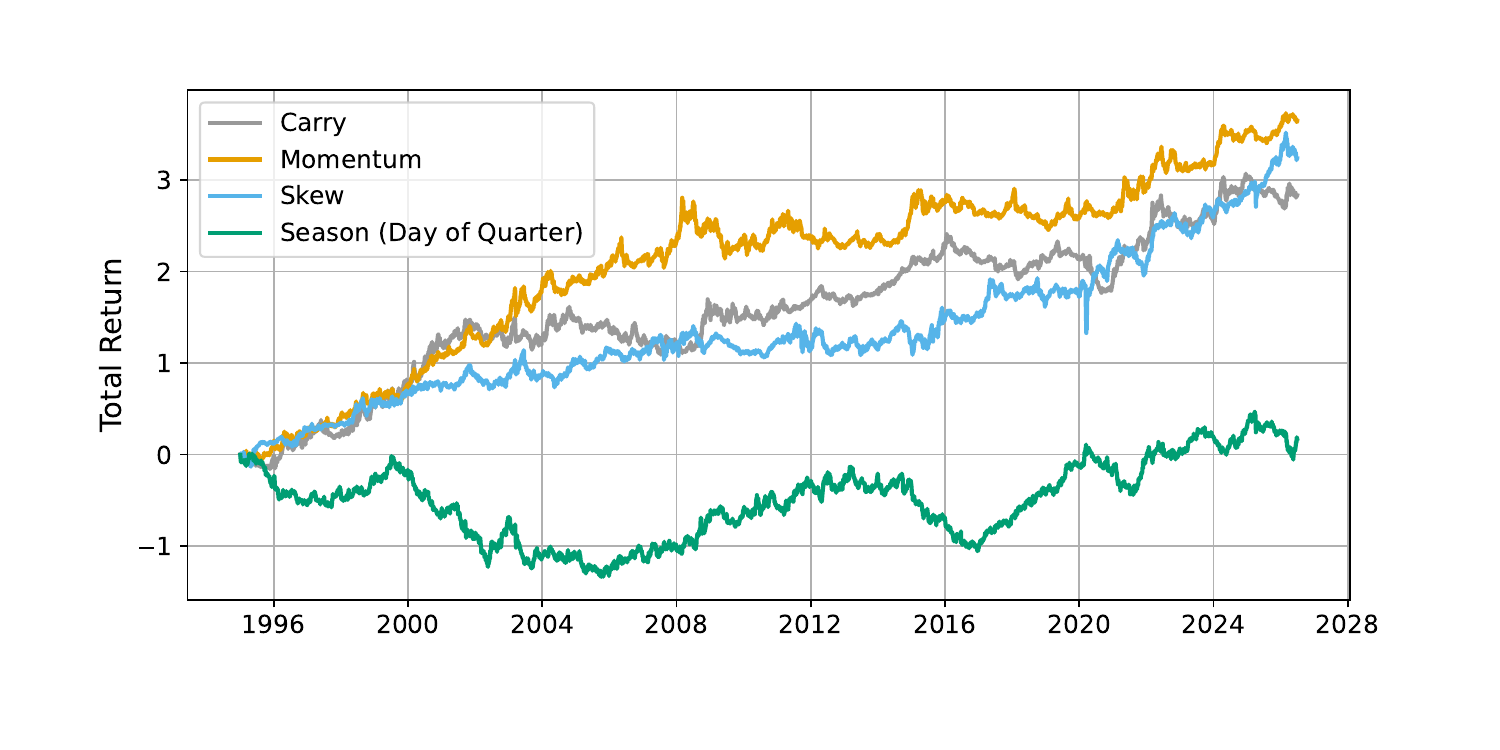}
    \caption{Net out-of-sample total return.}
    \label{fig:day_of_quarter_eq_net}
  \end{subfigure}
  \begin{subfigure}[b]{0.49\linewidth}
    \includegraphics[width=\linewidth ]{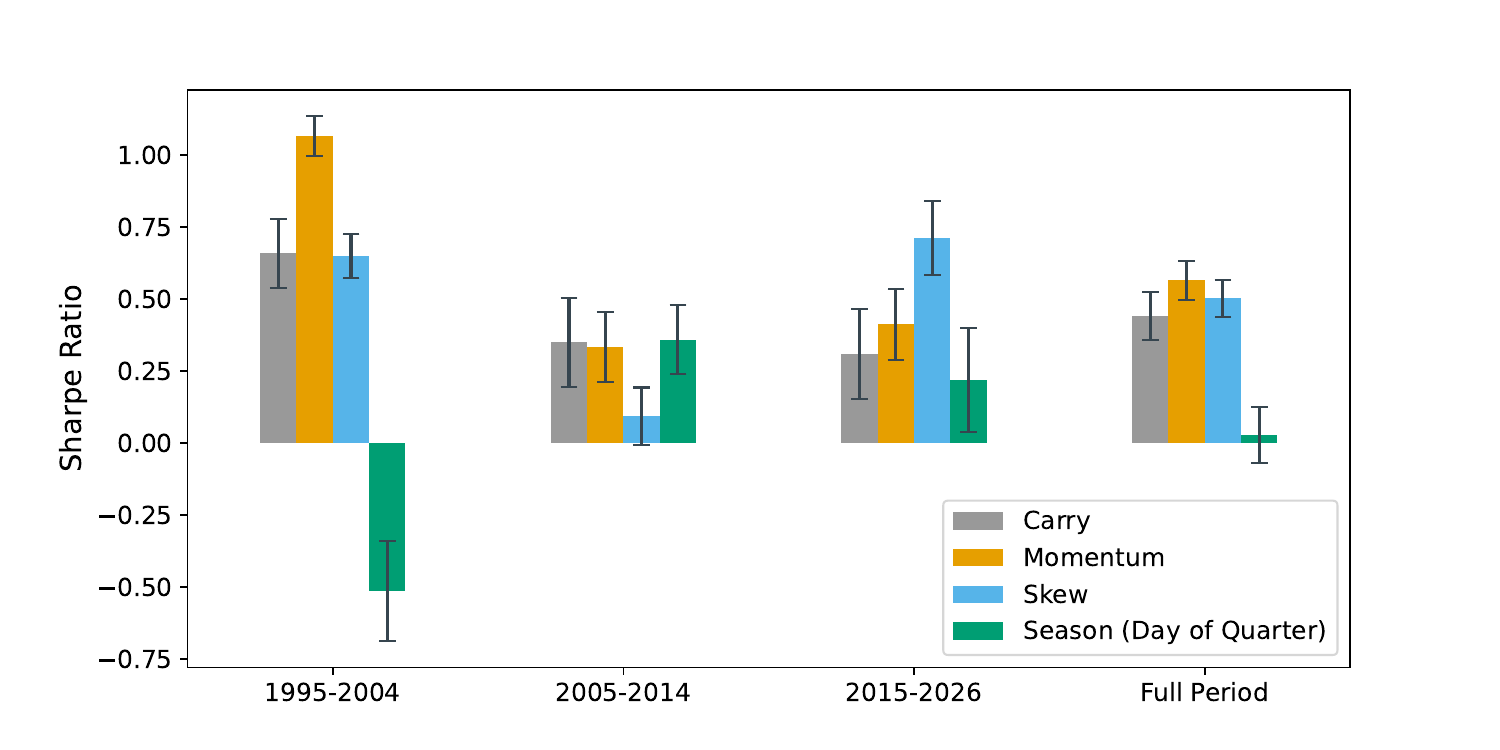}
    \caption{Net out-of-sample Sharpe ratio.}
    \label{fig:day_of_quarter_sharpe_net}
  \end{subfigure}

  \caption{Net out-of-sample total return and Sharpe ratio for models fit using the day-of-quarter seasonality factor.
  The out-of-sample period is January $1995$ through June $2026$.}
  \label{fig:day_of_quarter_net}
\end{figure}

Figure \ref{fig:day_of_quarter_gross} reports the gross out-of-sample total return and Sharpe ratio for the day-of-quarter seasonality factor portfolio. 
We observe a gross Sharpe ratio of approximately $0.51$ over the full evaluation period,
which is modestly weaker than the day-of-month factor but remains competitive with the gross Sharpe ratios of the other factor portfolios.
Figure \ref{fig:day_of_quarter_net} reports a net Sharpe ratio of approximately zero over the full period. However, we do observe a positive net Sharpe ratio of approximately $0.30$ since $2005$, 
largely attributable  to lower turnover.
\begin{table}[H]
\centering
\small

\begin{tabular}{llllll}
\toprule
 & Alpha & Carry & Momentum & Skew & R2 \\
\midrule
Agriculture & 0.0232* (0.012)  & 0.0085 (0.0137)  & -0.0133 (0.0136)  & 0.0044 (0.0112)  & 0.0001 \\
Bond & 0.0185 (0.0145)  & 0.102*** (0.0133)  & 0.0232* (0.0123)  & 0.0058 (0.0124)  & 0.0137 \\
Currency & 0.0207 (0.0137)  & 0.0953*** (0.0131)  & -0.0716*** (0.0121)  & -0.0052 (0.0123)  & 0.0084 \\
Energy & 0.0078 (0.0174)  & 0.0301* (0.0163)  & -0.0464*** (0.016)  & 0.0272** (0.0115)  & 0.0017 \\
Equity & 0.021* (0.0124)  & -0.0002 (0.0114)  & 0.0205* (0.0115)  & -0.0121 (0.0116)  & 0.0004 \\
Metal & 0.011 (0.0071)  & 0.0803*** (0.0111)  & -0.037*** (0.0115)  & -0.0431*** (0.0115)  & 0.0115 \\
Portfolio & 0.1087*** (0.0354)  & 0.0057 (0.0123)  & -0.0149 (0.0125)  & 0.022* (0.0114)  & 0.0005 \\
\bottomrule
\end{tabular}

\caption{Multi-factor regression analysis for the day-of-quarter seasonality factor regressed on time series 
carry, momentum and skewness factors. The intercept term is annualized. Statistical significance
at the $0.10$, $0.05$ and $0.01$ level is denoted by $*$, $**$ and $***$, respectively. 
The analysis period is January $1995$ through June $2026$. }
\label{table:day_of_quarter}
\end{table}

\subsection{Seasonality Anomaly: Day-of-Half-Year} \label{sec:day_of_half_year}
\begin{figure}[H]
  \centering
  \begin{subfigure}[b]{0.43\linewidth}
    \includegraphics[width=\linewidth ]{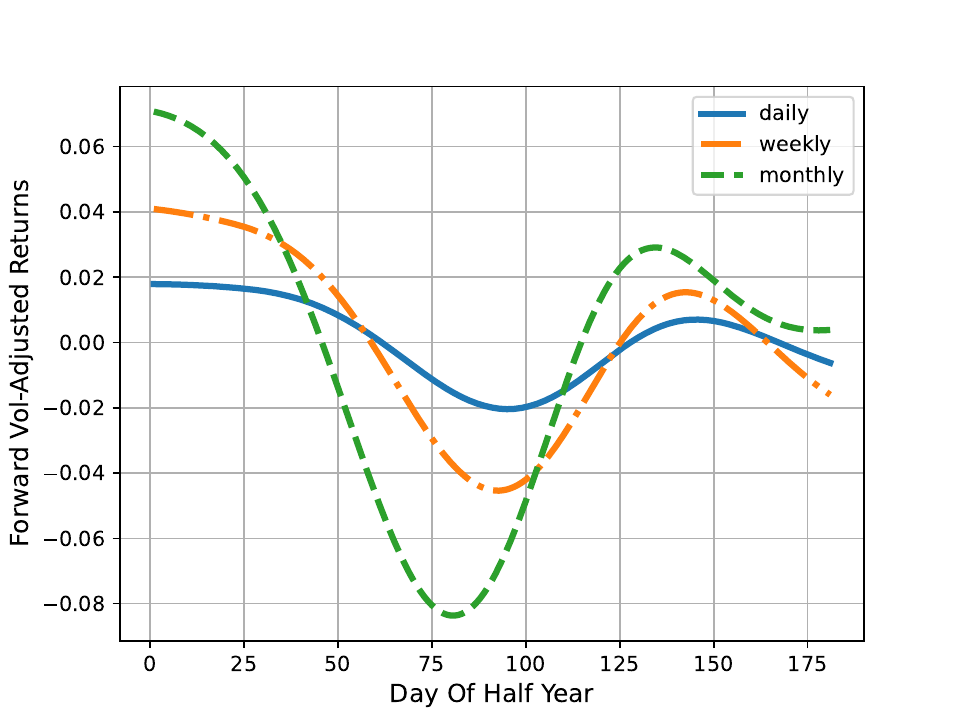}
    \caption{Bond Sector}
  \end{subfigure}
  \begin{subfigure}[b]{0.43\linewidth}
    \includegraphics[width=\linewidth ]{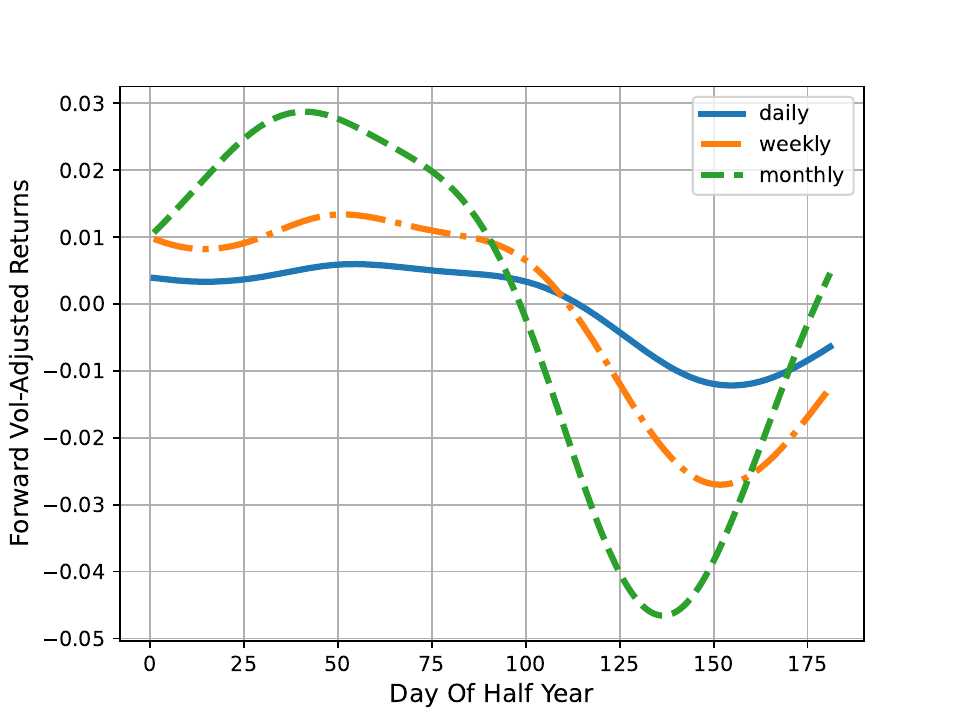}
    \caption{Commodity Sector}
  \end{subfigure}
  \begin{subfigure}[b]{0.43\linewidth}
    \includegraphics[width=\linewidth ]{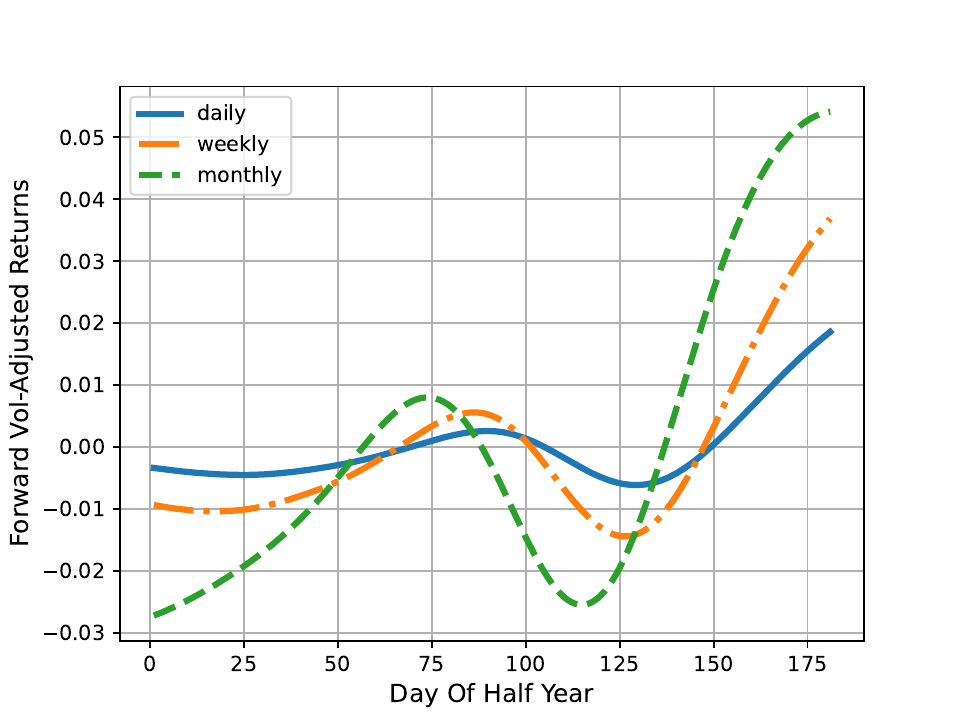}
    \caption{Currency Sector}
  \end{subfigure}
  \begin{subfigure}[b]{0.43\linewidth}
    \includegraphics[width=\linewidth ]{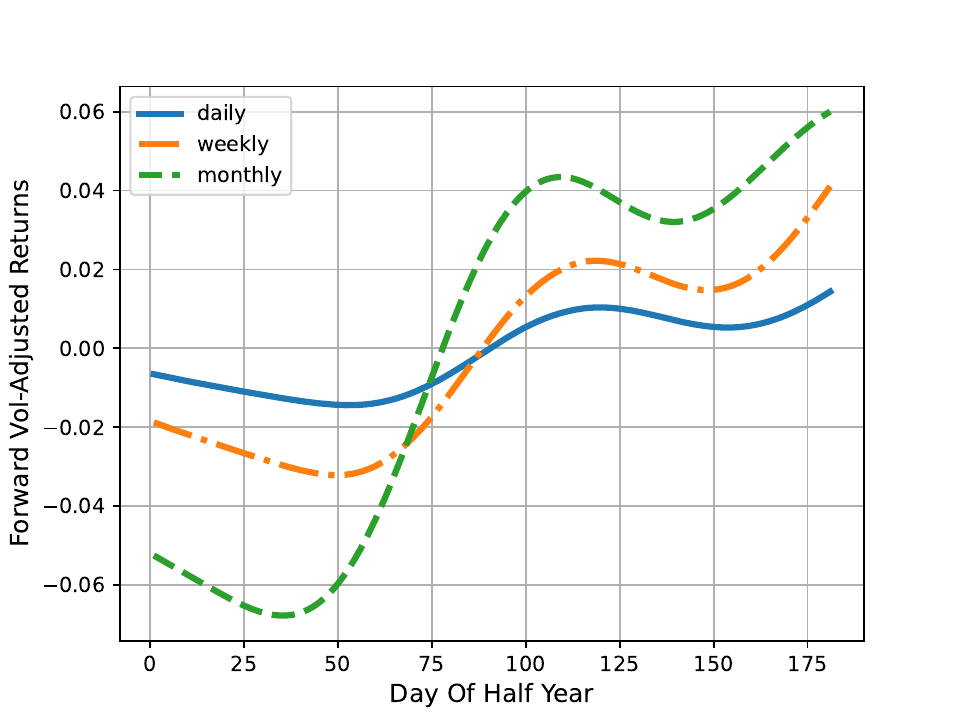}
    \caption{Equity Sector}
  \end{subfigure}
  
  \caption{Estimated forward seasonality response as a function of the day-of-half-year seasonality factor. 
  Models are fit using all available training data from January $1950$ through June $2026$. }
  \label{fig:day-of-half-year}
\end{figure}

\begin{figure}[ht]
  \centering
  \begin{subfigure}[b]{0.49\linewidth}
    \includegraphics[trim={0.5cm 1.cm 1cm 1.cm}, clip, width=\linewidth ]{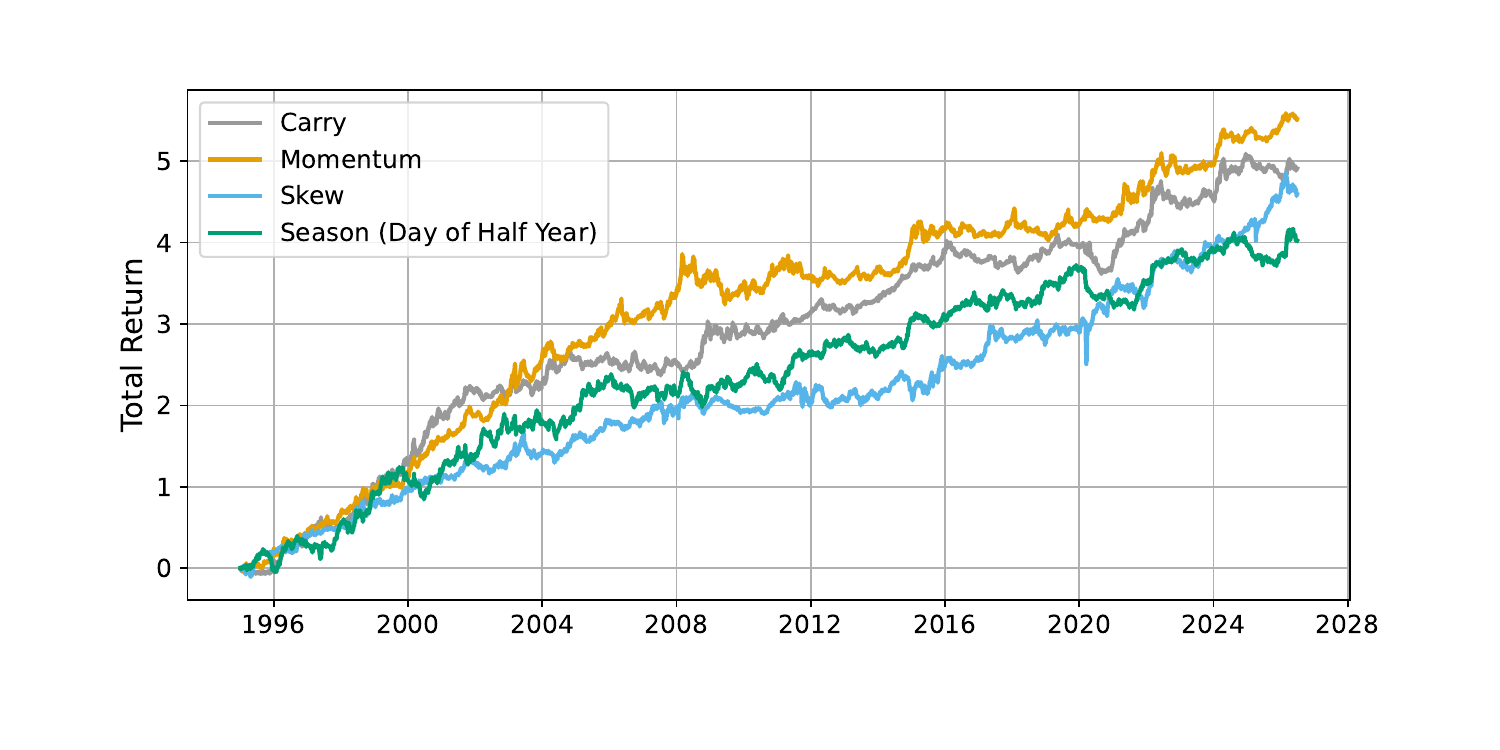}
    \caption{Gross out-of-sample total return.}
    \label{fig:day_of_half_year_eq_gross}
  \end{subfigure}
  \begin{subfigure}[b]{0.49\linewidth}
    \includegraphics[width=\linewidth ]{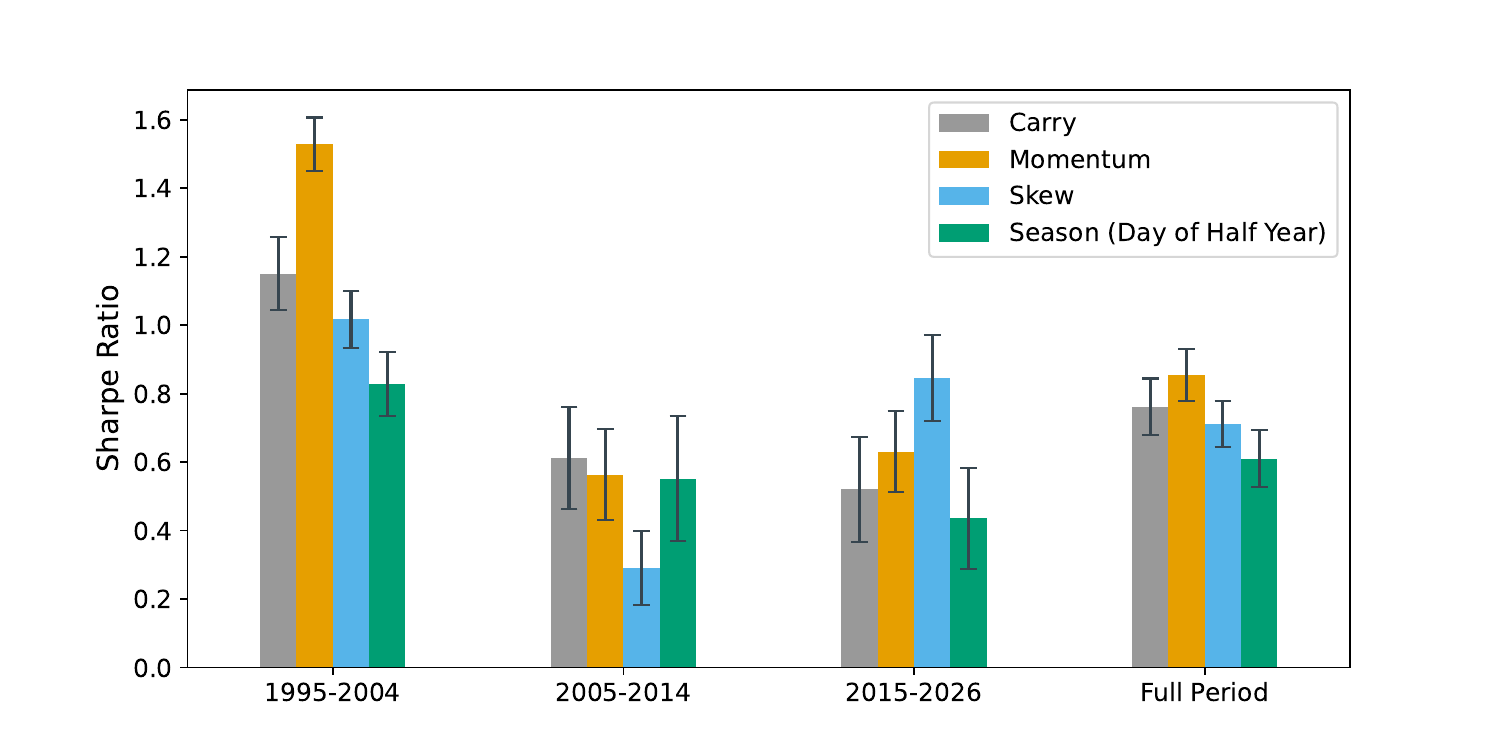}
    \caption{Gross out-of-sample Sharpe ratio.}
    \label{fig:day_of_half_year_sharpe_gross}
  \end{subfigure}

  \caption{Gross out-of-sample total return and Sharpe ratio for models fit using the day-of-half-year seasonality factor.
  The out-of-sample period is January $1995$ through June $2026$.}
  \label{fig:day_of_half_year_gross}
\end{figure}

\begin{figure}[ht]
  \centering
  \begin{subfigure}[b]{0.49\linewidth}
    \includegraphics[trim={0.5cm 1.cm 1cm 1.cm}, clip, width=\linewidth ]{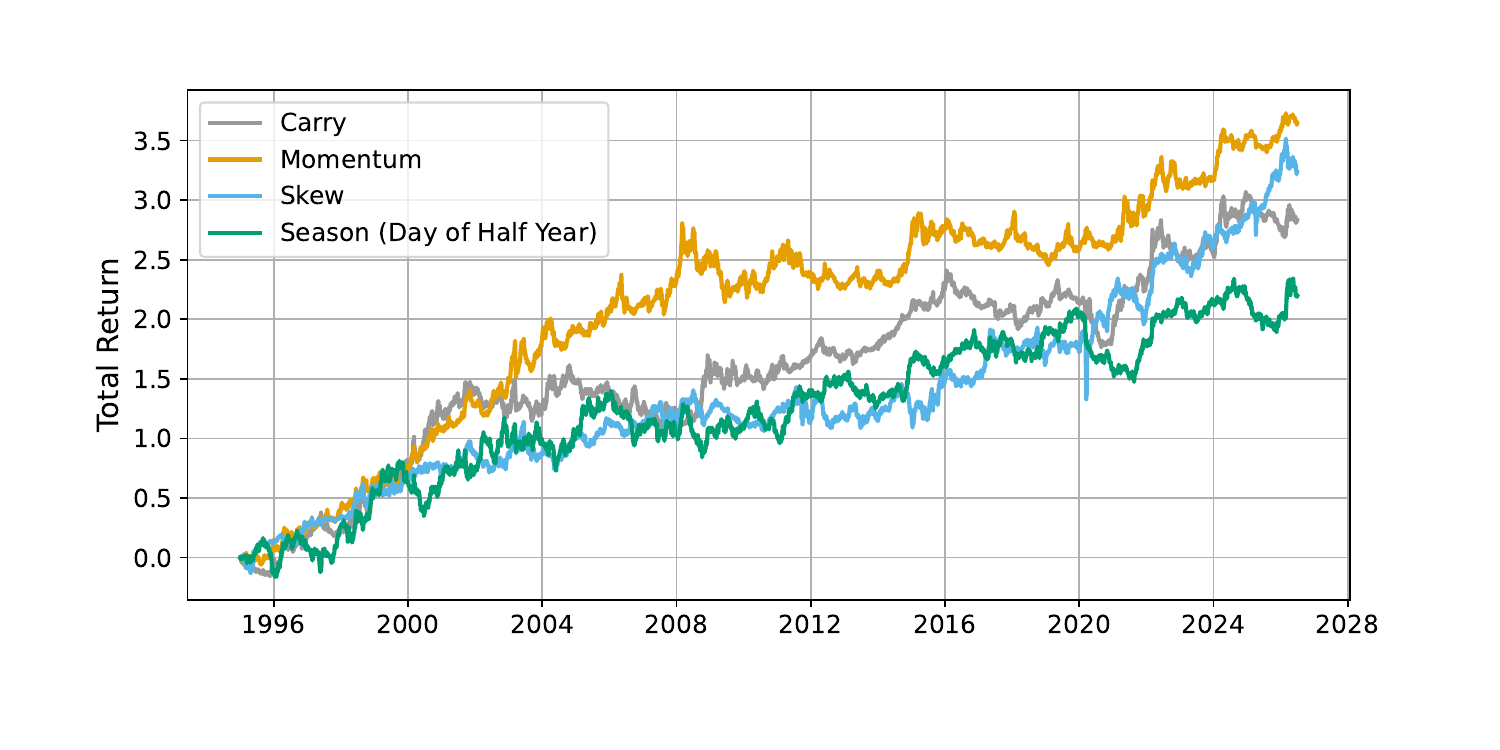}
    \caption{Net out-of-sample total return.}
    \label{fig:day_of_half_year_eq_net}
  \end{subfigure}
  \begin{subfigure}[b]{0.49\linewidth}
    \includegraphics[width=\linewidth ]{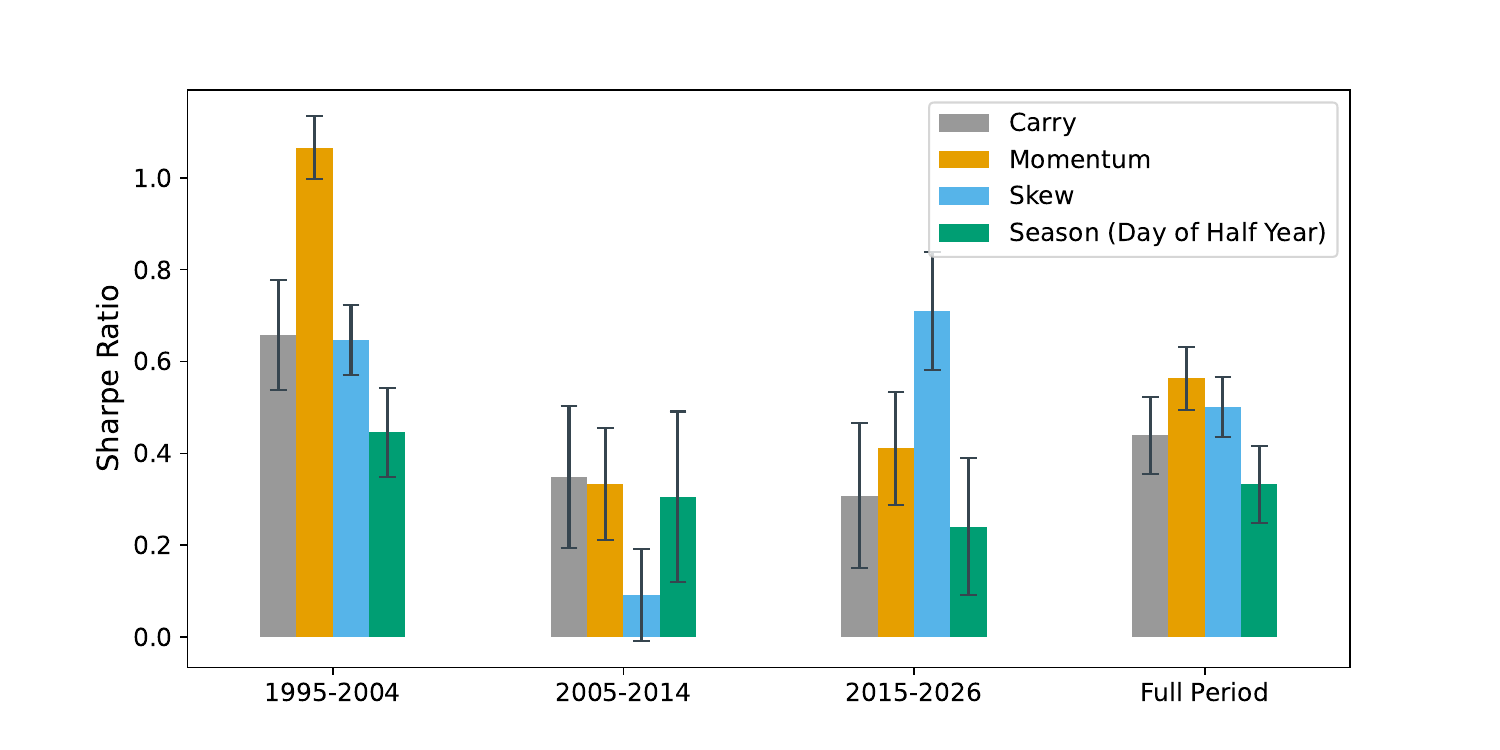}
    \caption{Net out-of-sample Sharpe ratio.}
    \label{fig:day_of_half_year_sharpe_net}
  \end{subfigure}

  \caption{Net out-of-sample total return and Sharpe ratio for models fit using the day-of-half-year seasonality factor.
  The out-of-sample period is January $1995$ through June $2026$.}
  \label{fig:day_of_half_year_net}
\end{figure}

Table \ref{table:day_of_half_year} reports the results of the multi-factor regression analysis for the day-of-half-year seasonality factor.
Here, we observe that agriculture $(2.90\%)$ and energy $(6.58\%)$ sectors exhibit statistically significant annualized alphas, with energy markets
accounting for over half of the total portfolio alpha of $11.65\%$. Once again, we observe no strong statistical 
dependency on the other time series factors, with all  $R^2$ values less than $1\%$. Indeed, the forward response function for commodity markets,
reported in Figure \ref{fig:day-of-half-year}, illustrates a potential seasonal cycle in energy markets: we observe positive returns
during peak winter and summer demand periods, followed by negative returns over the lower fall and spring demand seasons. A similar seasonal 
pattern, coinciding with the turnover of agricultural crops, may also explain the statistically significant alpha observed in agriculture markets.
Moreover, Figures \ref{fig:day_of_half_year_gross} and \ref{fig:day_of_half_year_net} report significantly positive gross and net returns with Sharpe ratios 
of $0.62$ and $0.33$, respectively, over the full period. Unlike the previous seasonal factors, the day-of-half-year factor displays positive
net Sharpe ratios over every evaluation period. However, as indicated by the regression analysis, the returns are not pervasive and instead are largely limited
to the energy and agriculture sectors. 

\begin{table}[H]
\centering
\small

\begin{tabular}{llllll}
\toprule
 & Alpha & Carry & Momentum & Skew & R2 \\
\midrule
Agriculture & 0.029* (0.0148)  & 0.0255* (0.0137)  & -0.0148 (0.0136)  & 0.0537*** (0.0112)  & 0.0036 \\
Bond & 0.0072 (0.0099)  & 0.106*** (0.0133)  & -0.0035 (0.0123)  & -0.0355*** (0.0125)  & 0.0088 \\
Currency & 0.0027 (0.0063)  & 0.0469*** (0.0131)  & -0.0466*** (0.0121)  & -0.016 (0.0123)  & 0.0025 \\
Energy & 0.0658** (0.0276)  & 0.1244*** (0.0162)  & -0.081*** (0.0159)  & -0.028** (0.0115)  & 0.0073 \\
Equity & 0.0135 (0.0107)  & -0.012 (0.0114)  & 0.0011 (0.0115)  & -0.0353*** (0.0116)  & 0.0015 \\
Metal & 0.0064 (0.0061)  & 0.0284** (0.0111)  & -0.0249** (0.0115)  & -0.0675*** (0.0116)  & 0.0074 \\
Portfolio & 0.1165*** (0.0356)  & 0.0712*** (0.0123)  & -0.0387*** (0.0125)  & 0.0206* (0.0113)  & 0.0046 \\
\bottomrule
\end{tabular}

\caption{Multi-factor regression analysis for the day-of-half-year seasonality factor regressed on time series 
carry, momentum and skewness factors. The intercept term is annualized. Statistical significance
at the $0.10$, $0.05$ and $0.01$ level is denoted by $*$, $**$ and $***$, respectively. 
The analysis period is January $1995$ through June $2026$. }
\label{table:day_of_half_year}
\end{table}

\subsection{Seasonality Anomaly: Day-of-Year} \label{sec:day_of_year}
\begin{figure}[H]
  \centering
  \begin{subfigure}[b]{0.43\linewidth}
    \includegraphics[width=\linewidth ]{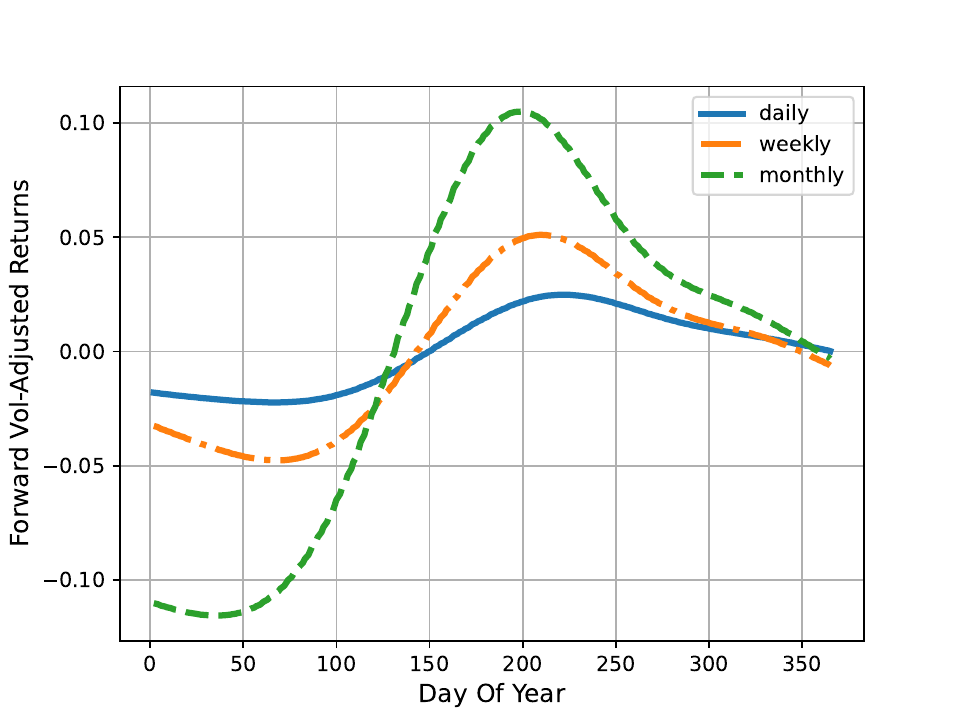}
    \caption{Bond Sector}
  \end{subfigure}
  \begin{subfigure}[b]{0.43\linewidth}
    \includegraphics[width=\linewidth ]{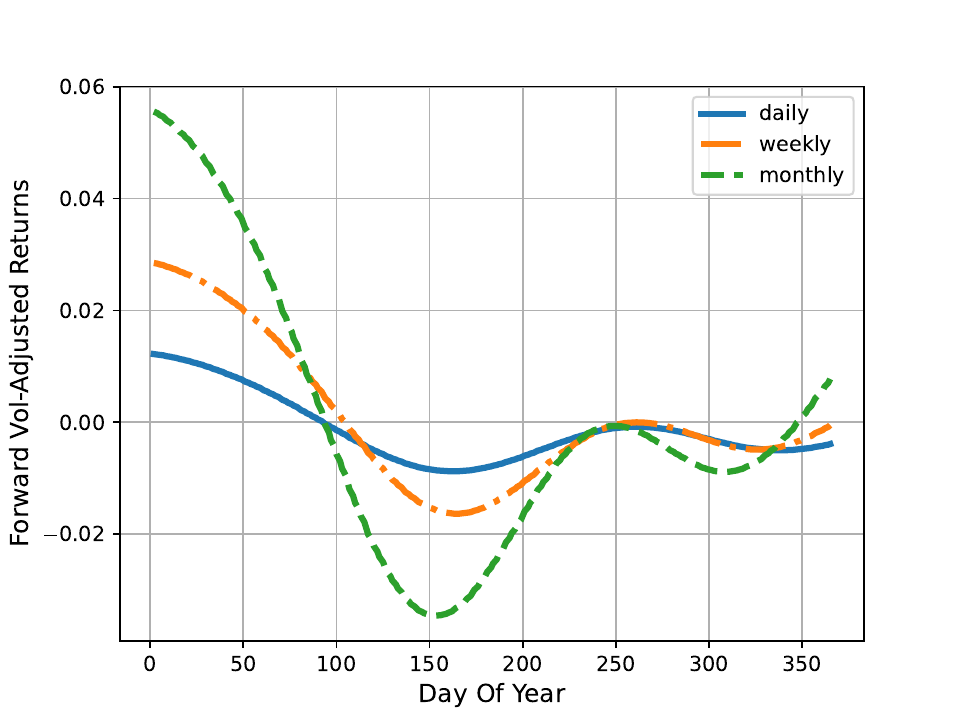}
    \caption{Commodity Sector}
  \end{subfigure}
  \begin{subfigure}[b]{0.43\linewidth}
    \includegraphics[width=\linewidth ]{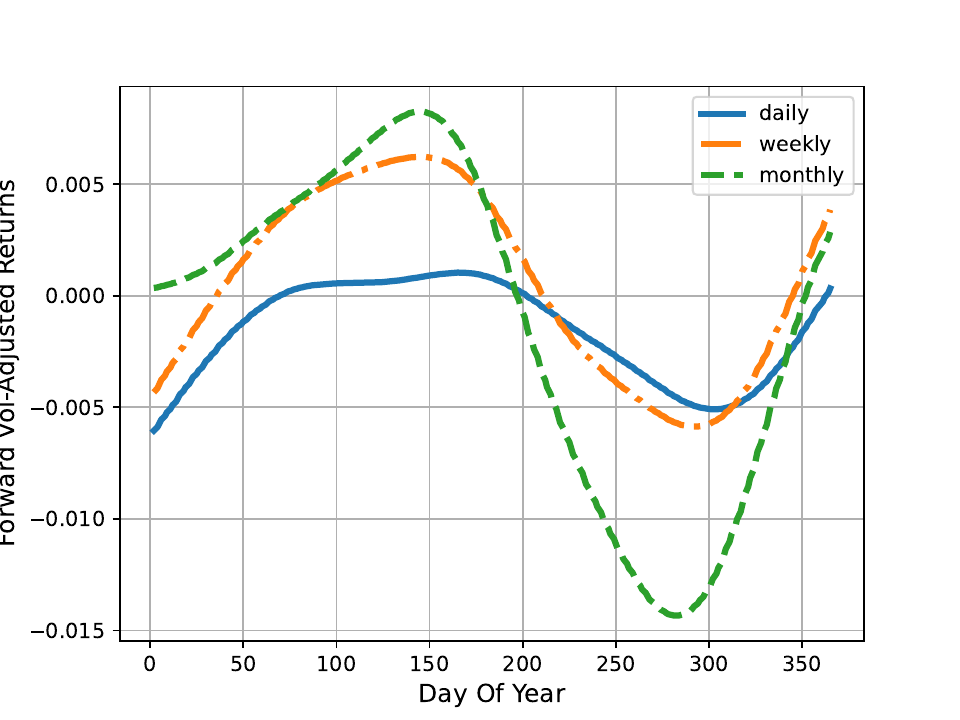}
    \caption{Currency Sector}
  \end{subfigure}
  \begin{subfigure}[b]{0.43\linewidth}
    \includegraphics[width=\linewidth ]{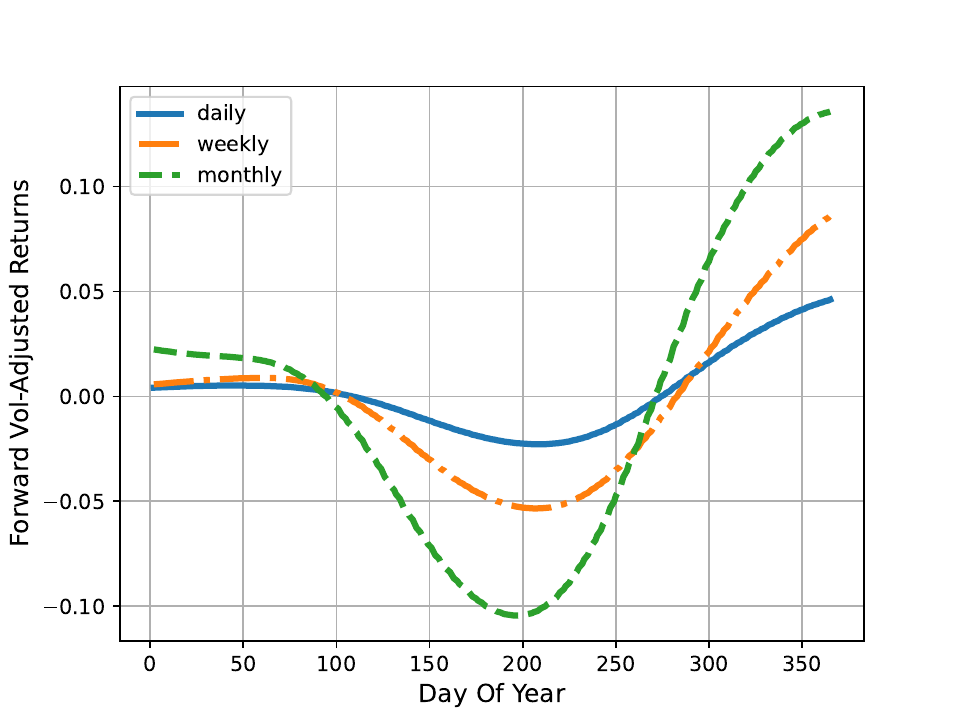}
    \caption{Equity Sector}
  \end{subfigure}
  
  \caption{Estimated forward seasonality response as a function of the day-of-year seasonality factor. 
  Models are fit using all available training data from January $1950$ through June $2026$. }
  \label{fig:day-of-year}
\end{figure}

Figure \ref{fig:day-of-year} reports the estimated forward response as a function of the day-of-year seasonality feature. The seasonal pattern
in equity markets is clear: we observe the well-documented `sell in May and go away' effect followed by the aptly named `Santa Claus rally' effect,
with positive expected returns in the last two months of the year. Conversely, bond markets tend to exhibit an inverse seasonal pattern to equities, 
with the period of May through September displaying strongly positive forward returns. Commodity markets display positive response 
values in the first quarter but are generally mixed thereafter, perhaps driven by transitions into new production, inventory, weather and demand cycles 
that become more heterogeneous amongst markets as the year progresses. Lastly, currency markets display a cyclical response pattern, with 
forward returns peaking in the spring and becoming generally negative in the fall.

\begin{figure}[ht]
  \centering
  \begin{subfigure}[b]{0.49\linewidth}
    \includegraphics[trim={0.5cm 1.cm 1cm 1.cm}, clip, width=\linewidth ]{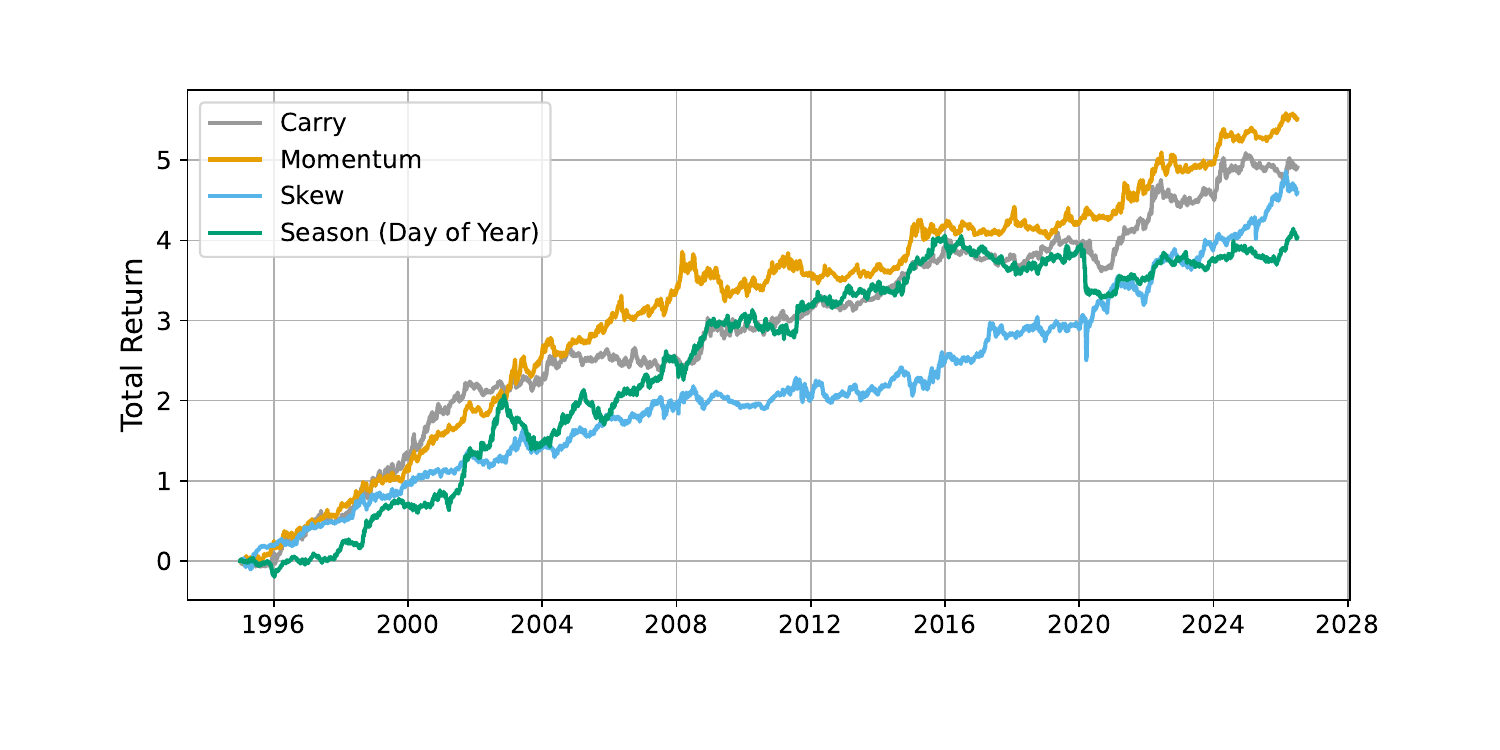}
    \caption{Gross out-of-sample total return.}
    \label{fig:day_of_year_eq_gross}
  \end{subfigure}
  \begin{subfigure}[b]{0.49\linewidth}
    \includegraphics[width=\linewidth ]{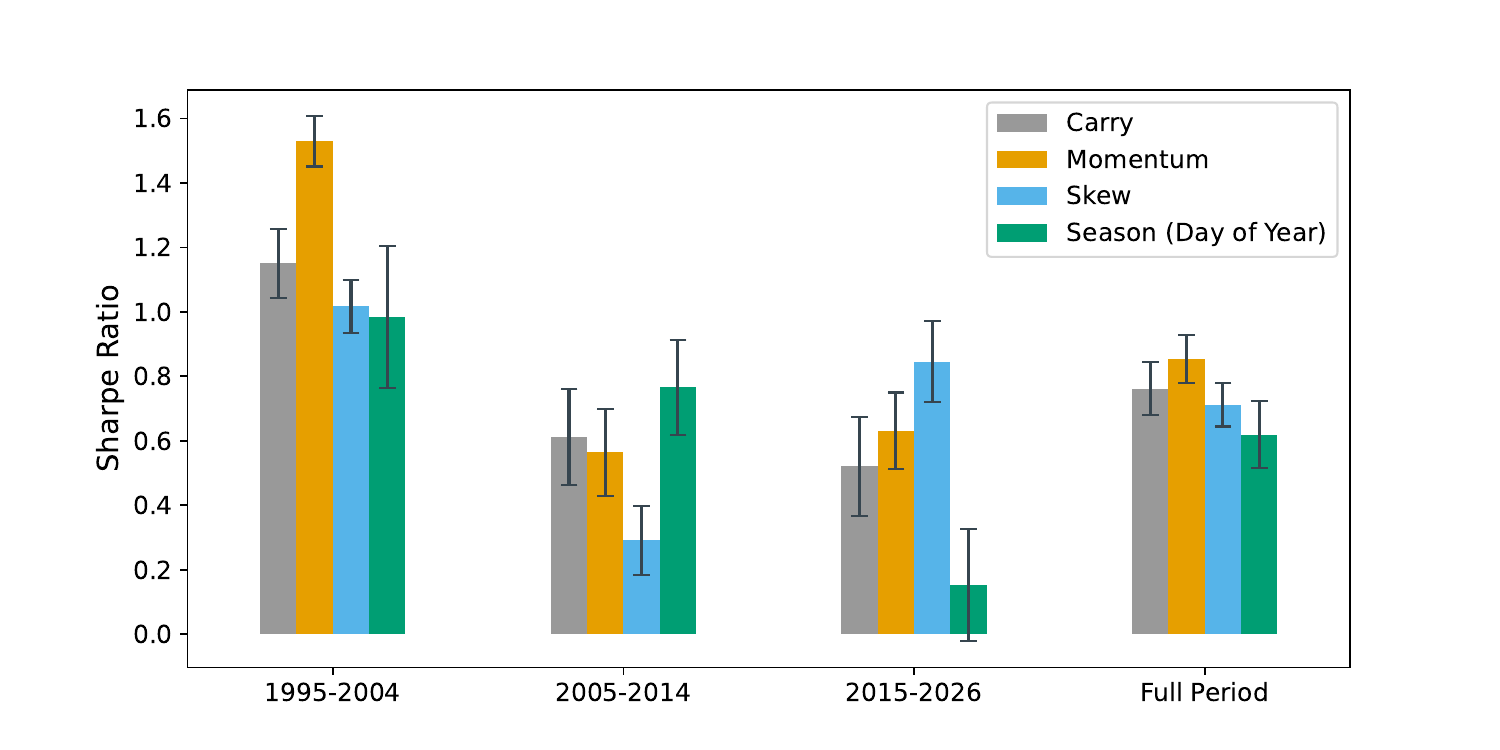}
    \caption{Gross out-of-sample Sharpe ratio.}
    \label{fig:day_of_year_sharpe_gross}
  \end{subfigure}

  \caption{Gross out-of-sample total return and Sharpe ratio for models fit using the day-of-year seasonality factor.
  The out-of-sample period is January $1995$ through June $2026$.}
  \label{fig:day_of_year_gross}
\end{figure}

\begin{figure}[ht]
  \centering
  \begin{subfigure}[b]{0.49\linewidth}
    \includegraphics[trim={0.5cm 1.cm 1cm 1.cm}, clip, width=\linewidth ]{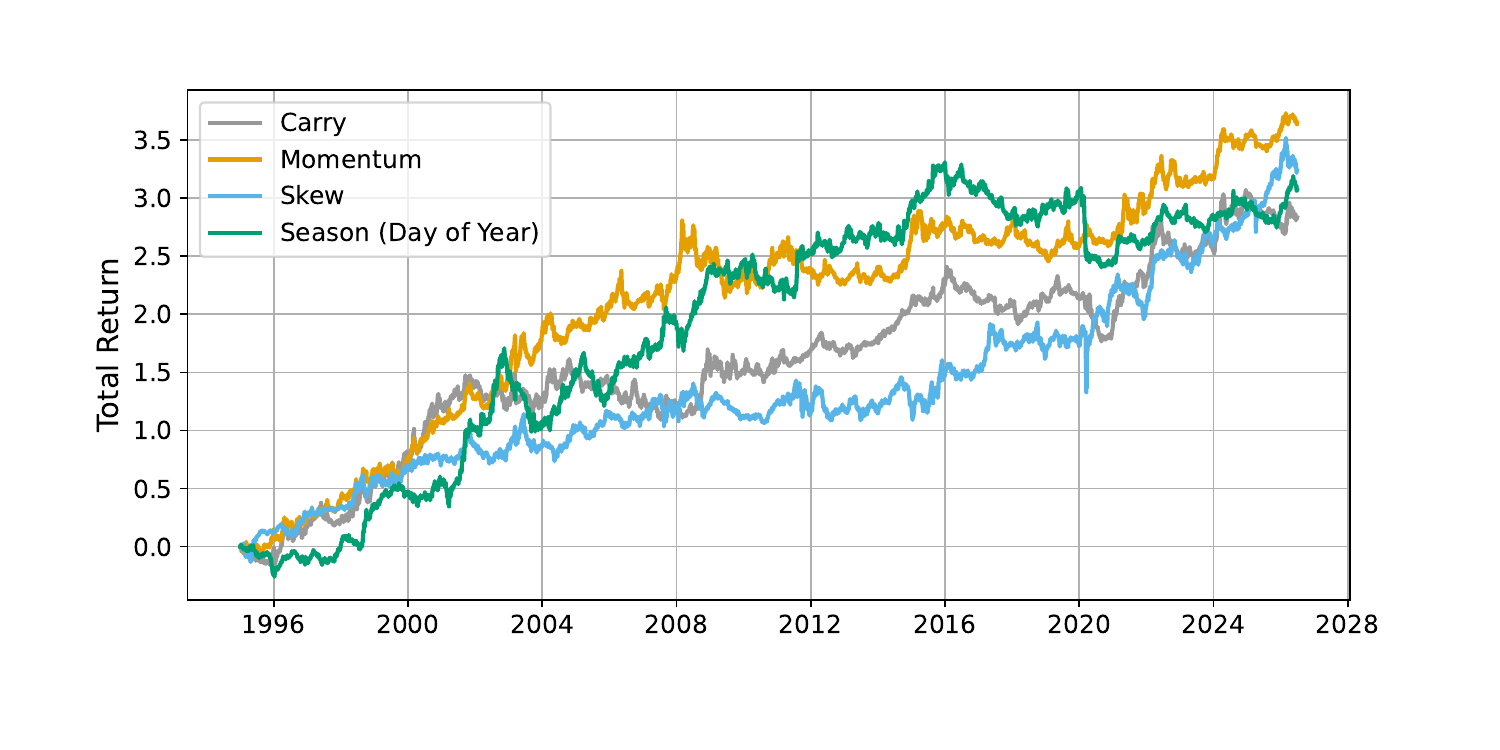}
    \caption{Net out-of-sample total return.}
    \label{fig:day_of_year_eq_net}
  \end{subfigure}
  \begin{subfigure}[b]{0.49\linewidth}
    \includegraphics[width=\linewidth ]{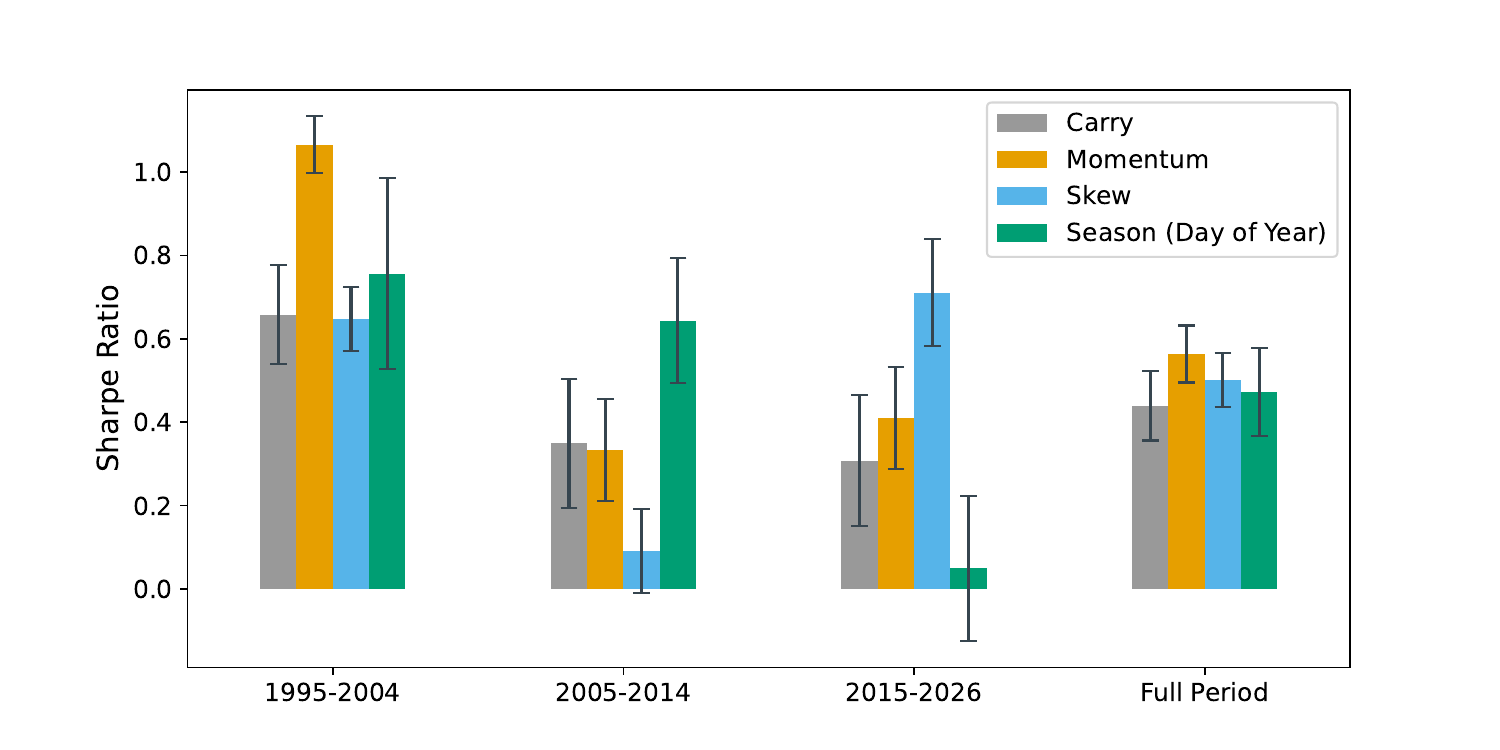}
    \caption{Net out-of-sample Sharpe ratio.}
    \label{fig:day_of_year_sharpe_net}
  \end{subfigure}

  \caption{Net out-of-sample total return and Sharpe ratio for models fit using the day-of-year seasonality factor.
  The out-of-sample period is January $1995$ through June $2026$.}
  \label{fig:day_of_year_net}
\end{figure}

Table \ref{table:day_of_year} reports the results of the multi-factor regression analysis for the day-of-year factor.
We observe modest factor loadings on the carry and skewness factors, which are statistically significant across
most sectors but do not meaningfully explain any variance, with the majority of $R^2$ values reported as less
than $1\%$. Once again, we report a statistically significant 
portfolio alpha of $11.35\%$, driven by the alpha in bond $(2.78\%)$, energy $(3.29\%)$, equity $(2.63\%)$ and metal $(1.79\%)$ sectors.
Lastly, Figures \ref{fig:day_of_year_gross} and \ref{fig:day_of_year_net} report
significant gross and net Sharpe ratios of $0.64$ and $0.39$, respectively, the highest net Sharpe ratio of any individual seasonality factor, indicating a strong day-of-year effect in excess of transaction costs. 

\begin{table}[H]
\centering
\small

\begin{tabular}{llllll}
\toprule
 & Alpha & Carry & Momentum & Skew & R2 \\
\midrule
Agriculture & 0.0066 (0.0078)  & 0.0876*** (0.0137)  & -0.0184 (0.0135)  & 0.016 (0.0112)  & 0.0068 \\
Bond & 0.0278 (0.0196)  & 0.096*** (0.0133)  & 0.0093 (0.0123)  & 0.0146 (0.0124)  & 0.0117 \\
Currency & -0.0016 (0.0045)  & 0.0426*** (0.0131)  & 0.0064 (0.0121)  & -0.0687*** (0.0123)  & 0.004 \\
Energy & 0.0329** (0.015)  & 0.0203 (0.0163)  & 0.0136 (0.016)  & -0.0529*** (0.0115)  & 0.0029 \\
Equity & 0.0263* (0.0156)  & 0.1141*** (0.0113)  & -0.0011 (0.0114)  & 0.0611*** (0.0115)  & 0.0197 \\
Metal & 0.0179** (0.0071)  & -0.0382*** (0.0111)  & 0.037*** (0.0115)  & -0.0119 (0.0116)  & 0.0026 \\
Portfolio & 0.1135*** (0.0352)  & 0.016 (0.0123)  & 0.0392*** (0.0125)  & 0.0096 (0.0114)  & 0.0026 \\
\bottomrule
\end{tabular}

\caption{Multi-factor regression analysis for the day-of-year seasonality factor regressed on time series 
carry, momentum and skewness factors. The intercept term is annualized. Statistical significance
at the $0.10$, $0.05$ and $0.01$ level is denoted by $*$, $**$ and $***$, respectively. 
The analysis period is January $1995$ through June $2026$. }
\label{table:day_of_year}
\end{table}

\subsection{Seasonality Ensemble} \label{sec:ensemble}
The individual seasonality factor analysis revealed several empirical findings. First, individual seasonality factors typically display higher turnover 
and lower net Sharpe ratios than existing global factor portfolios. Second, individual seasonality factors exhibit sector return attributions
that are not pervasive and instead tend to concentrate in a few sectors. Therefore, from a practitioner's perspective, it would be difficult to 
justify the implementation of any individual seasonality factor as a stand-alone factor strategy. We attempt to overcome these weaknesses 
by considering an ensemble average of the individual seasonality factors. The ensemble is formed by taking an equal-weight average across each 
seasonality feature and forecast horizon pair. Note that, given the extreme turnover profile, we do not include the day-of-week factor. Moreover,
the proposed weighting scheme will naturally overweight seasonality features with longer seasonal periodicity (such as day-of-year) as they
have more valid forecast horizons. In general, we find that the final seasonality ensemble produces an aggregate turnover profile that is
consistent with  the turnover of the established time series factor portfolios.
\begin{table}[H]
\centering
\small

\begin{tabular}{llllll}
\toprule
 & Alpha & Carry & Momentum & Skew & R2 \\
\midrule
Agriculture & 0.0183* (0.0097)  & 0.0734*** (0.0137)  & -0.0203 (0.0135)  & 0.04*** (0.0112)  & 0.0066 \\
Bond & 0.0308* (0.0178)  & 0.1251*** (0.0132)  & 0.0117 (0.0122)  & 0.0079 (0.0124)  & 0.0181 \\
Currency & 0.0027 (0.0054)  & 0.0751*** (0.0131)  & -0.0387*** (0.0121)  & -0.052*** (0.0123)  & 0.0049 \\
Energy & 0.0513*** (0.0187)  & 0.0754*** (0.0162)  & -0.0336** (0.016)  & -0.0459*** (0.0115)  & 0.004 \\
Equity & 0.0335** (0.0144)  & 0.1081*** (0.0113)  & 0.0052 (0.0114)  & 0.0425*** (0.0115)  & 0.0158 \\
Metal & 0.0194*** (0.007)  & -0.0158 (0.0111)  & 0.0223* (0.0115)  & -0.036*** (0.0116)  & 0.0015 \\
Portfolio & 0.1577*** (0.0353)  & 0.038*** (0.0123)  & 0.0203 (0.0125)  & 0.016 (0.0114)  & 0.0031 \\
\bottomrule
\end{tabular}

\caption{Multi-factor regression analysis for the seasonality ensemble factor regressed on time series 
carry, momentum and skewness factors. The intercept term is annualized. Statistical significance
at the $0.10$, $0.05$ and $0.01$ level is denoted by $*$, $**$ and $***$, respectively. 
The analysis period is January $1995$ through June $2026$. }
\label{table:ensemble}
\end{table}

Table \ref{table:ensemble} reports the results of the multi-factor regression analysis for the seasonality ensemble. In contrast to 
the majority of other seasonal factors, here we observe regression alphas that are evenly distributed across the various sectors. In fact, 
with the exception of the currency sector, all other sectors display statistically significant alpha values. Moreover, the total portfolio
exhibits an alpha of $15.77\%$, which is significant at the $1\%$ level and is not explained by the carry, momentum or skewness factors. The differentiated
returns of the seasonality ensemble portfolio are further reinforced in Table \ref{table:corr}, which demonstrates that the seasonality ensemble is 
nearly uncorrelated with all other factor portfolios, with correlation values less than $0.05$ over the full out-of-sample period.

\begin{figure}[ht]
  \centering
  \begin{subfigure}[b]{0.49\linewidth}
    \includegraphics[trim={0.5cm 1.cm 1cm 1.cm}, clip, width=\linewidth ]{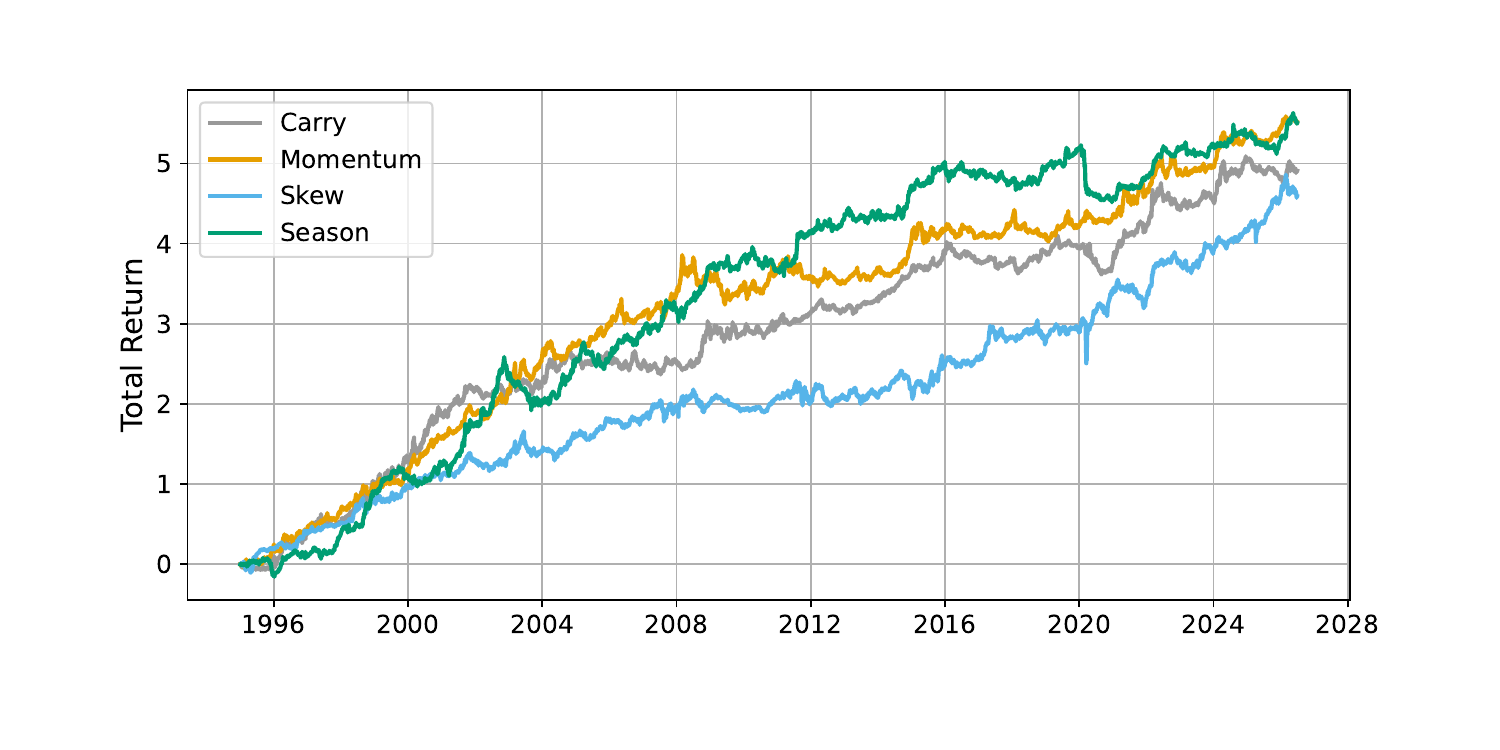}
    \caption{Gross out-of-sample total return.}
    \label{fig:season_eq_gross}
  \end{subfigure}
  \begin{subfigure}[b]{0.49\linewidth}
    \includegraphics[width=\linewidth ]{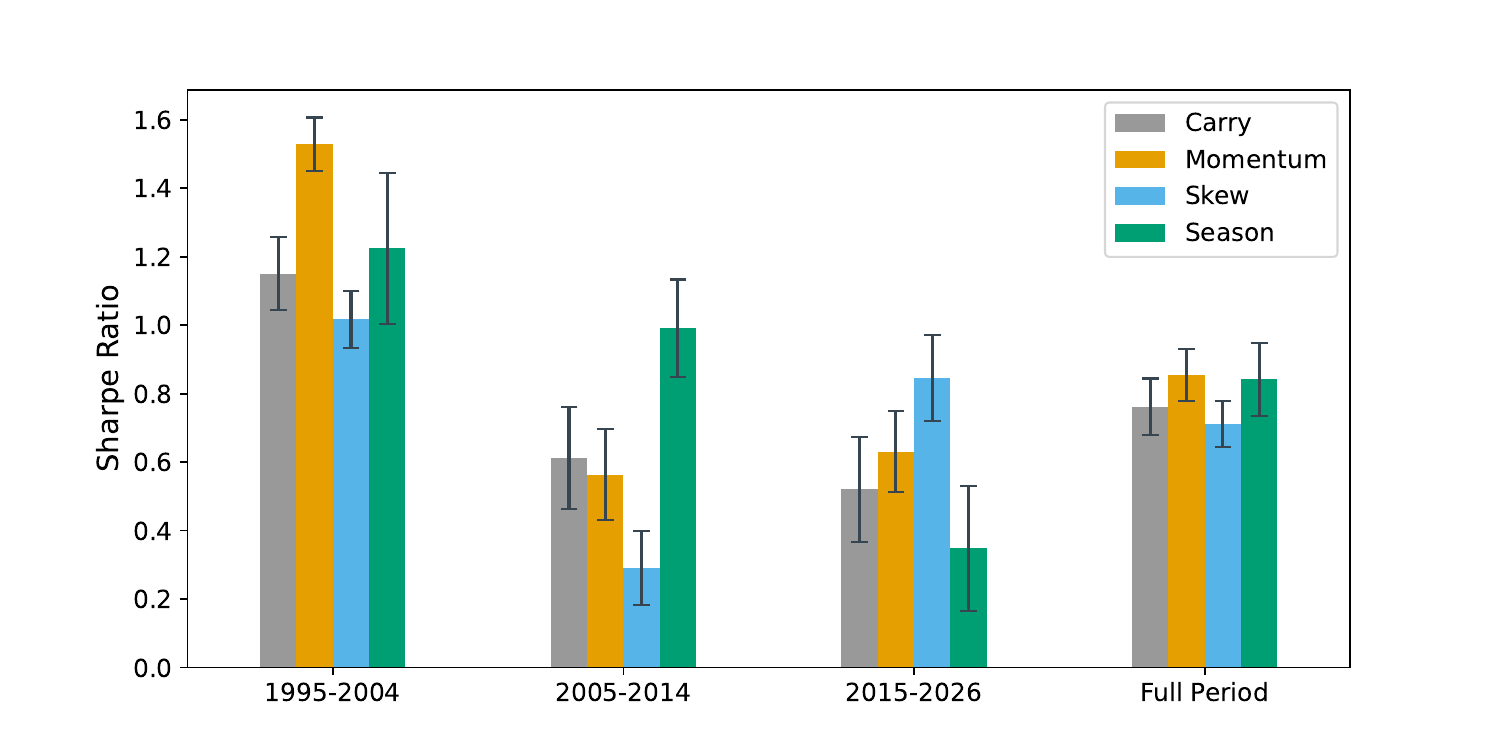}
    \caption{Gross out-of-sample Sharpe ratio.}
    \label{fig:season_sharpe_gross}
  \end{subfigure}

  \caption{Gross out-of-sample total return and Sharpe ratio for models fit using the seasonality ensemble factor.
  The out-of-sample period is January $1995$ through June $2026$.}
  \label{fig:season_gross}
\end{figure}

\begin{figure}[ht]
  \centering
  \begin{subfigure}[b]{0.49\linewidth}
    \includegraphics[trim={0.5cm 1.cm 1cm 1.cm}, clip, width=\linewidth ]{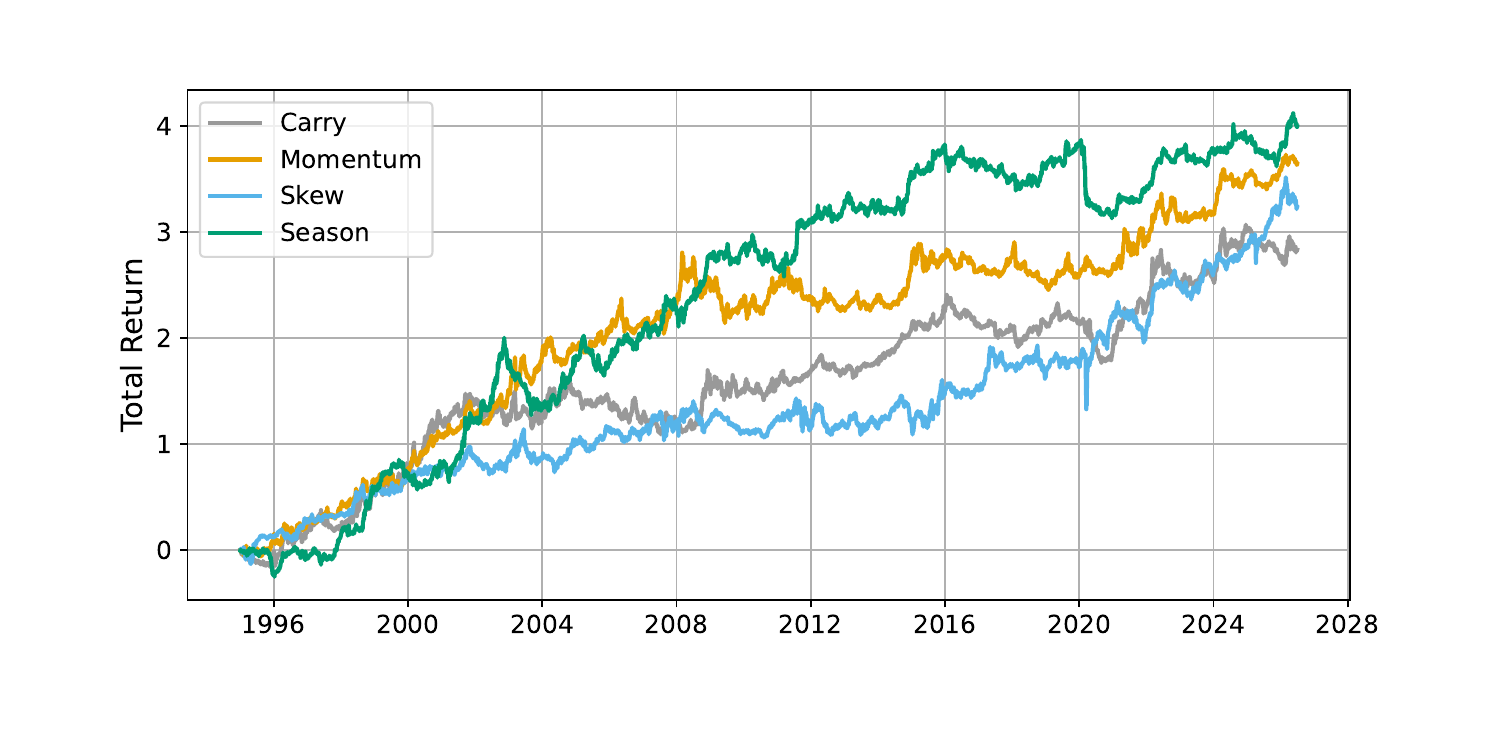}
    \caption{Net out-of-sample total return.}
    \label{fig:season_eq_net}
  \end{subfigure}
  \begin{subfigure}[b]{0.49\linewidth}
    \includegraphics[width=\linewidth ]{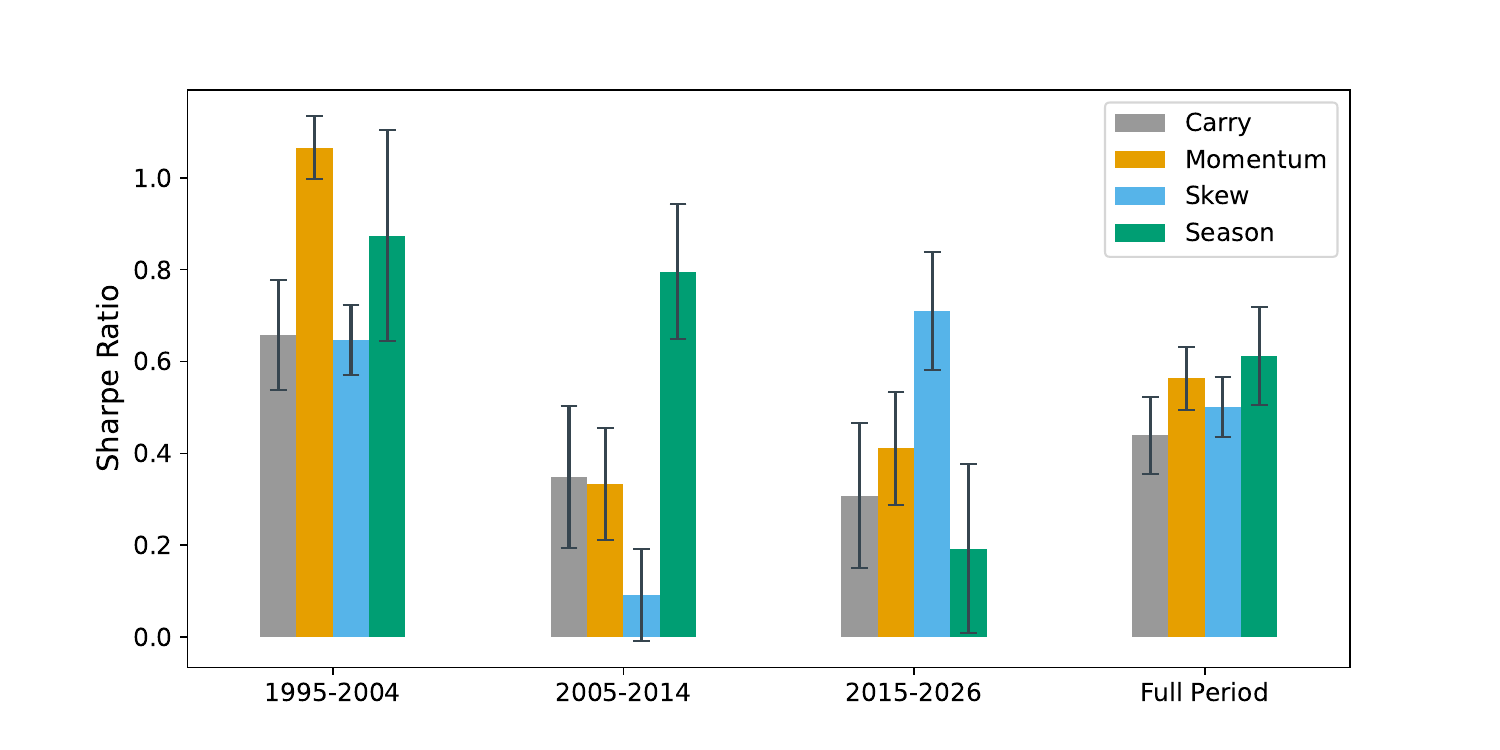}
    \caption{Net out-of-sample Sharpe ratio.}
    \label{fig:season_sharpe_net}
  \end{subfigure}

  \caption{Net out-of-sample total return and Sharpe ratio for models fit using the seasonality ensemble factor.
  The out-of-sample period is January $1995$ through June $2026$.}
  \label{fig:season_net}
\end{figure}

\begin{figure}[H]
  \centering
  \begin{subfigure}[b]{0.49\linewidth}
    \includegraphics[trim={0.5cm 1.cm 1cm 1.cm}, clip, width=\linewidth ]{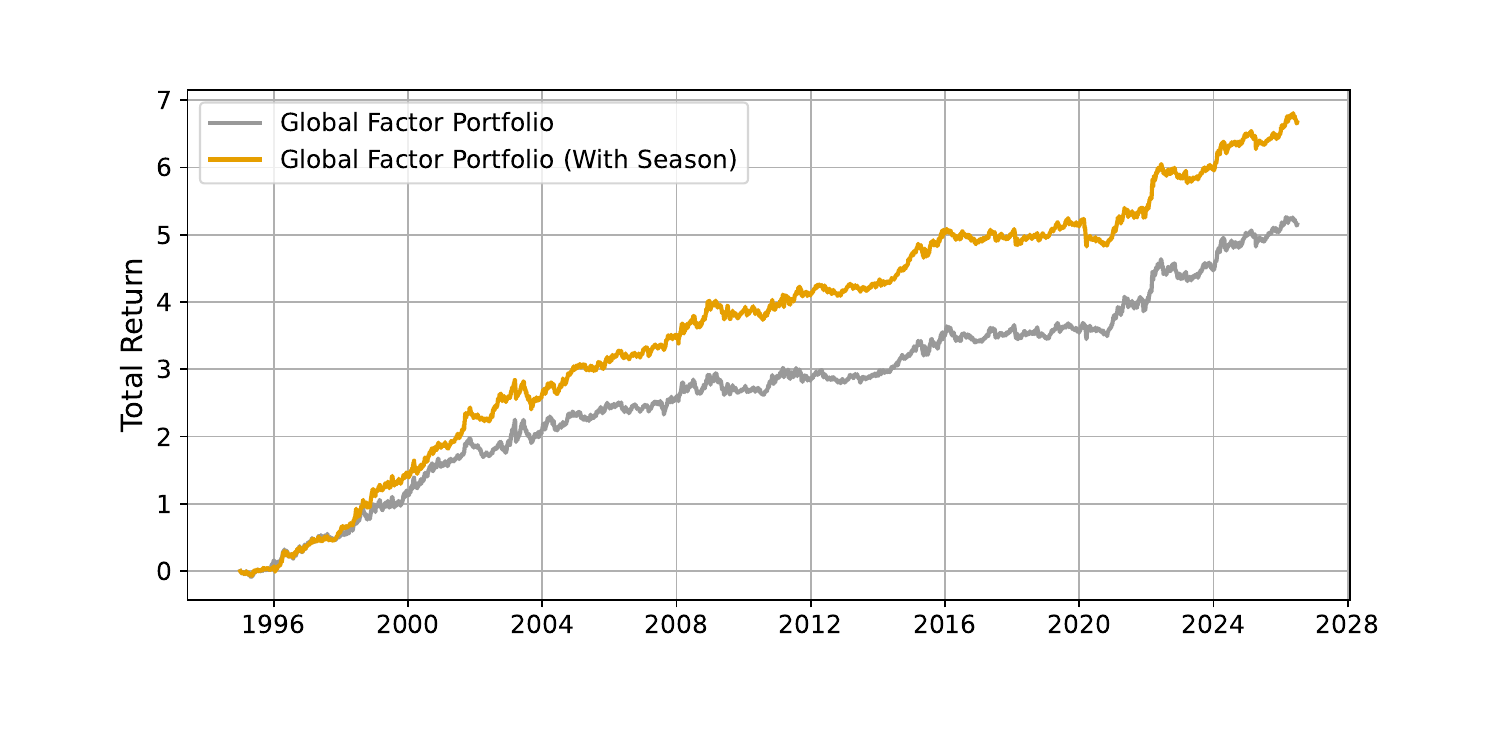}
    \caption{Net out-of-sample total return.}
    \label{fig:global_eq_net}
  \end{subfigure}
  \begin{subfigure}[b]{0.49\linewidth}
    \includegraphics[width=\linewidth ]{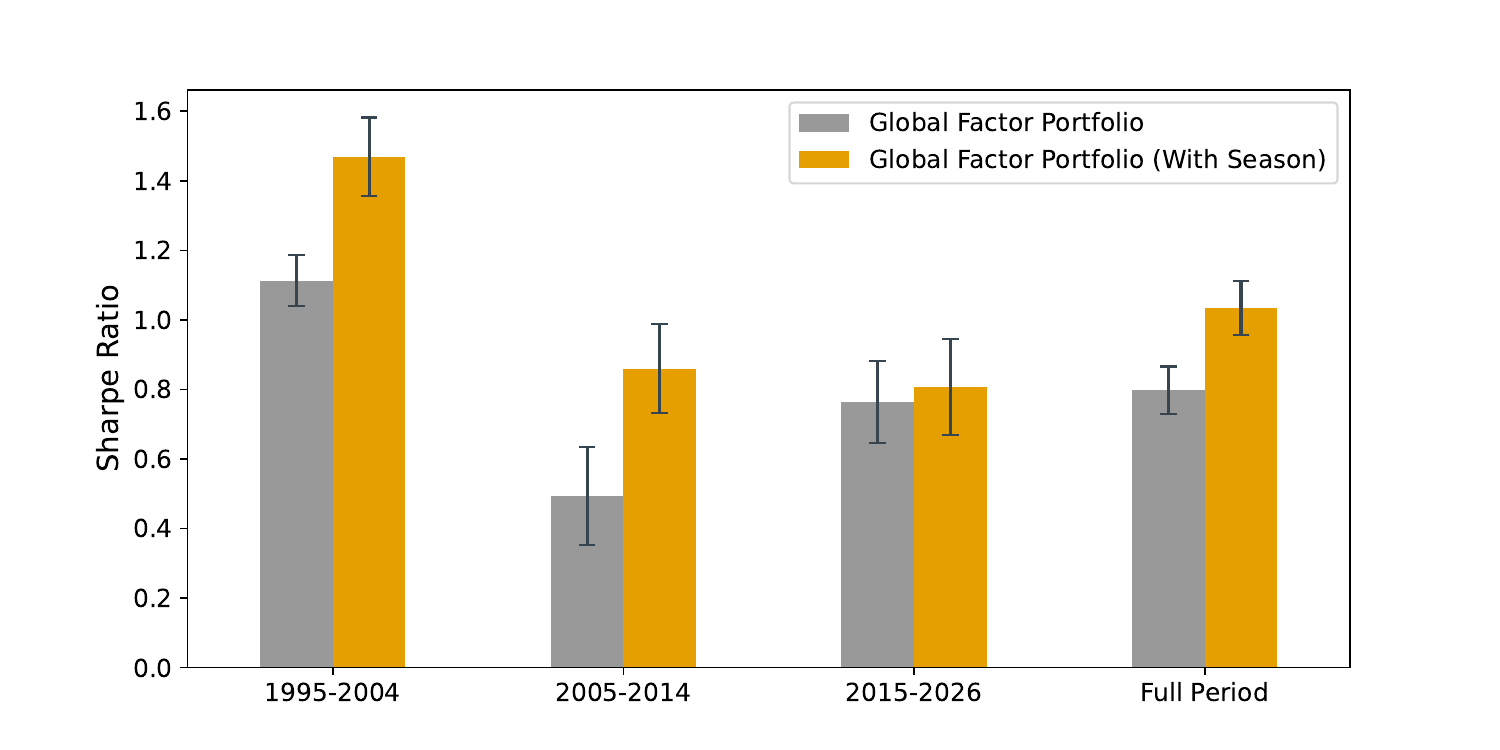}
    \caption{Net out-of-sample Sharpe ratio.}
    \label{fig:global_sharpe_net}
  \end{subfigure}

  \caption{Net out-of-sample total return and Sharpe ratio for the global factor portfolio (a simple average of carry, momentum
  and skew) and the global factor portfolio with seasonality. The out-of-sample period is January $1995$ through June $2026$.}
  \label{fig:global_net}
\end{figure}

Figures \ref{fig:season_gross} and \ref{fig:season_net} demonstrate that the seasonality ensemble produces gross and net Sharpe ratios 
of $0.85$ and $0.62$, respectively, which are comparable to the Sharpe ratios generated by the carry, 
momentum and skewness factor portfolios. The global seasonality ensemble portfolio therefore overcomes the weaknesses of the individual seasonality 
factor portfolios and displays strong out-of-sample net Sharpe ratios with
return drivers that are distributed across the majority of asset classes. Moreover, the seasonality factor returns are uncorrelated with existing factor portfolios
and exhibit differentiated periods of performance, notably outperforming all other factor portfolios during 
the 2005--2014 period. Finally, in Figure \ref{fig:global_net} we evaluate the value-add of incorporating the seasonality ensemble as 
an equal-weight component of a global factor portfolio, consisting of an equal-weight allocation to carry, momentum and skewness factors.
We find that by adding seasonality, the net Sharpe ratio over the full period improves by approximately $30\%$, going from a Sharpe ratio of $0.80$ to $1.04$. 
Additionally, we observe an improvement in the net Sharpe ratio over every evaluation period, including the most recent decade, where the seasonality factor has 
underperformed relative to the other time series factors. These results establish seasonality as a competitive global factor and,
for practitioners, highlight the benefit of adding an uncorrelated predictive model to a global factor portfolio.

\begin{table}[H]
\centering
\begin{tabular}{lrrrr}
\toprule
 & Carry & Momentum & Skew & Season \\
\midrule
Carry & 1.0000 & 0.4428 & 0.1306 & 0.0415 \\
Momentum & 0.4428 & 1.0000 & 0.2533 & 0.0092 \\
Skew & 0.1306 & 0.2533 & 1.0000 & 0.0221 \\
Season & 0.0415 & 0.0092 & 0.0221 & 1.0000 \\
\bottomrule
\end{tabular}

\caption{Correlation matrix across factor portfolios. The analysis period is January $1995$ through June $2026$.}
\label{table:corr}
\end{table}

\section{Conclusion}
We studied seasonality as a time series factor for predicting forward risk-adjusted returns across global markets. 
In contrast to traditional factor modelling, where a single  model is fit across all markets, we instead followed the work of \citet{Butler2026}
and fit seasonality models at the individual market level. Seasonality factor models were cast as a conditional expectation problem and
fit using nonlinear kernel regression, thereby enabling each feature-market pair to express its potentially unique seasonal pattern. 

We considered five seasonality features: day-of-week, day-of-month, day-of-quarter, day-of-half-year and day-of-year.
Historical walk-forward analysis, spanning more than 30 years of market data, in combination with multi-factor regression analysis, demonstrated
that each of the five individual seasonality factor portfolios produced statistically significant gross alpha that was not explained
by existing time series factors. However, individual seasonality factor portfolios have several shortcomings. First, high-turnover
seasonality anomalies, such as day-of-week and day-of-month, incur exorbitant transaction costs, resulting in net Sharpe ratios that are
strongly negative. Second, lower-turnover seasonality features, such as day-of-quarter, day-of-half-year and day-of-year, produce Sharpe ratios that overcome
trading costs but are historically weaker than the established time series carry, momentum and skewness factors. Moreover, the multi-factor 
regression analysis revealed that individual seasonality factor returns displayed sector concentration that is generally inconsistent with the 
spirit of a global factor that is pervasive across multiple asset classes.  

To overcome these shortcomings, we considered a seasonality ensemble, constructed as a simple equal-weight average across each seasonality feature and
forecast horizon pair. By construction, our seasonality ensemble overweights the lower-turnover factors, thereby providing an ensemble
portfolio turnover that is comparable to that of existing factor portfolios. With the exception of the currency sector, all other sectors exhibited
statistically significant excess returns (alpha), along with a portfolio-level alpha that is significant at the $1\%$ level. From an economic perspective, the seasonality ensemble
portfolio produced out-of-sample net Sharpe ratios that are comparable to the net Sharpe ratios exhibited by the other time 
series factors. Additionally, the seasonality factor returns are nearly uncorrelated with established time series factors and exhibit
periods of strong differentiated outperformance. Indeed, adding the seasonality ensemble factor to an equal-weight portfolio of carry, momentum and skewness
improved the net Sharpe ratio by approximately $30\%$ over the full out-of-sample period. 

This paper consolidates many of the findings of previous work on market seasonality into a generalized framework for constructing a global
seasonality factor portfolio. Moreover, the methodological framework expands on the previous work of \citet{MoskowitzNonlinear} and \citet{Butler2026},
demonstrating the potential for statistical learning techniques to further the exploration and discovery of time series factors in financial markets.
Nonetheless, there are many areas for future exploration within the global seasonality body of literature. For example, a potential weakness of our
analysis relates to the procedure used for constructing continuous adjusted time series, which rolls each futures market to the nearest, most liquid 
contract. The potential drawback of this approach, particularly in commodity markets, is that the total time series is concatenated from prices
related to differing inventory, production or crop lifecycles. Instead, constructing continuous adjusted time series
from identical expiration months or forming calendar spreads may help to isolate stronger seasonality signals in futures markets.
Alternatively, seasonality may be optimally suited as a conditional feature to further enhance the predictive power of existing time series factor 
relationships. Moreover, there are likely several other seasonality anomalies, beyond the five seasonal features explored in this paper, that are worth 
investigating, including derivative expiration anomalies or event-driven seasonality effects. Overall, however, the empirical results provided in 
this paper indicate persistent seasonality anomalies that, in aggregate, are pervasive across
global bond, commodity and equity markets and help to provide a starting point for 
the consideration of seasonality as a differentiated and competitive global time series factor.

\appendix
\setcounter{table}{0}
\renewcommand{\thetable}{A\arabic{table}}
\setcounter{figure}{0}                       
\renewcommand\thefigure{A\arabic{figure}}  

\section{Seasonality Models} \label{sec:appendix}

\subsection{Bond Markets}
\begin{figure}[H]
  \centering
  \begin{subfigure}[b]{0.30\linewidth}
    \includegraphics[width=\linewidth ]{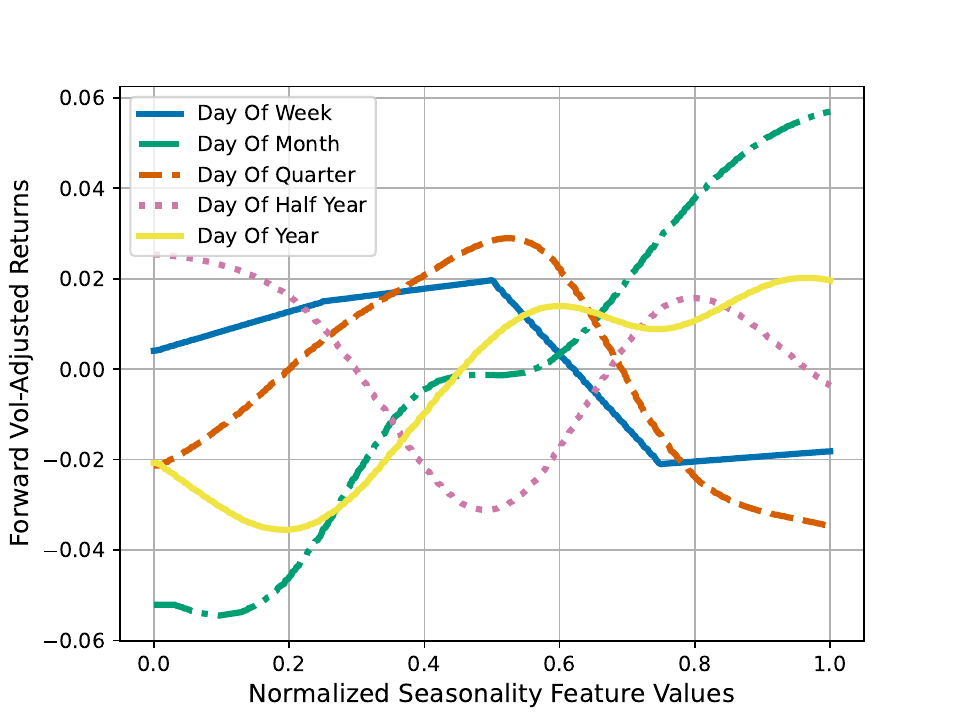}
    \caption{Canada 10-Year Bond}
  \end{subfigure}
  \begin{subfigure}[b]{0.30\linewidth}
    \includegraphics[width=\linewidth ]{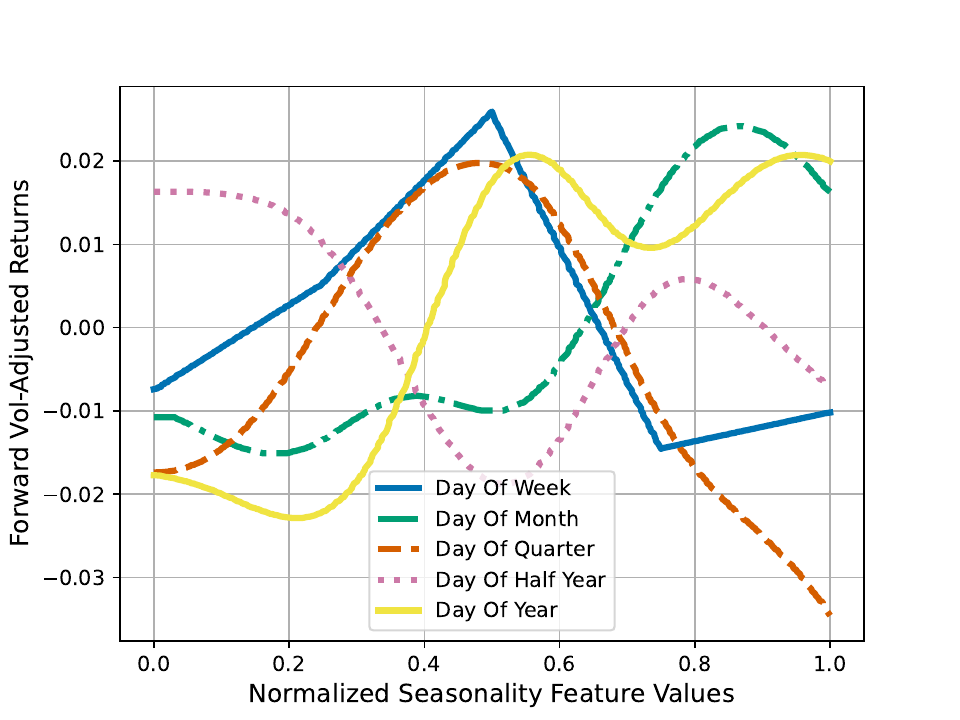}
    \caption{UK Gilts}
  \end{subfigure}
  \begin{subfigure}[b]{0.30\linewidth}
    \includegraphics[width=\linewidth ]{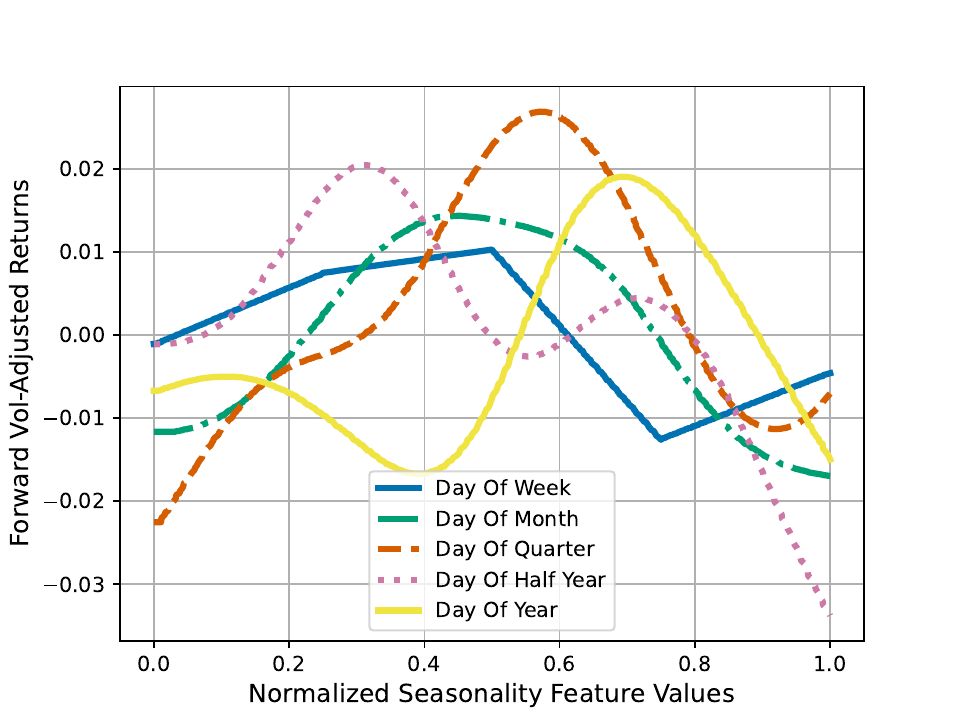}
    \caption{Japan 10-Year Bond}
  \end{subfigure}
  \begin{subfigure}[b]{0.30\linewidth}
    \includegraphics[width=\linewidth ]{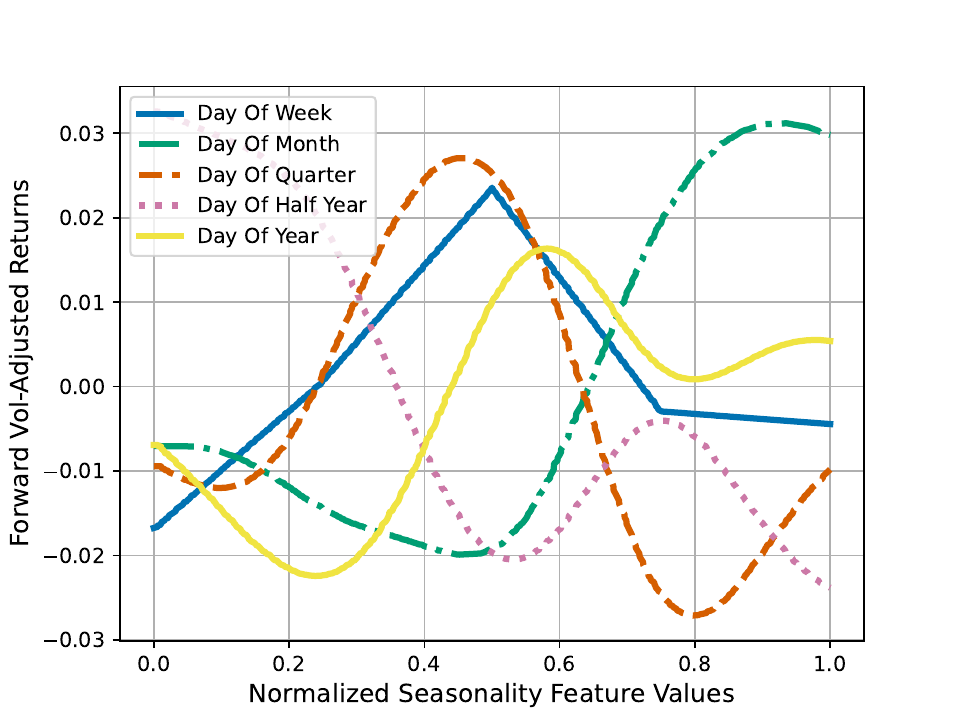}
    \caption{German Bobl}
  \end{subfigure}
  \begin{subfigure}[b]{0.30\linewidth}
    \includegraphics[width=\linewidth ]{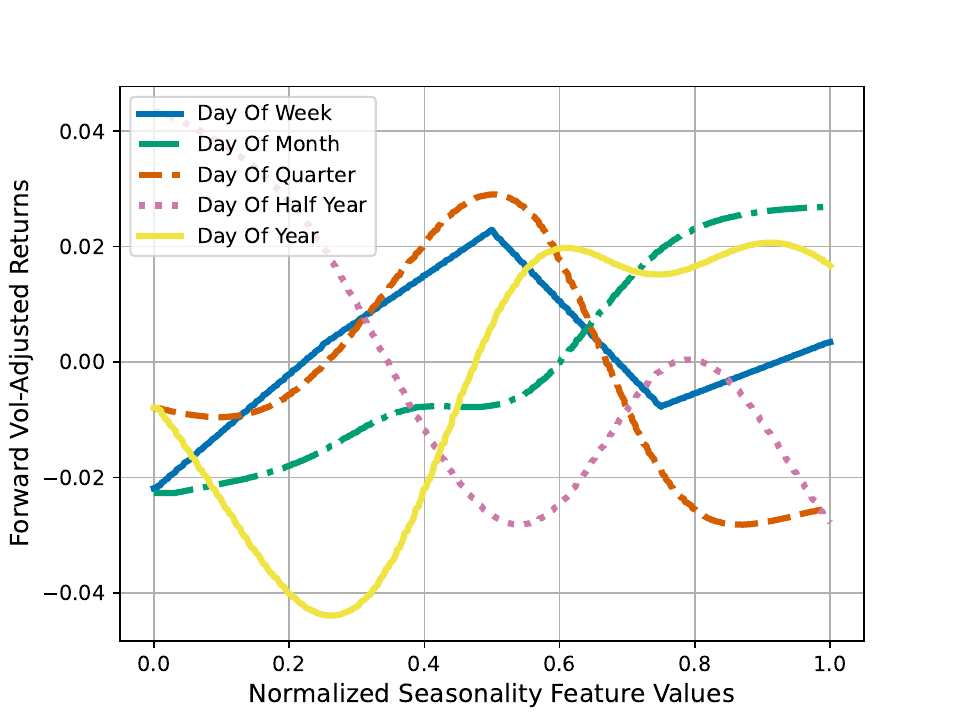}
    \caption{German Bund}
  \end{subfigure}
  \begin{subfigure}[b]{0.30\linewidth}
    \includegraphics[width=\linewidth ]{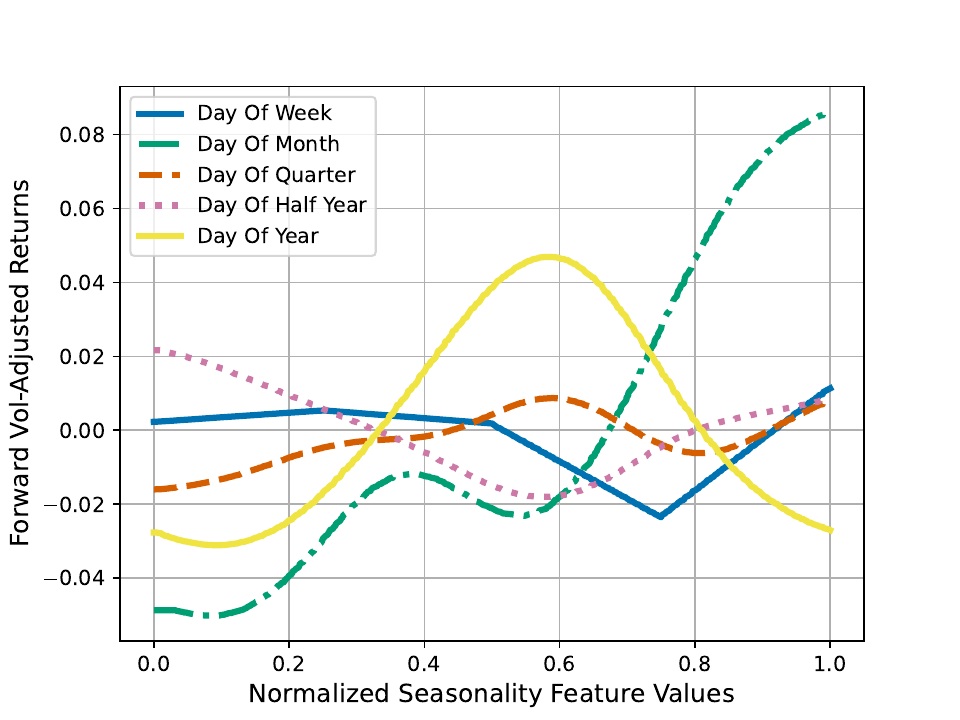}
    \caption{US 5-Year Bond}
  \end{subfigure}
  \begin{subfigure}[b]{0.30\linewidth}
    \includegraphics[width=\linewidth ]{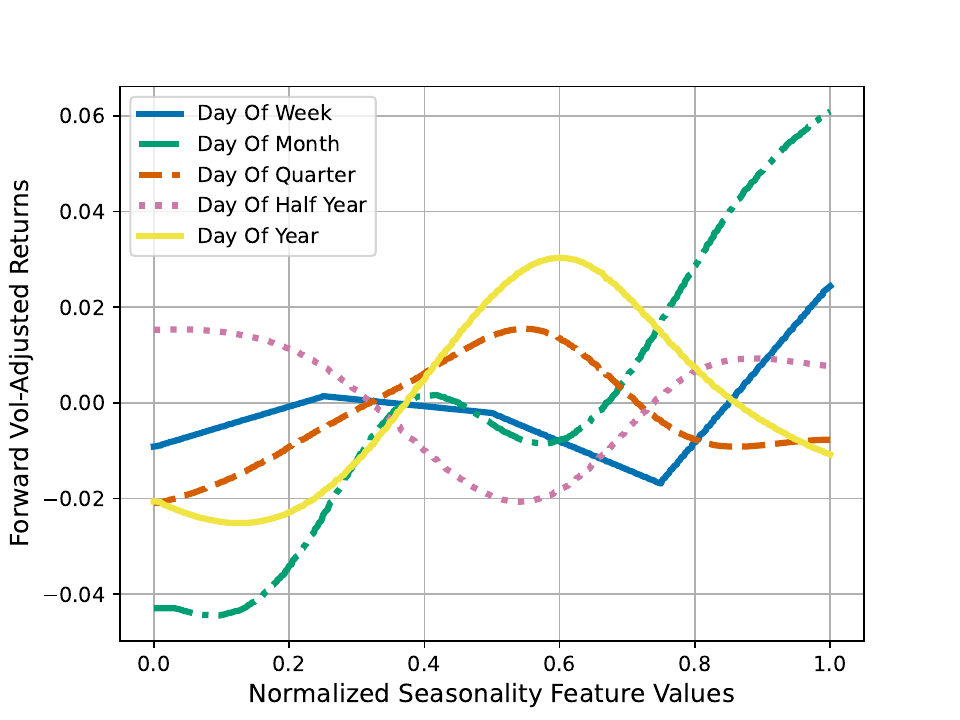}
    \caption{US 10-Year Bond}
  \end{subfigure}
  \begin{subfigure}[b]{0.30\linewidth}
    \includegraphics[width=\linewidth ]{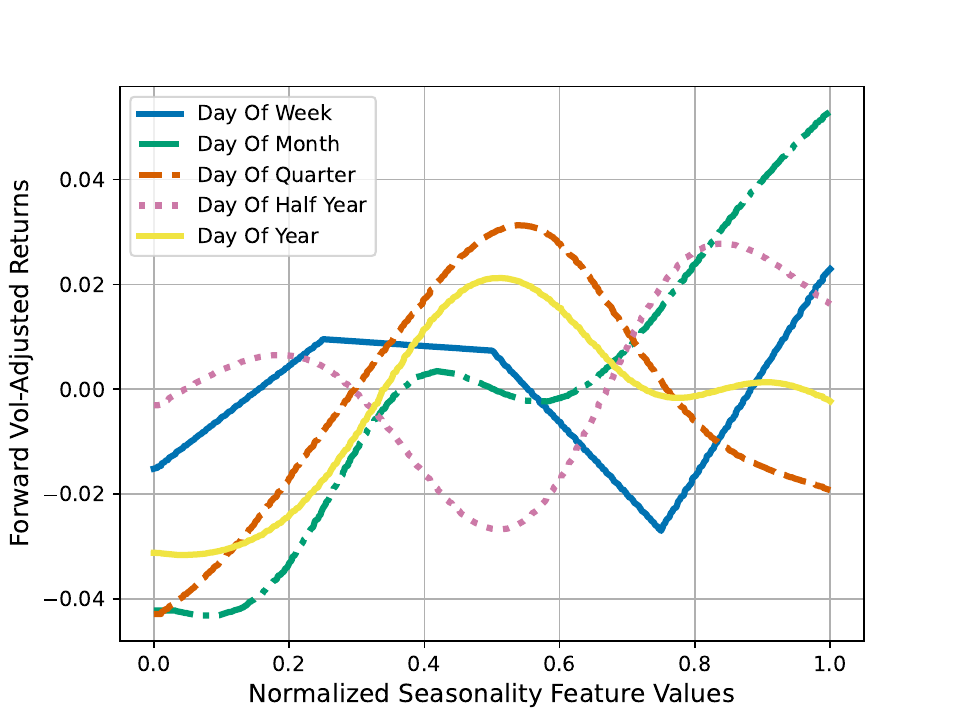}
    \caption{US 30-Year Bond}
  \end{subfigure}
  
  \caption{Estimated forward vol-adjusted returns as a function of the normalized seasonality features for global bond markets.
  Models are fit using all available training data from January $1950$ through June $2026$. }
\end{figure}

\subsection{Currency Markets}
\begin{figure}[H]
  \centering
  \begin{subfigure}[b]{0.30\linewidth}
    \includegraphics[width=\linewidth ]{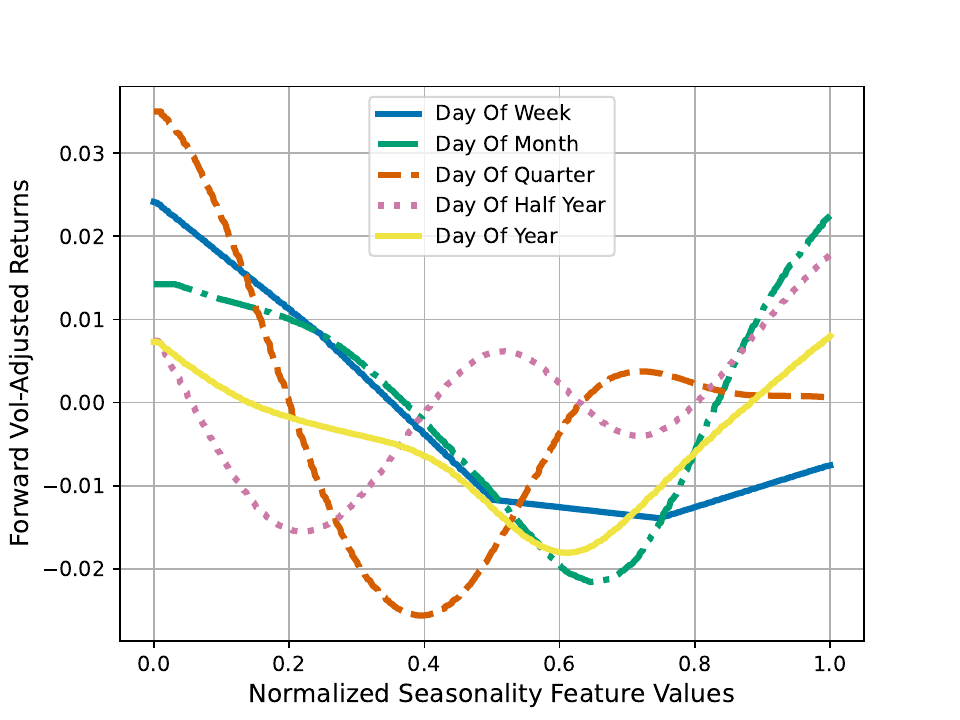}
    \caption{Australian Dollar}
  \end{subfigure}
  \begin{subfigure}[b]{0.30\linewidth}
    \includegraphics[width=\linewidth ]{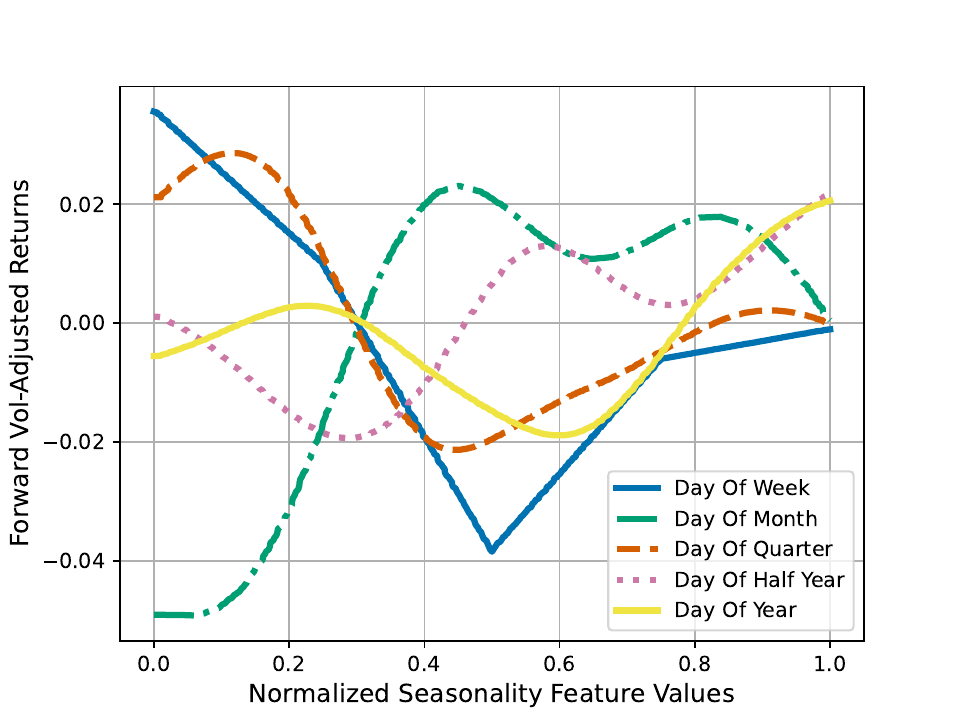}
    \caption{British Pound}
  \end{subfigure}
  \begin{subfigure}[b]{0.30\linewidth}
    \includegraphics[width=\linewidth ]{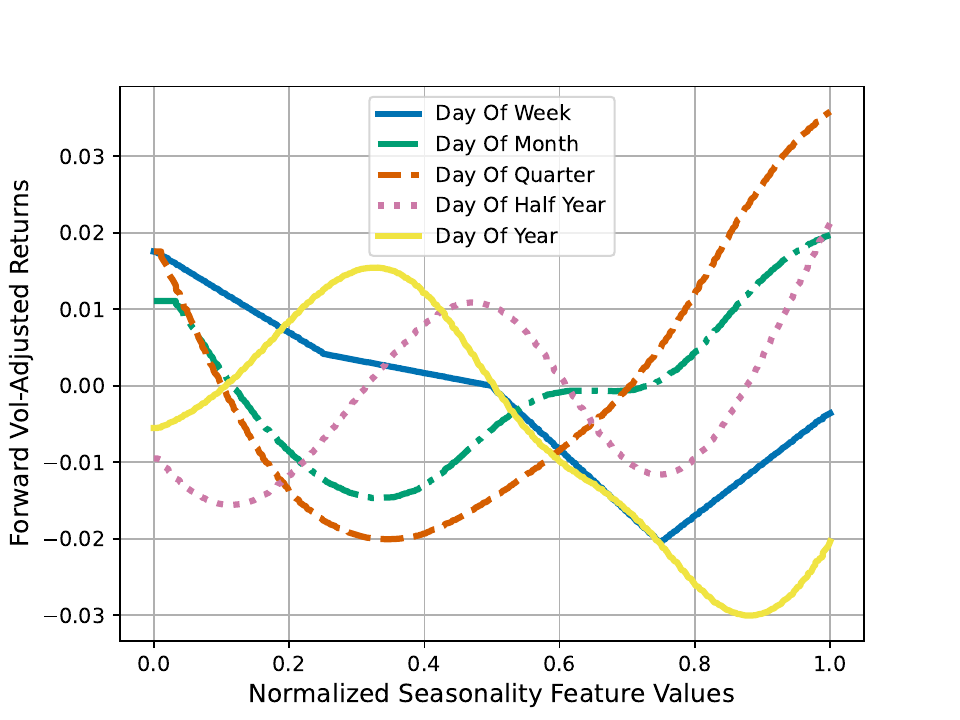}
    \caption{Canadian Dollar}
  \end{subfigure}
  \begin{subfigure}[b]{0.30\linewidth}
    \includegraphics[width=\linewidth ]{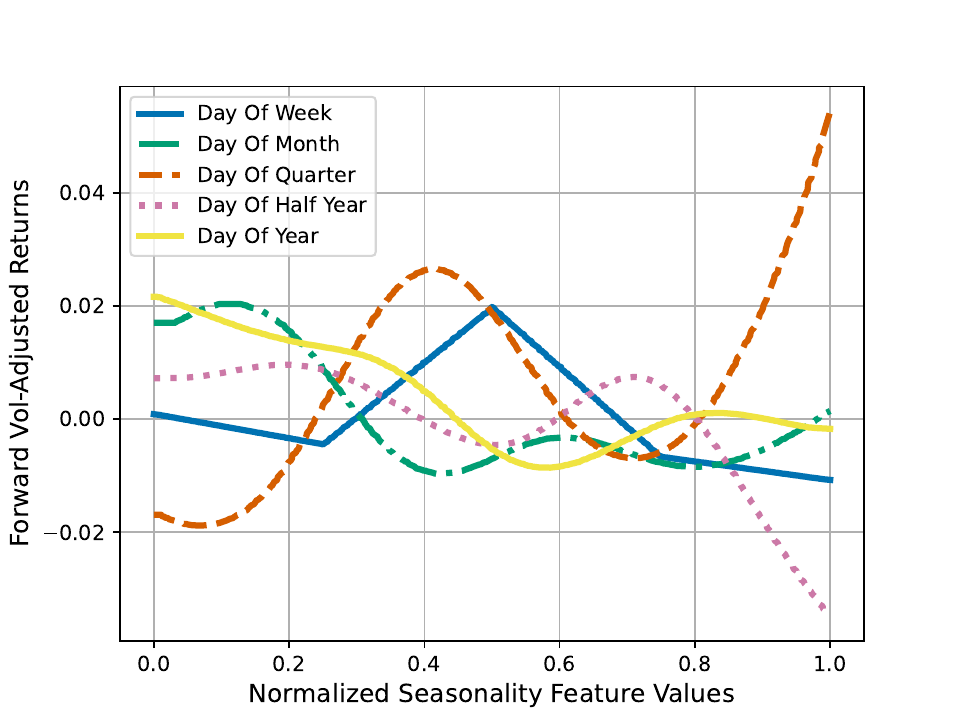}
    \caption{US Dollar}
  \end{subfigure}
  \begin{subfigure}[b]{0.30\linewidth}
    \includegraphics[width=\linewidth ]{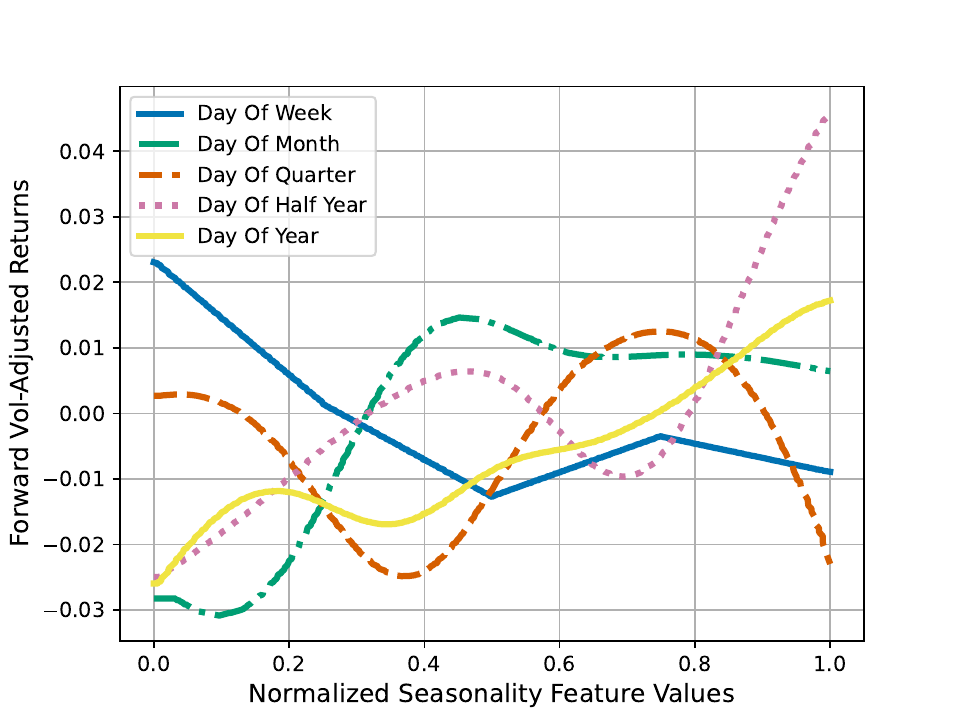}
    \caption{Euro}
  \end{subfigure}
  \begin{subfigure}[b]{0.30\linewidth}
    \includegraphics[width=\linewidth ]{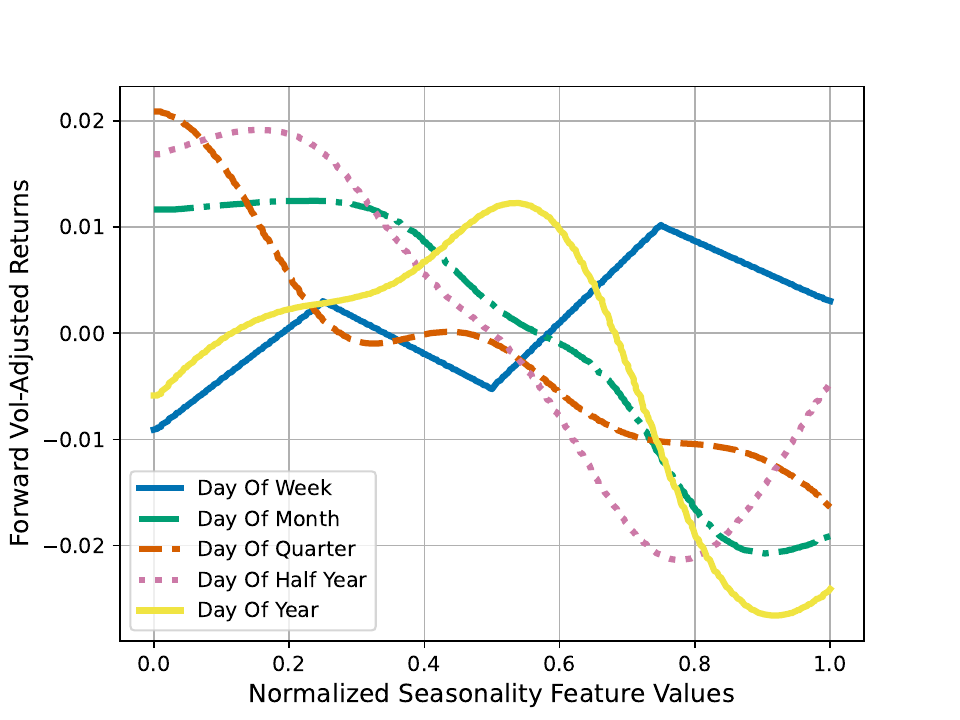}
    \caption{Japanese Yen}
  \end{subfigure}
  \begin{subfigure}[b]{0.30\linewidth}
    \includegraphics[width=\linewidth ]{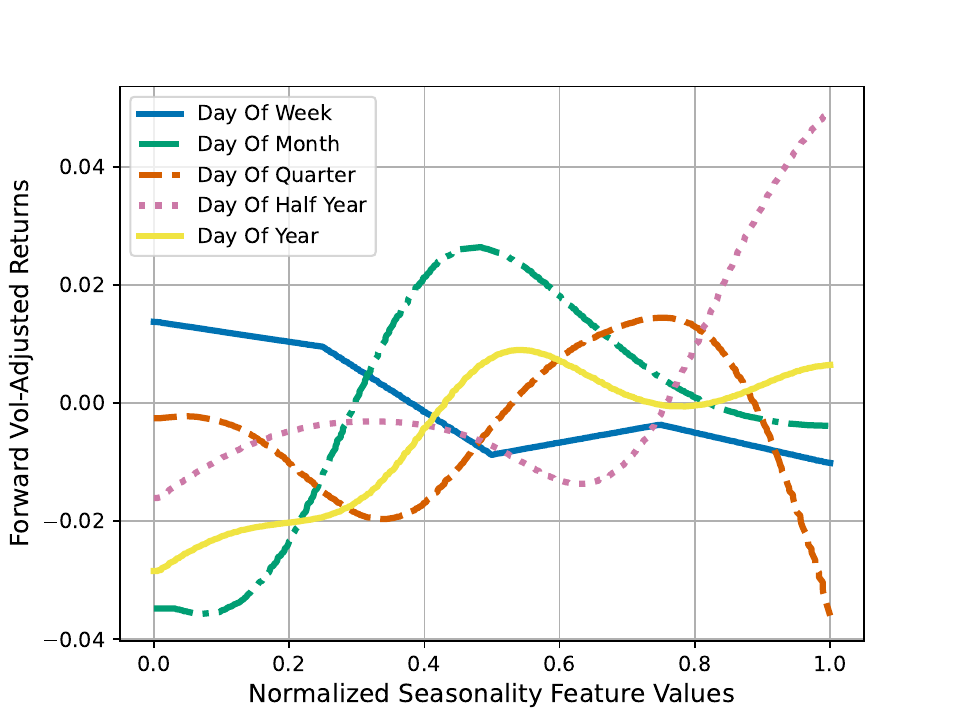}
    \caption{Swiss Franc}
  \end{subfigure}

  \caption{Estimated forward vol-adjusted returns as a function of the normalized seasonality features for global currency markets.
  Models are fit using all available training data from January $1950$ through June $2026$. }
\end{figure}

\subsection{Energy Markets}
\begin{figure}[H]
  \centering
  \begin{subfigure}[b]{0.30\linewidth}
    \includegraphics[width=\linewidth ]{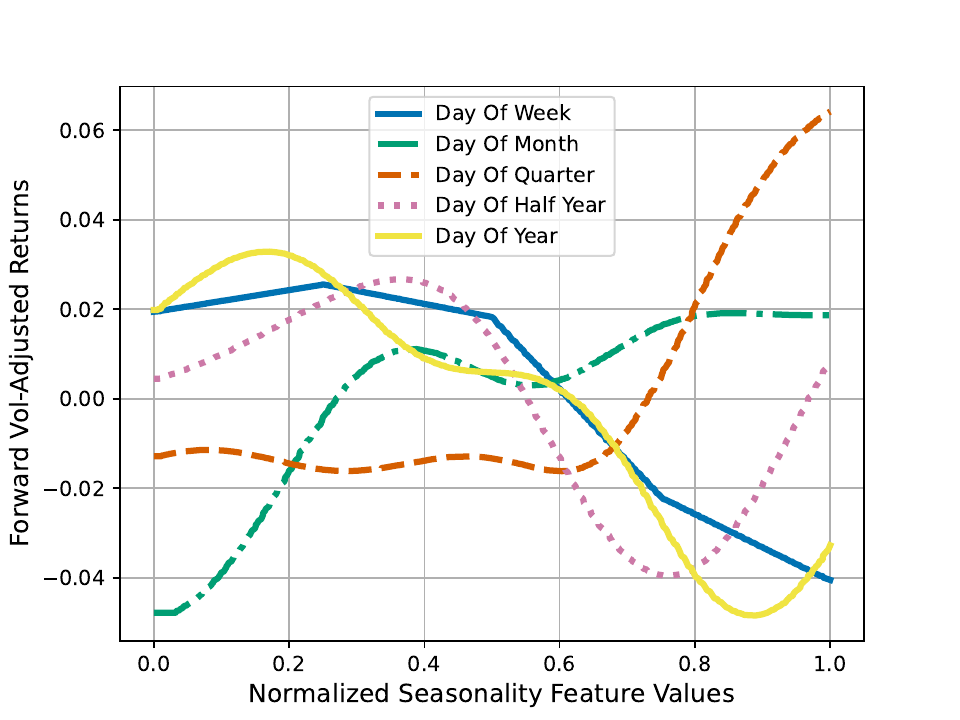}
    \caption{Crude Oild}
  \end{subfigure}
  \begin{subfigure}[b]{0.30\linewidth}
    \includegraphics[width=\linewidth ]{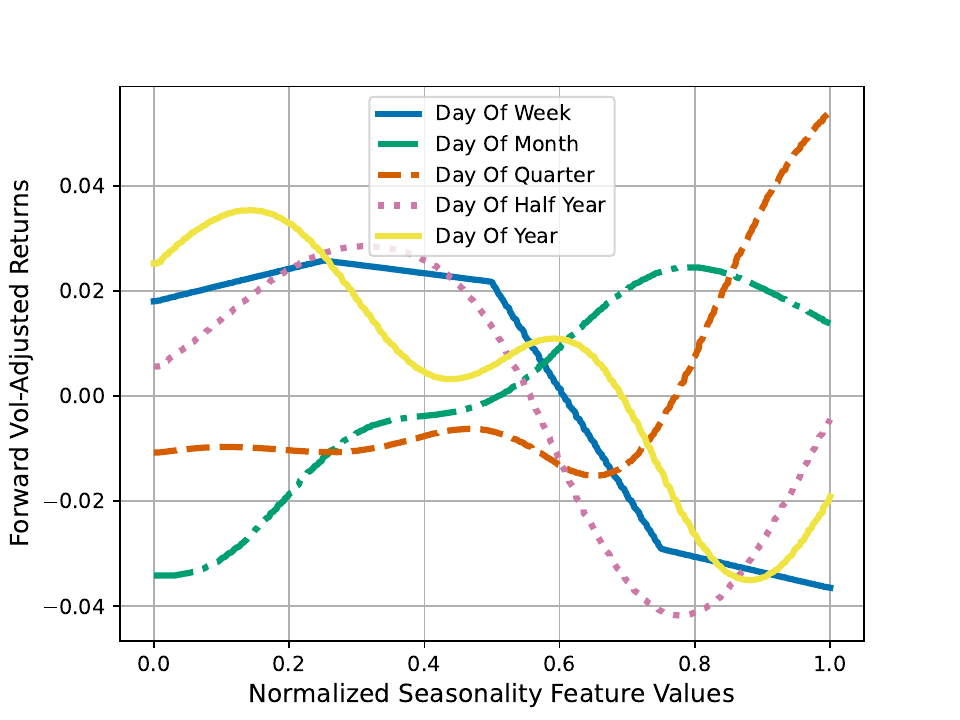}
    \caption{Brent Crude}
  \end{subfigure}
  \begin{subfigure}[b]{0.30\linewidth}
    \includegraphics[width=\linewidth ]{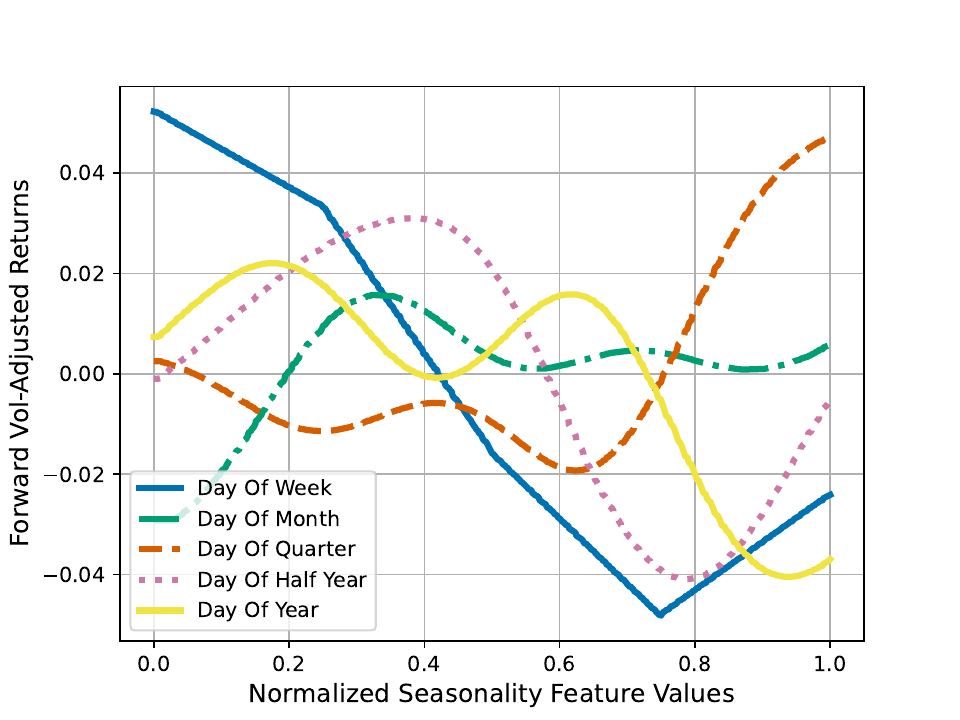}
    \caption{Heating Oil}
  \end{subfigure}
  \begin{subfigure}[b]{0.30\linewidth}
    \includegraphics[width=\linewidth ]{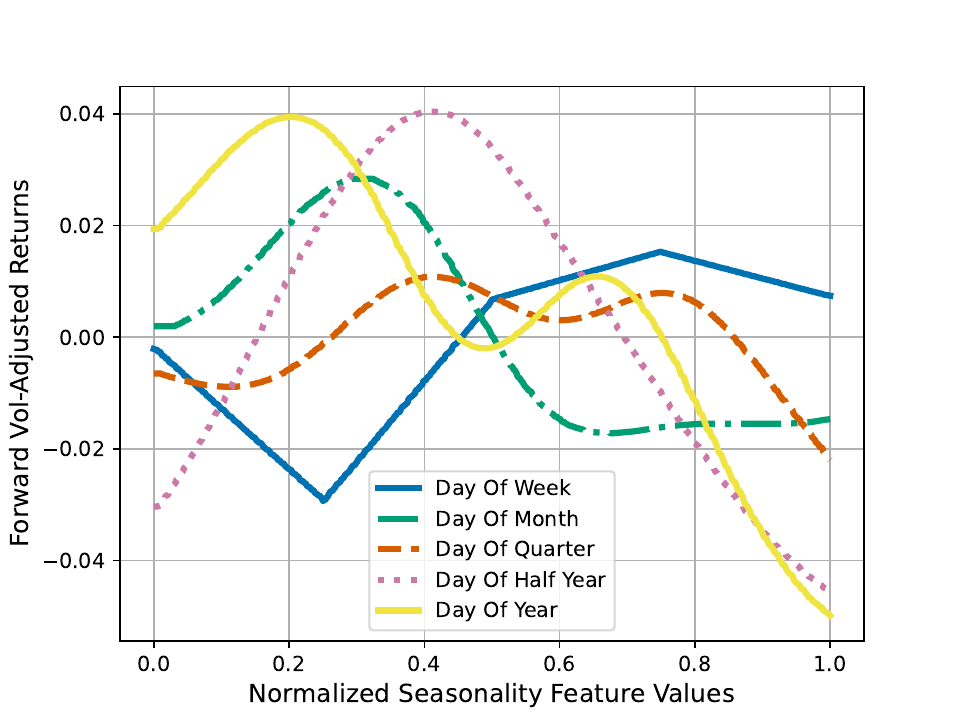}
    \caption{Natural Gas}
  \end{subfigure}
  \begin{subfigure}[b]{0.30\linewidth}
    \includegraphics[width=\linewidth ]{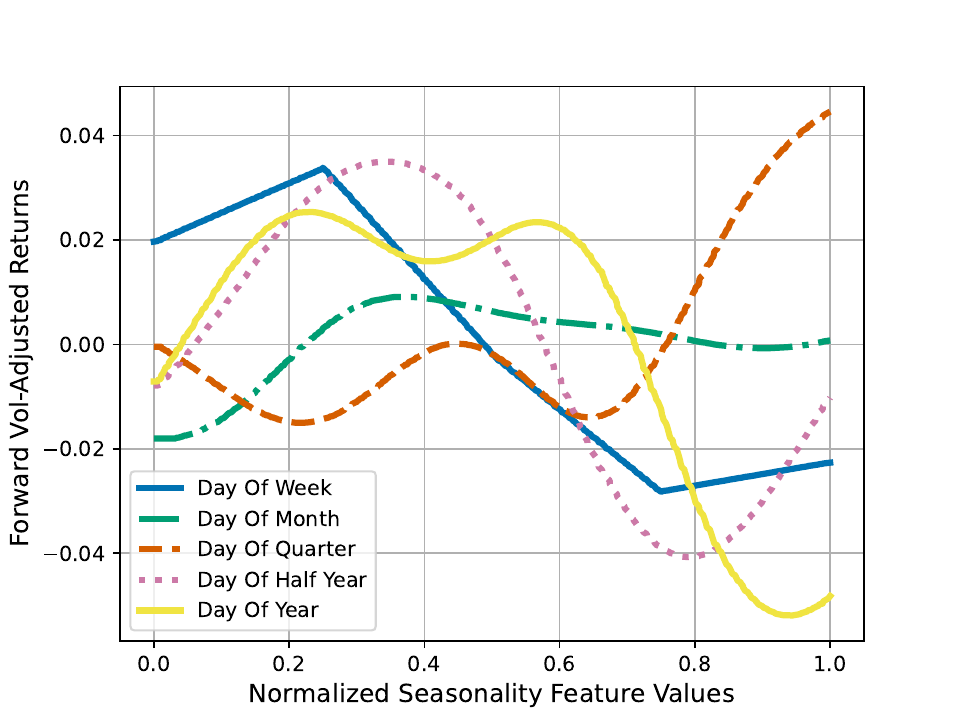}
    \caption{Gasoil}
  \end{subfigure}
  \begin{subfigure}[b]{0.30\linewidth}
    \includegraphics[width=\linewidth ]{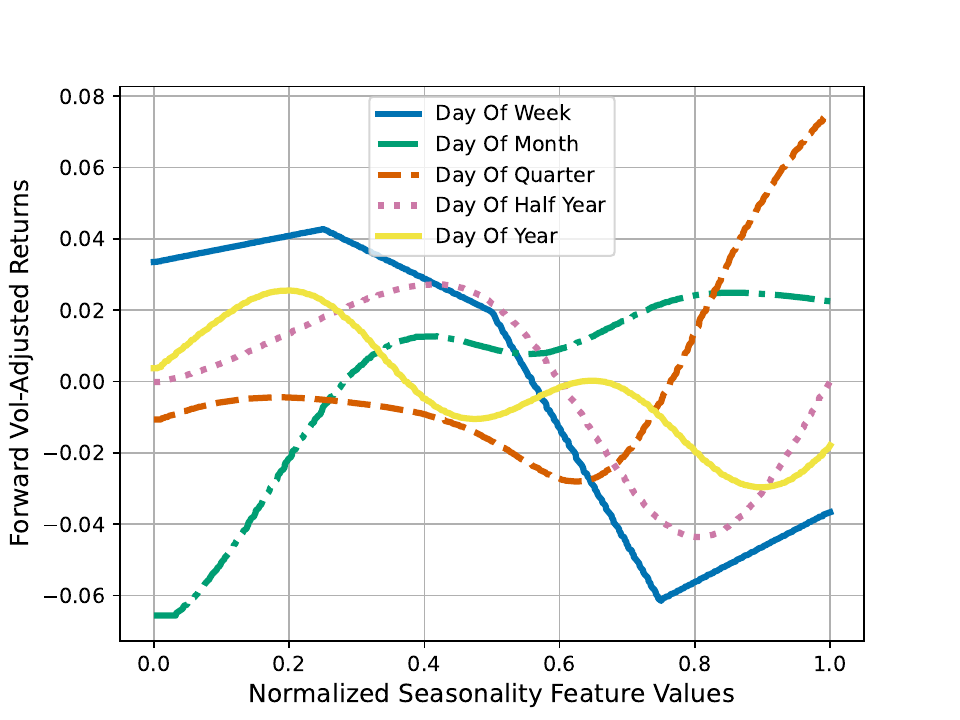}
    \caption{RBOB}
  \end{subfigure}

  \caption{Estimated forward vol-adjusted returns as a function of the normalized seasonality features for global energy markets.
  Models are fit using all available training data from January $1950$ through June $2026$. }
\end{figure}

\subsection{Equity Markets}
\begin{figure}[H]
  \centering
  \begin{subfigure}[b]{0.30\linewidth}
    \includegraphics[width=\linewidth ]{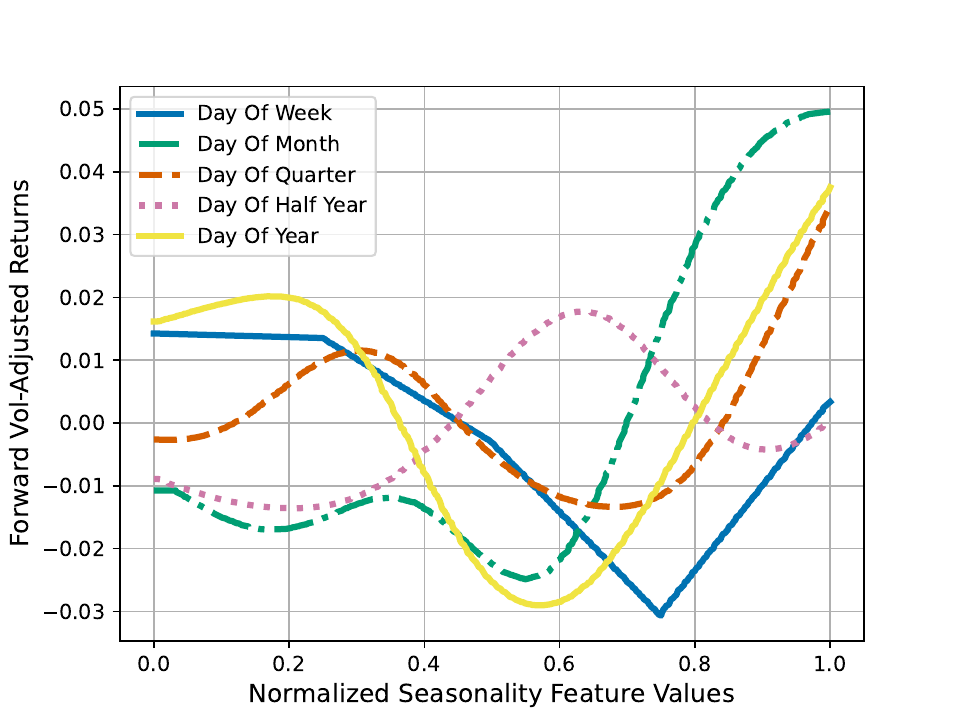}
    \caption{CAC 40}
  \end{subfigure}
  \begin{subfigure}[b]{0.30\linewidth}
    \includegraphics[width=\linewidth ]{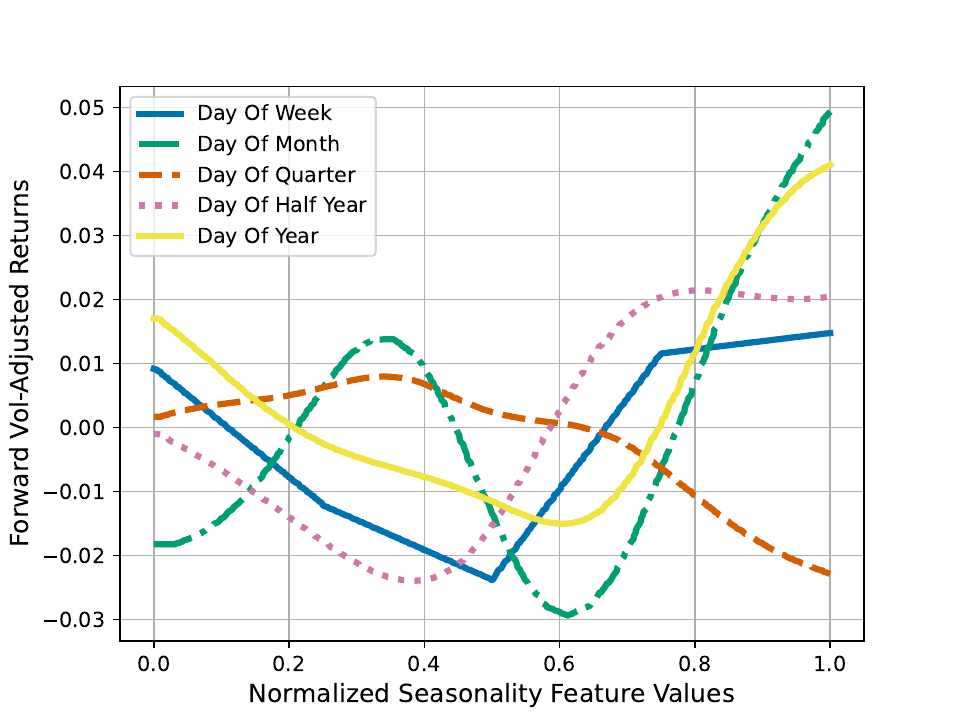}
    \caption{S\&P 500}
  \end{subfigure}
  \begin{subfigure}[b]{0.30\linewidth}
    \includegraphics[width=\linewidth ]{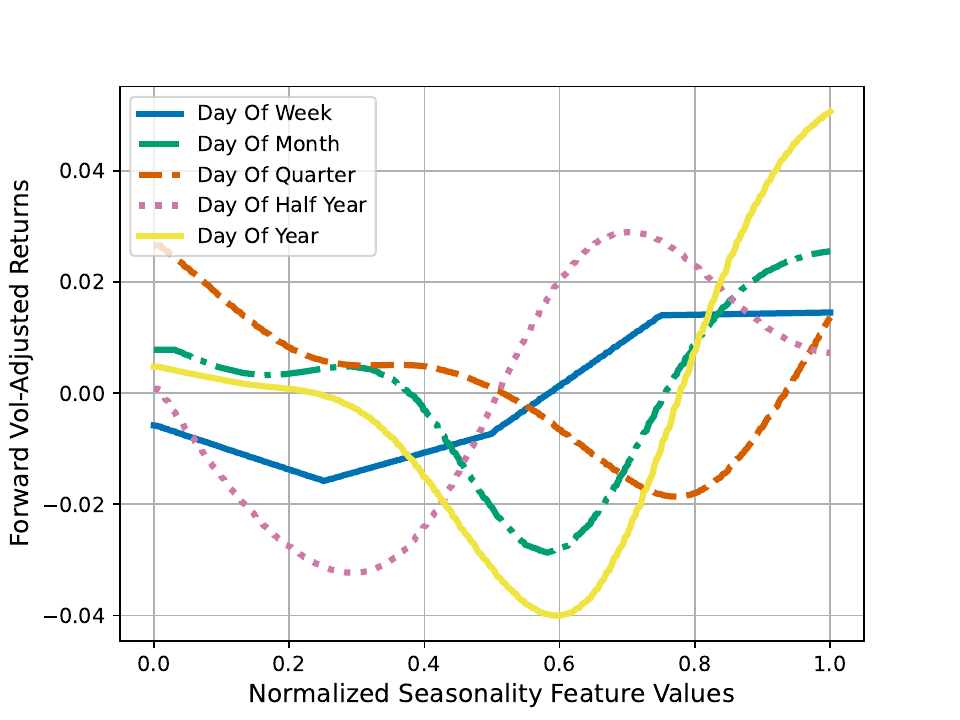}
    \caption{DAX}
  \end{subfigure}
  \begin{subfigure}[b]{0.30\linewidth}
    \includegraphics[width=\linewidth ]{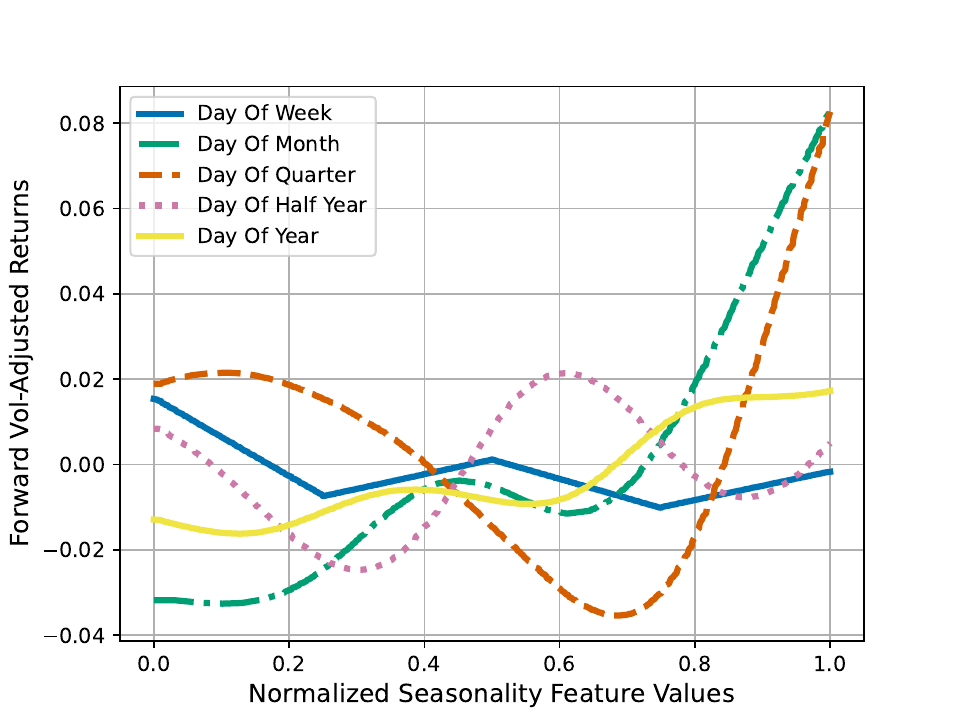}
    \caption{Hang Seng}
  \end{subfigure}
  \begin{subfigure}[b]{0.30\linewidth}
    \includegraphics[width=\linewidth ]{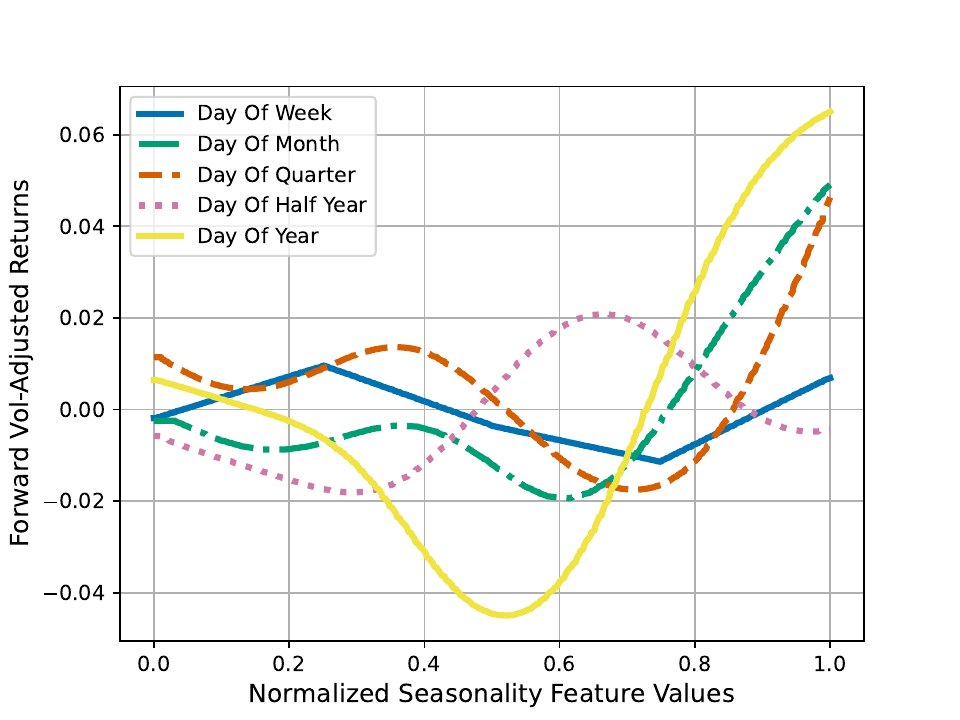}
    \caption{IBEX 35}
  \end{subfigure}
  \begin{subfigure}[b]{0.30\linewidth}
    \includegraphics[width=\linewidth ]{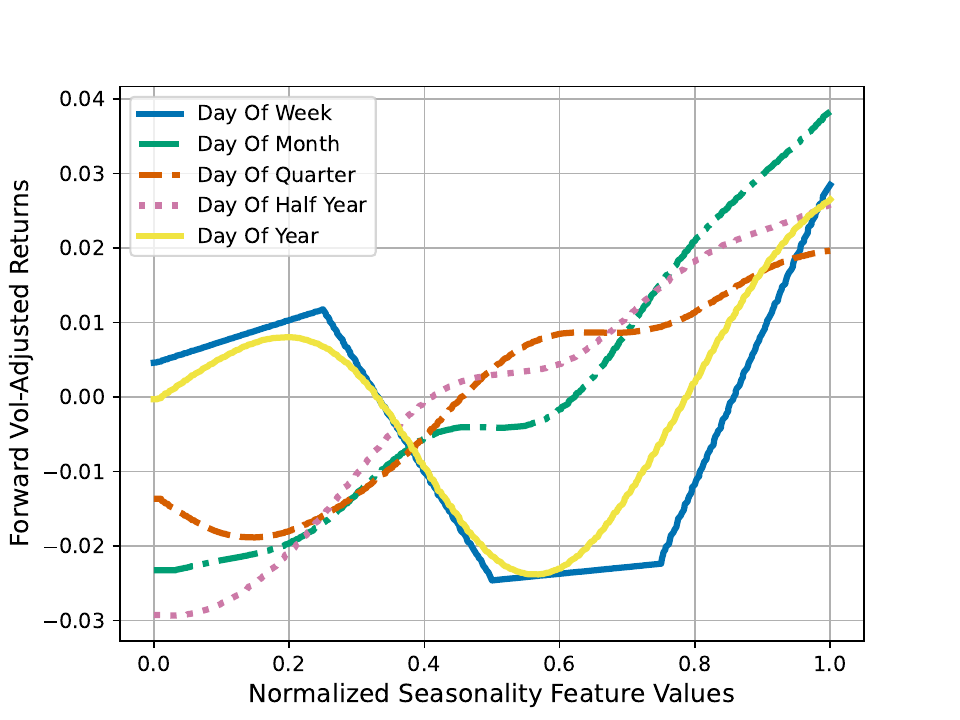}
    \caption{Nikkei}
  \end{subfigure}
  \begin{subfigure}[b]{0.30\linewidth}
    \includegraphics[width=\linewidth ]{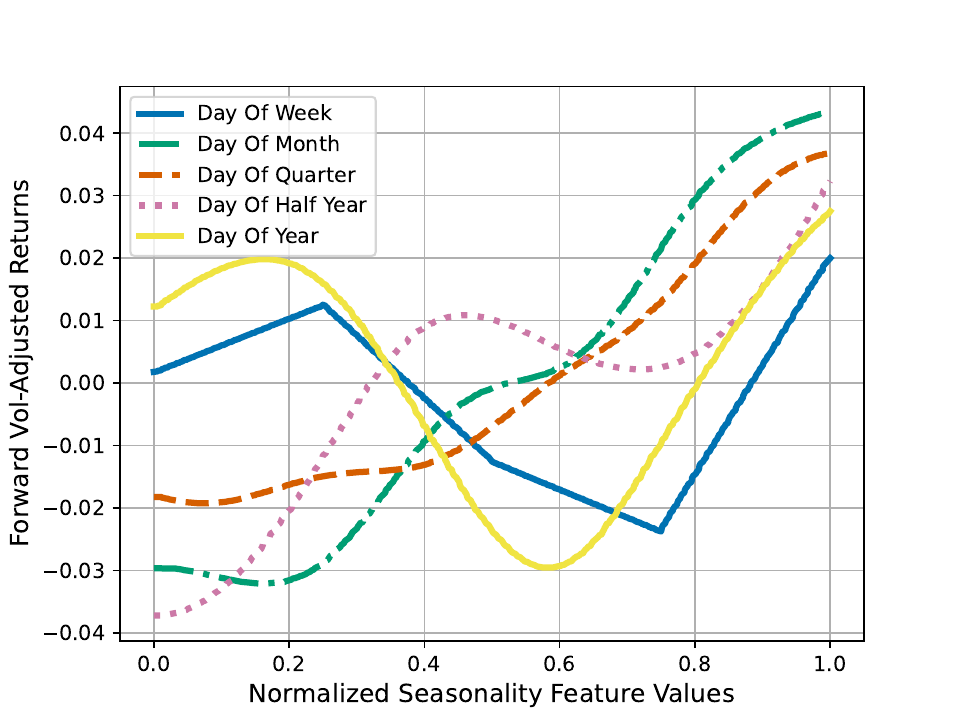}
    \caption{Topix}
  \end{subfigure}
  \begin{subfigure}[b]{0.30\linewidth}
    \includegraphics[width=\linewidth ]{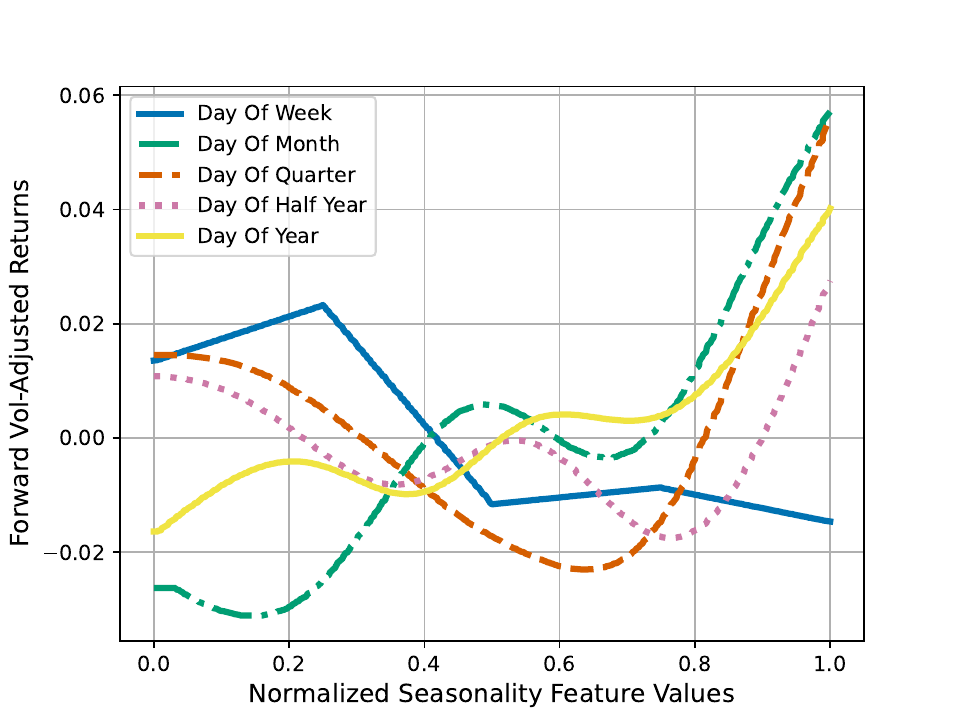}
    \caption{ASX 200}
  \end{subfigure}
  \begin{subfigure}[b]{0.30\linewidth}
    \includegraphics[width=\linewidth ]{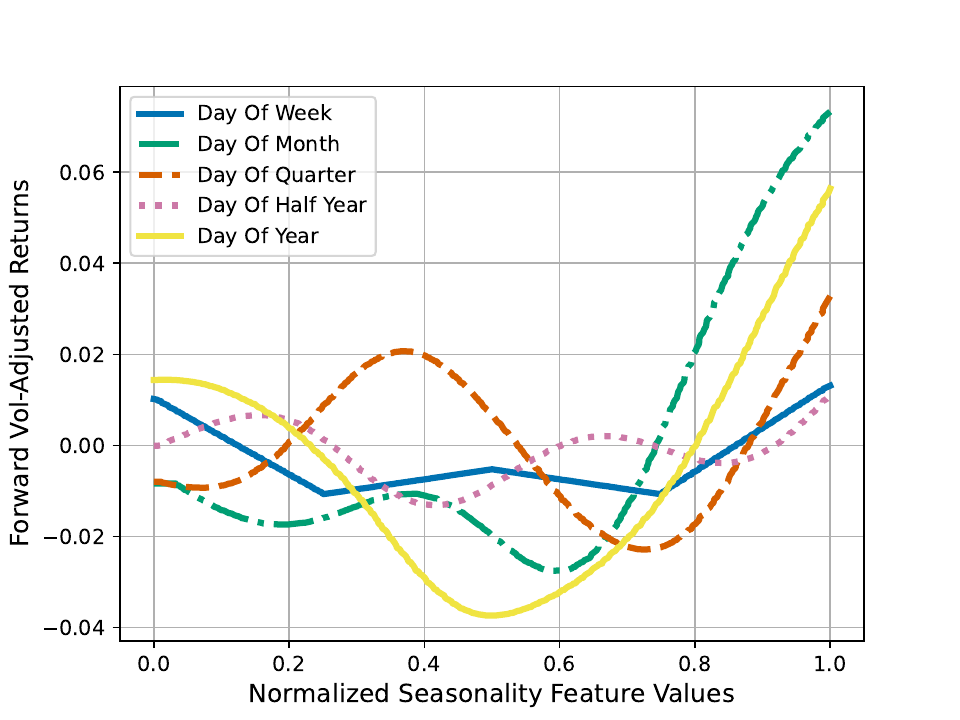}
    \caption{FTSE 100}
  \end{subfigure}
  
  \caption{Estimated forward vol-adjusted returns as a function of the normalized seasonality features for global equity markets.
  Models are fit using all available training data from January $1950$ through June $2026$. }
\end{figure}

\subsection{Grain Markets}
\begin{figure}[H]
  \centering
  \begin{subfigure}[b]{0.30\linewidth}
    \includegraphics[width=\linewidth ]{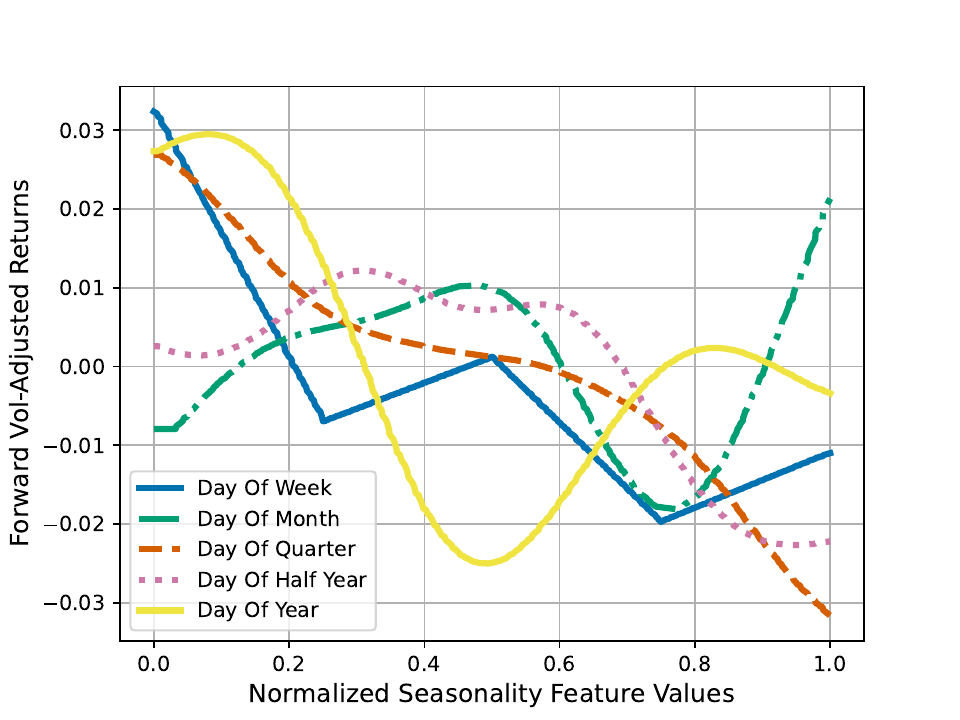}
    \caption{Bean Oil}
  \end{subfigure}
  \begin{subfigure}[b]{0.30\linewidth}
    \includegraphics[width=\linewidth ]{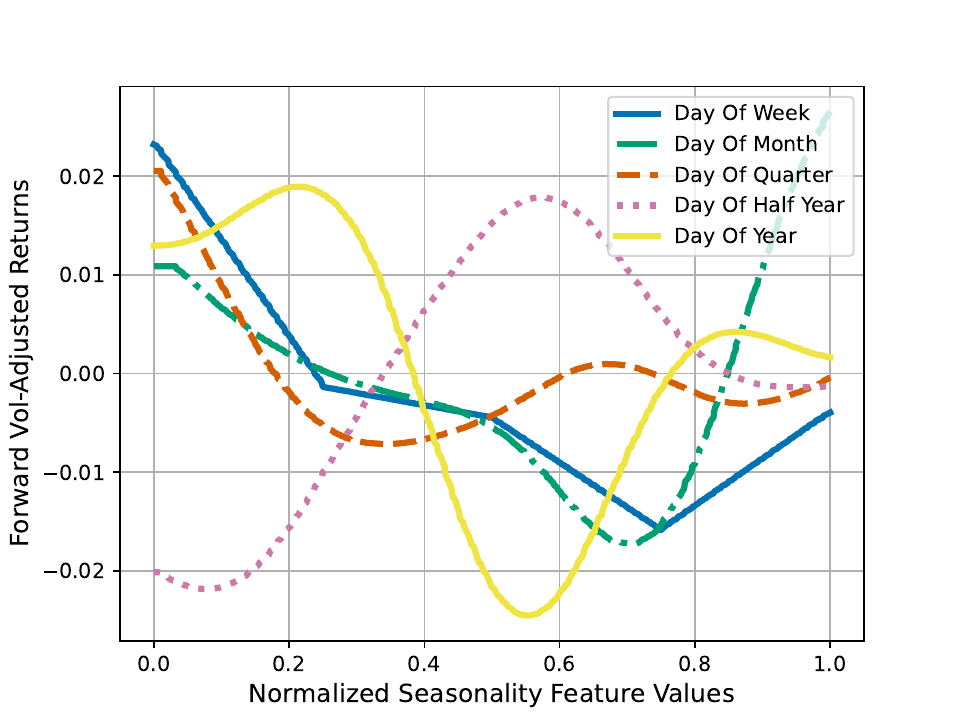}
    \caption{Corn}
  \end{subfigure}
  \begin{subfigure}[b]{0.30\linewidth}
    \includegraphics[width=\linewidth ]{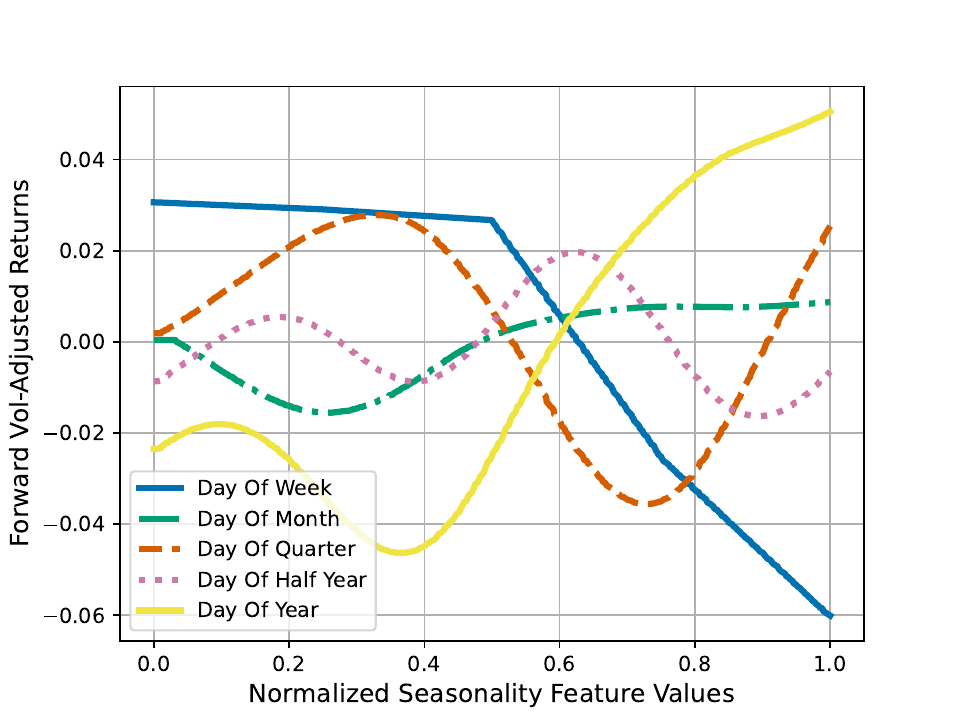}
    \caption{Palm Oil}
  \end{subfigure}
  \begin{subfigure}[b]{0.30\linewidth}
    \includegraphics[width=\linewidth ]{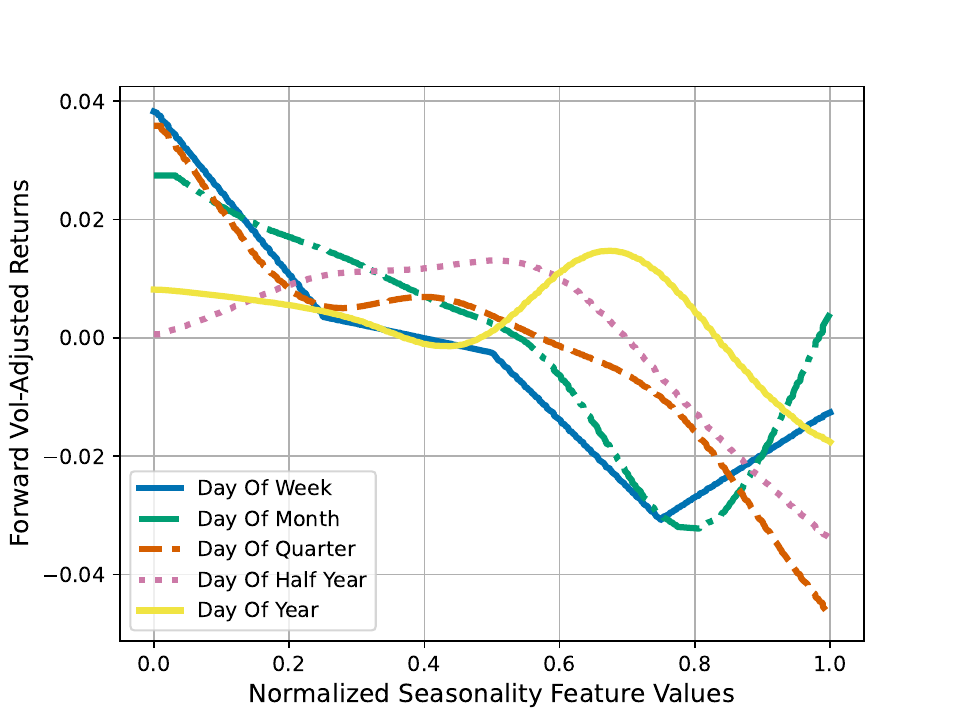}
    \caption{KC Wheat}
  \end{subfigure}
  \begin{subfigure}[b]{0.30\linewidth}
    \includegraphics[width=\linewidth ]{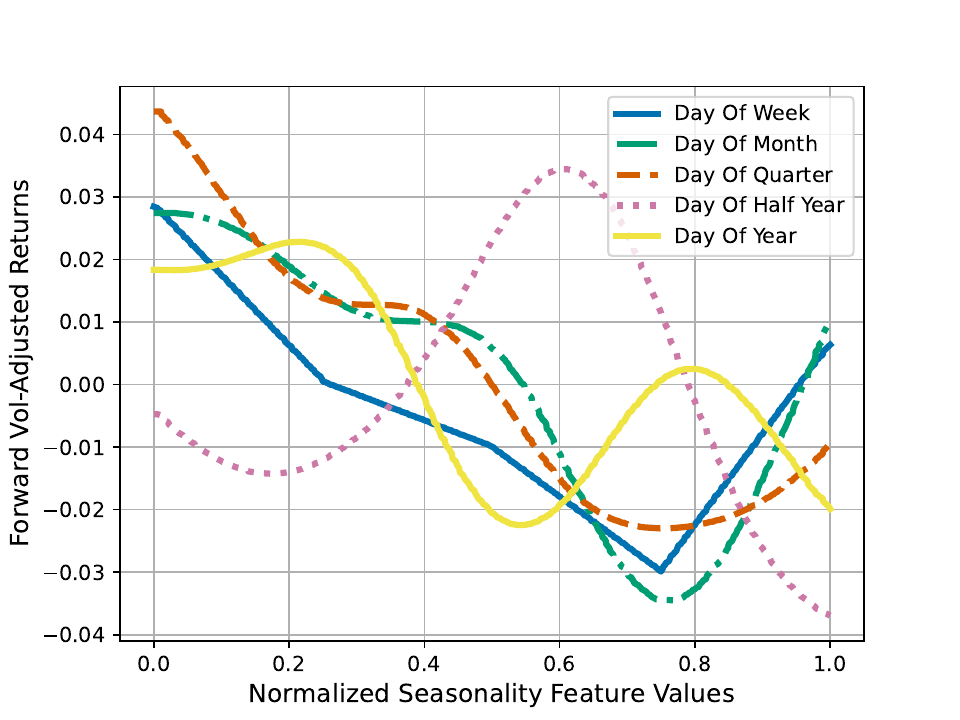}
    \caption{Canola}
  \end{subfigure}
  \begin{subfigure}[b]{0.30\linewidth}
    \includegraphics[width=\linewidth ]{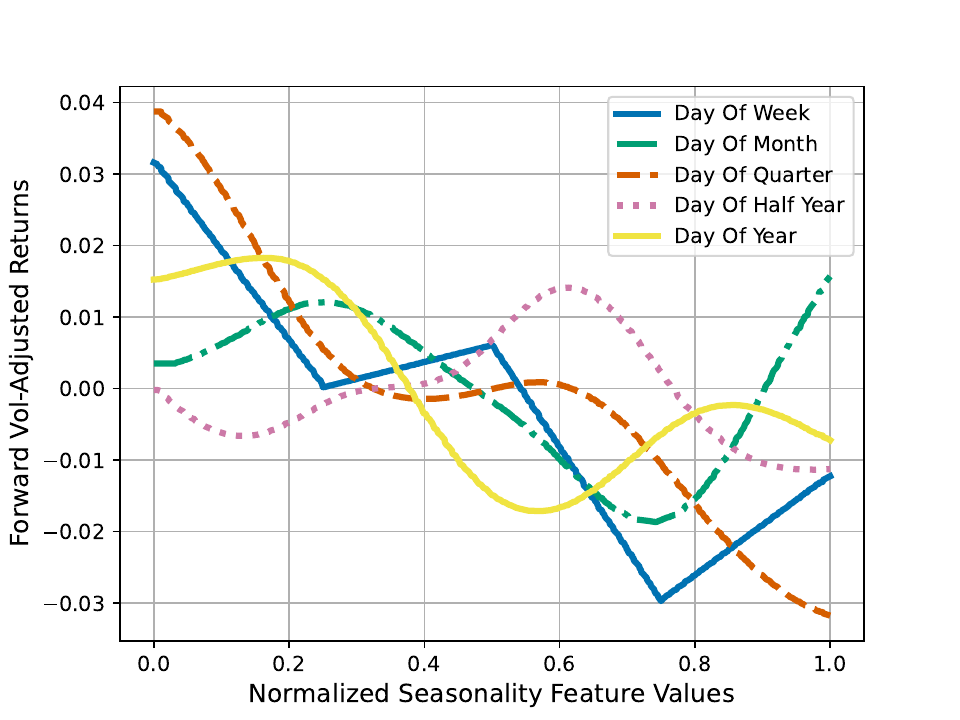}
    \caption{Soybean}
  \end{subfigure}
  \begin{subfigure}[b]{0.30\linewidth}
    \includegraphics[width=\linewidth ]{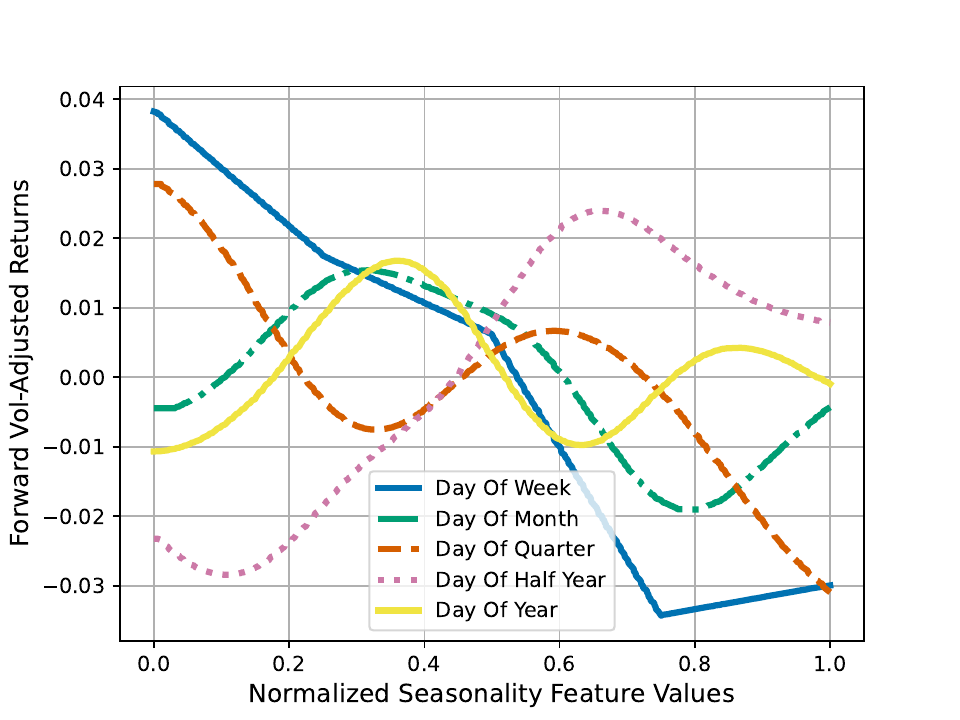}
    \caption{Soybean Meal}
  \end{subfigure}
  \begin{subfigure}[b]{0.30\linewidth}
    \includegraphics[width=\linewidth ]{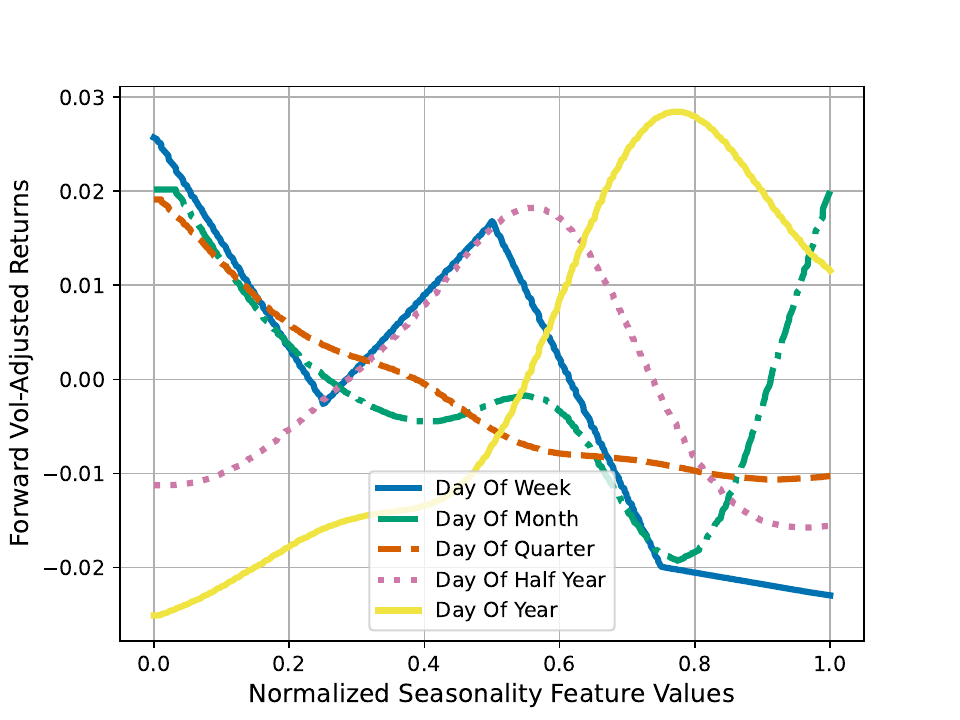}
    \caption{Wheat}
  \end{subfigure}

  \caption{Estimated forward vol-adjusted returns as a function of the normalized seasonality features for global grain markets.
  Models are fit using all available training data from January $1950$ through June $2026$. }
\end{figure}

\subsection{Metal Markets}
\begin{figure}[H]
  \centering
  \begin{subfigure}[b]{0.30\linewidth}
    \includegraphics[width=\linewidth ]{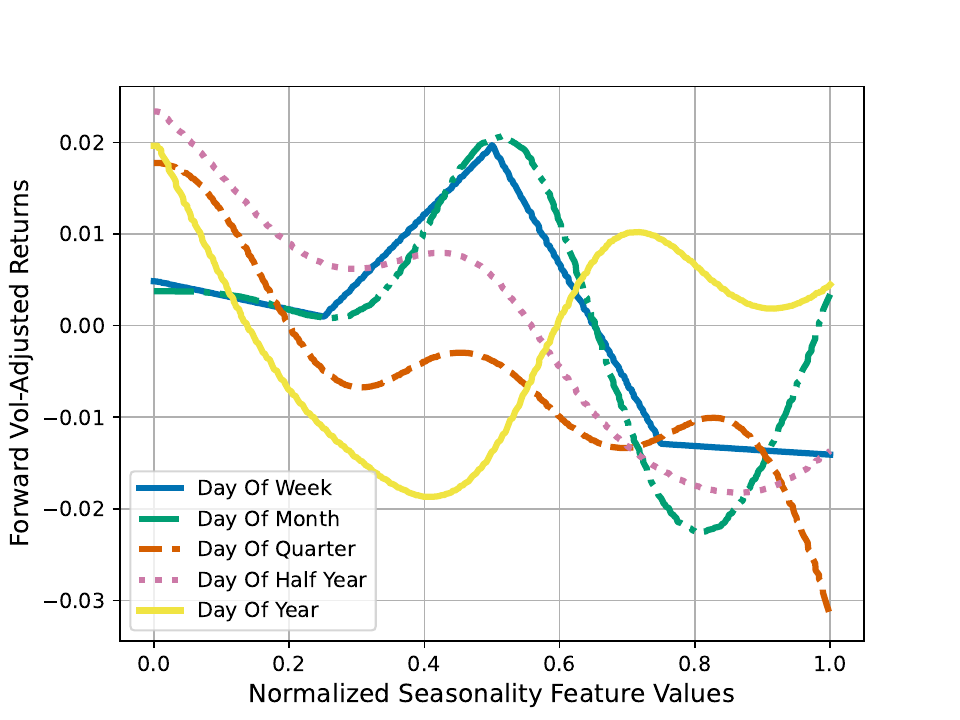}
    \caption{Gold}
  \end{subfigure}
  \begin{subfigure}[b]{0.30\linewidth}
    \includegraphics[width=\linewidth ]{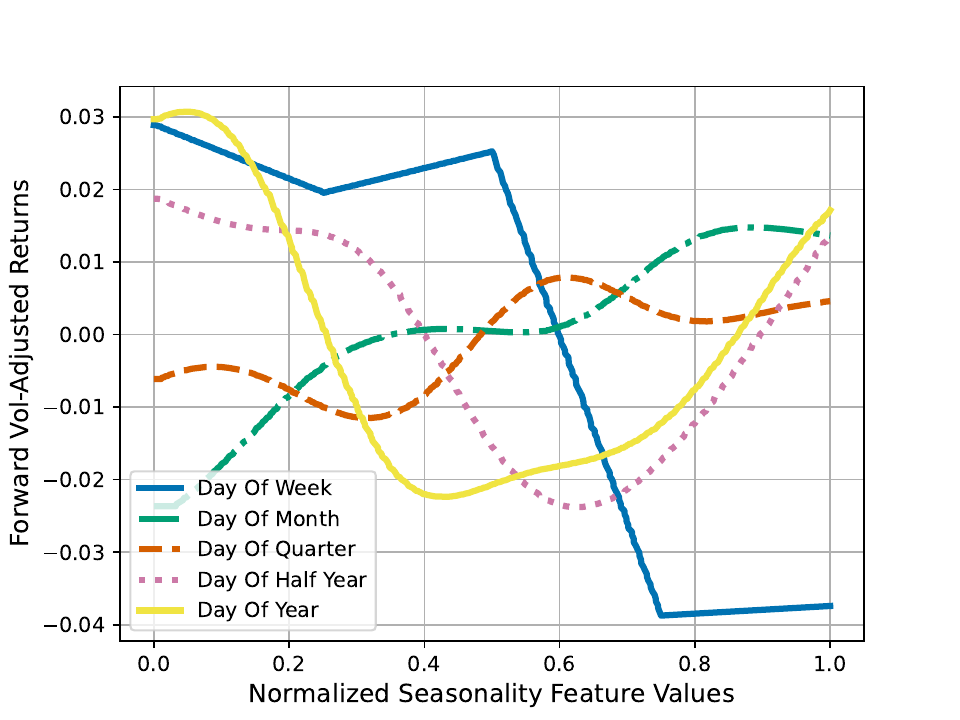}
    \caption{Copper}
  \end{subfigure}
  \begin{subfigure}[b]{0.30\linewidth}
    \includegraphics[width=\linewidth ]{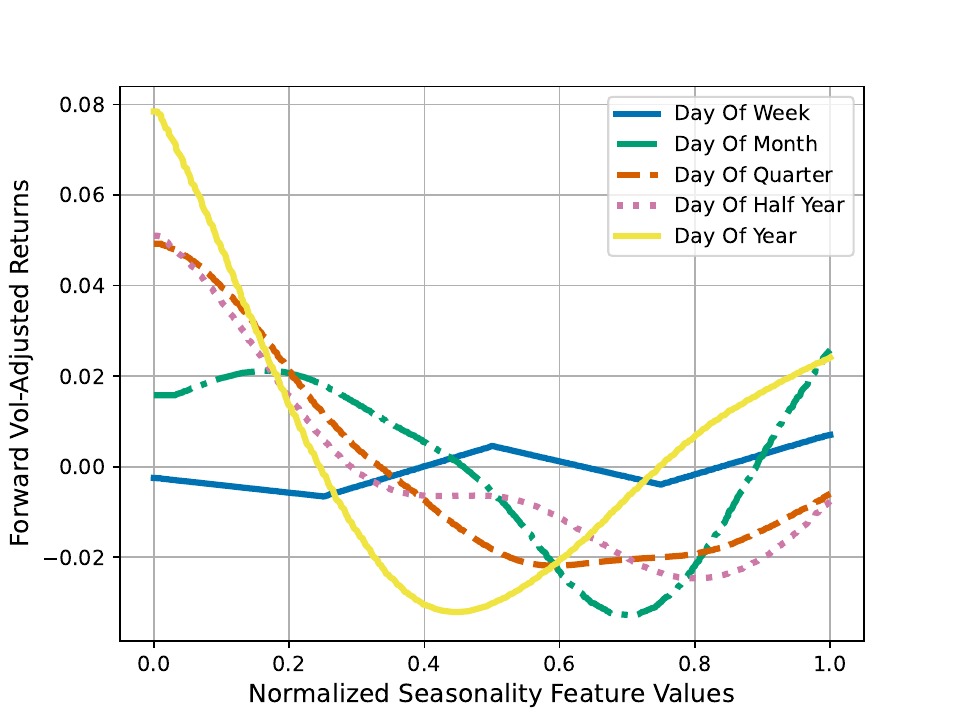}
    \caption{Palladium}
  \end{subfigure}
  \begin{subfigure}[b]{0.30\linewidth}
    \includegraphics[width=\linewidth ]{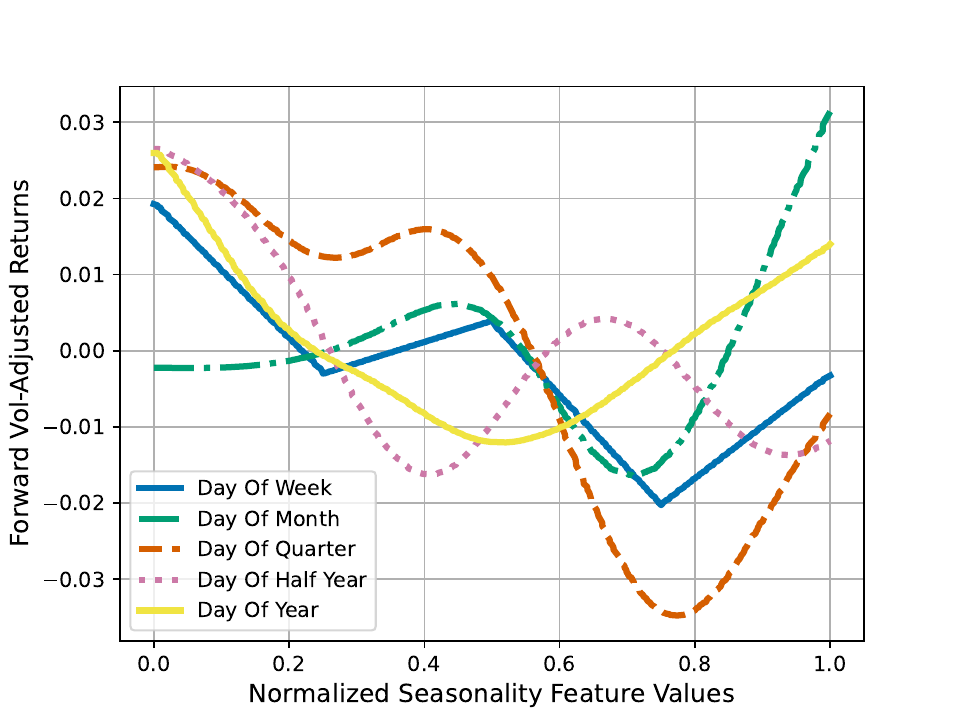}
    \caption{Platinum}
  \end{subfigure}
  \begin{subfigure}[b]{0.30\linewidth}
    \includegraphics[width=\linewidth ]{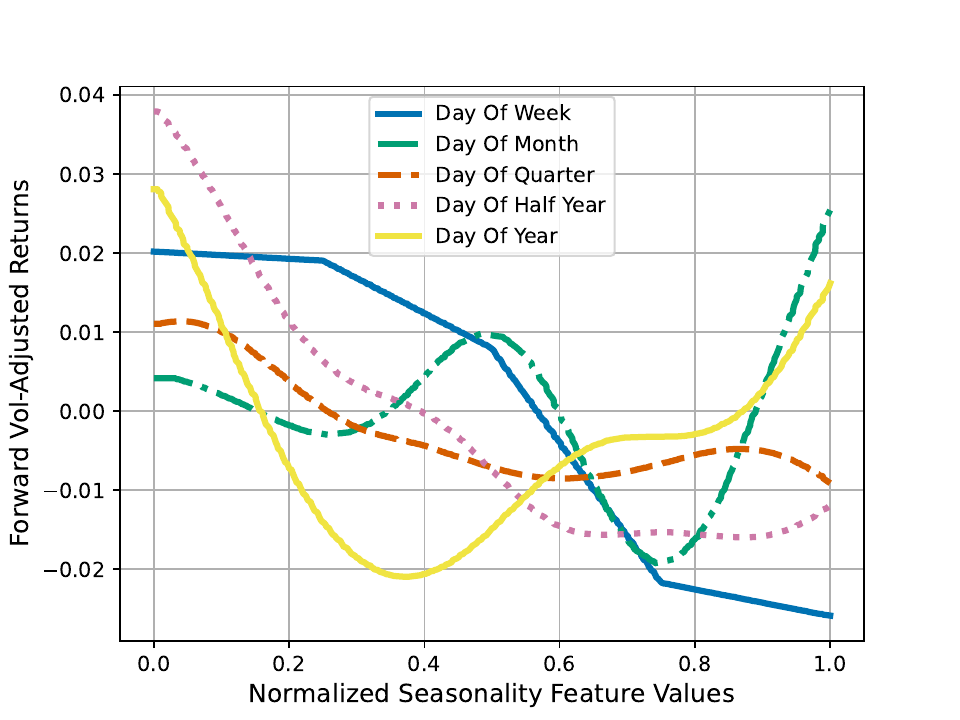}
    \caption{Silver}
  \end{subfigure}
  
  \caption{Estimated forward vol-adjusted returns as a function of the normalized seasonality features for global metal markets.
  Models are fit using all available training data from January $1950$ through June $2026$. }
\end{figure}

\subsection{Soft Markets}
\begin{figure}[H]
  \centering
  \begin{subfigure}[b]{0.30\linewidth}
    \includegraphics[width=\linewidth ]{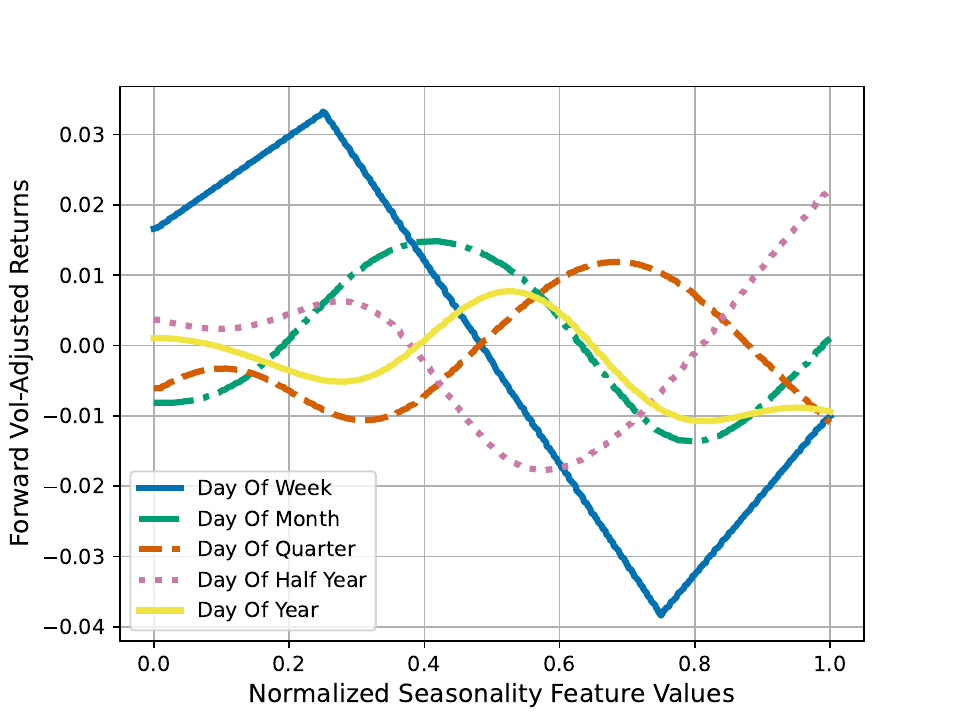}
    \caption{Cocoa}
  \end{subfigure}
  \begin{subfigure}[b]{0.30\linewidth}
    \includegraphics[width=\linewidth ]{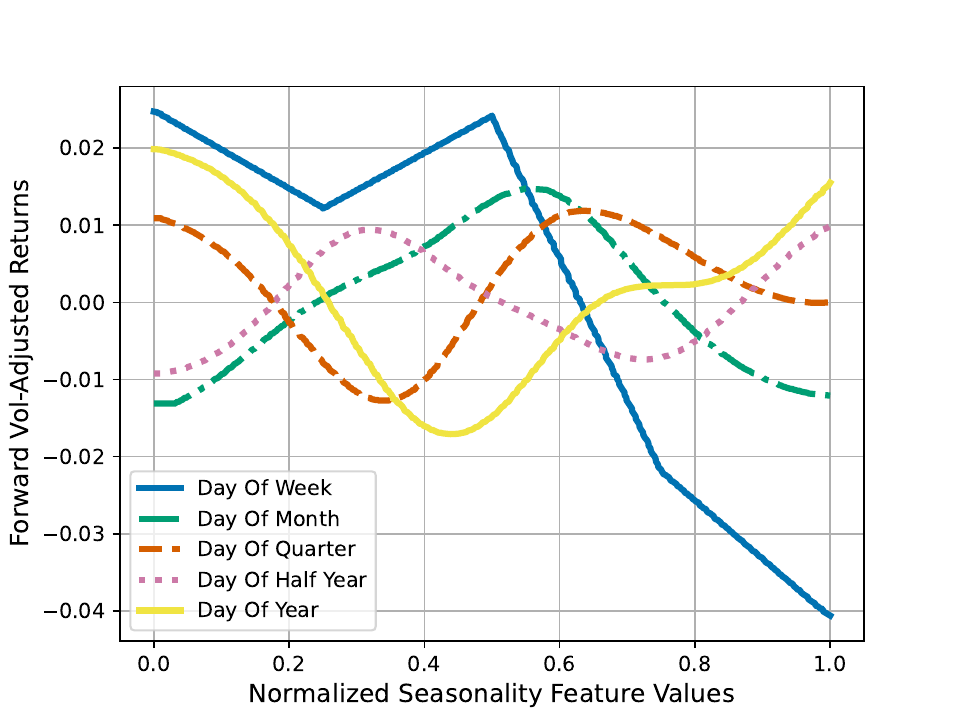}
    \caption{Cotton}
  \end{subfigure}
  \begin{subfigure}[b]{0.30\linewidth}
    \includegraphics[width=\linewidth ]{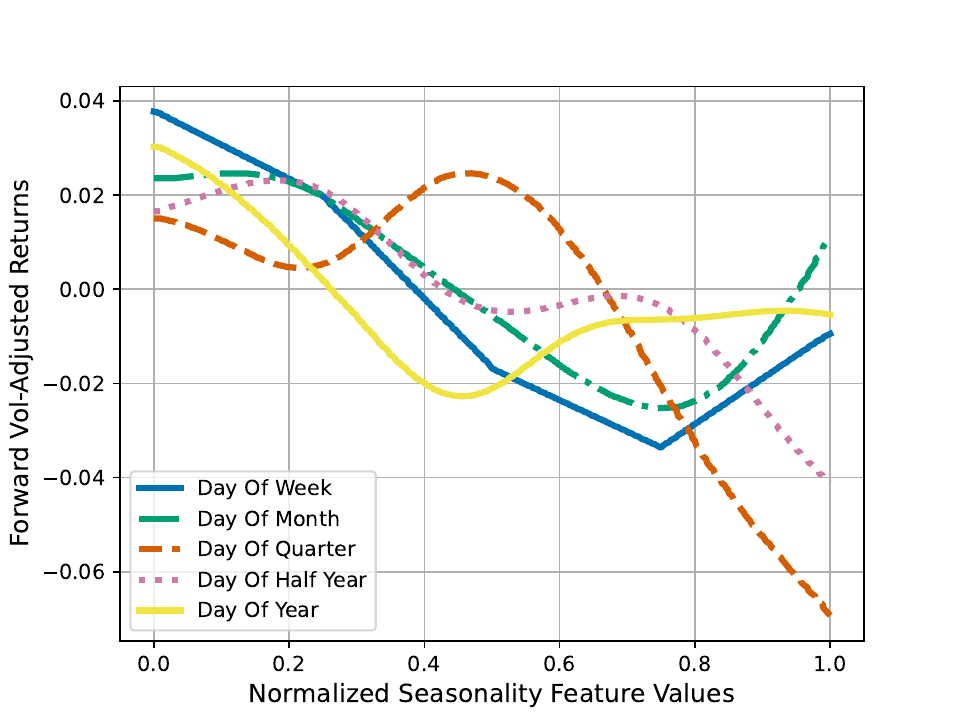}
    \caption{Robusta Coffee}
  \end{subfigure}
  \begin{subfigure}[b]{0.30\linewidth}
    \includegraphics[width=\linewidth ]{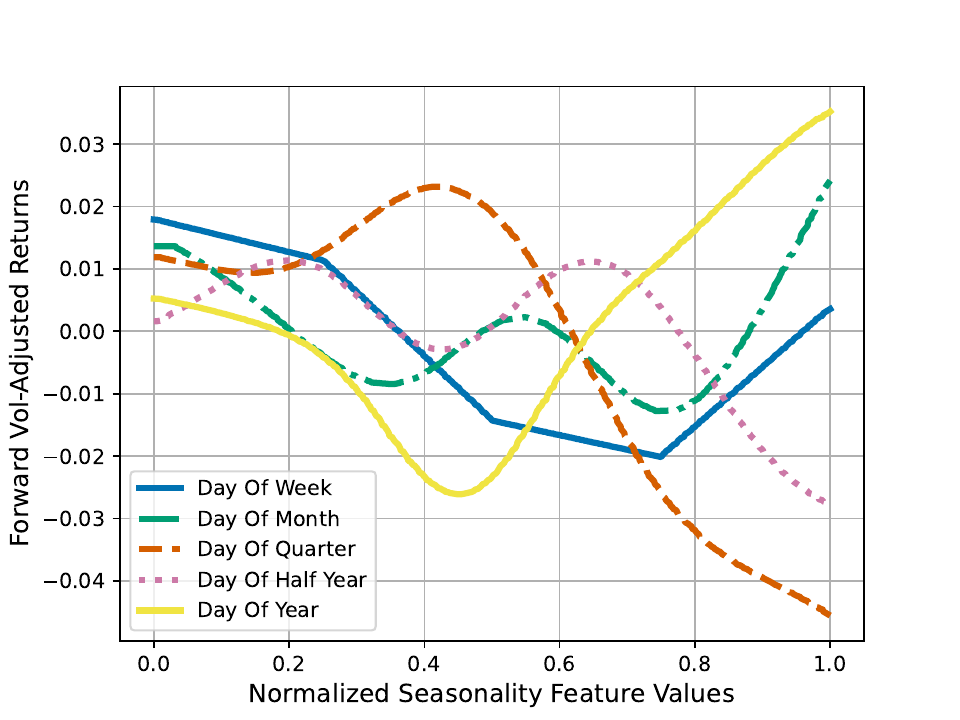}
    \caption{Coffee}
  \end{subfigure}
  \begin{subfigure}[b]{0.30\linewidth}
    \includegraphics[width=\linewidth ]{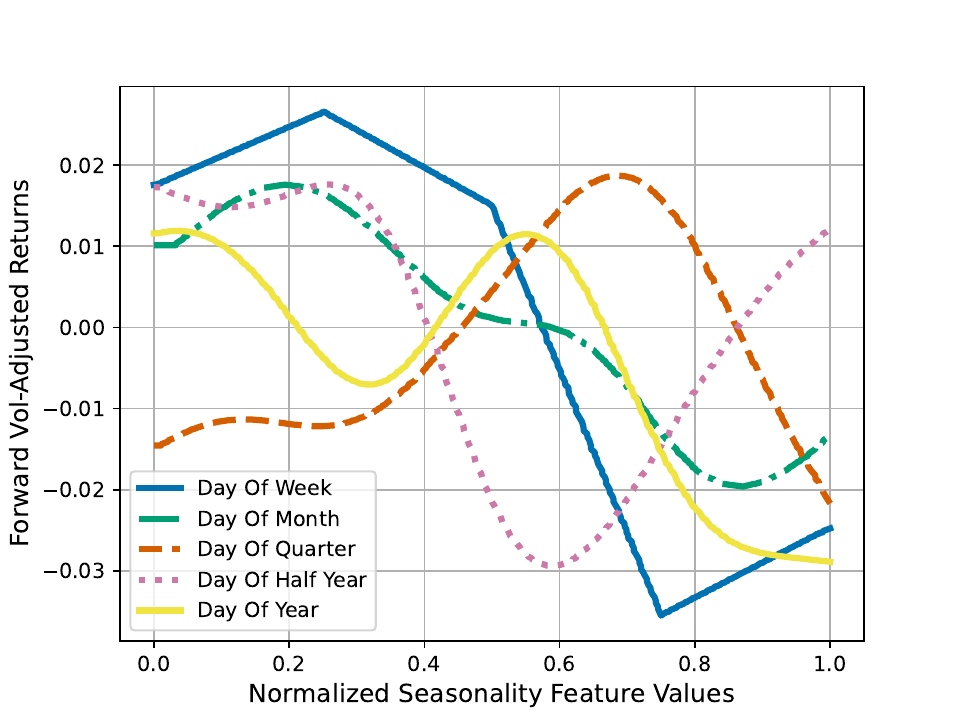}
    \caption{UK Cocoa}
  \end{subfigure}
  \begin{subfigure}[b]{0.30\linewidth}
    \includegraphics[width=\linewidth ]{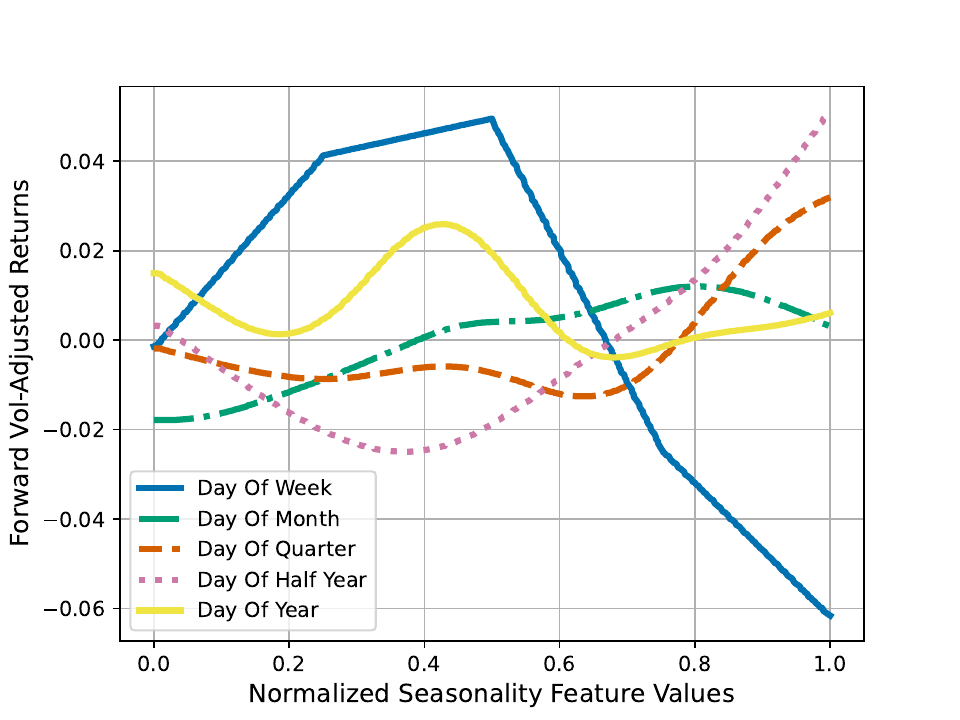}
    \caption{Uk Sugar}
  \end{subfigure}
  \begin{subfigure}[b]{0.30\linewidth}
    \includegraphics[width=\linewidth ]{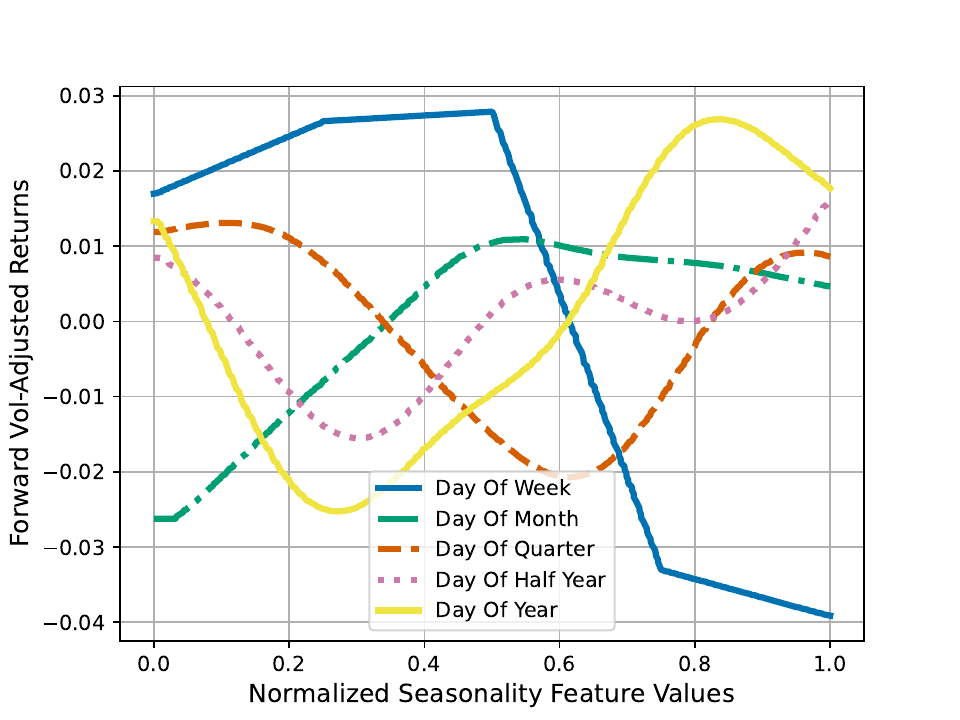}
    \caption{Sugar}
  \end{subfigure}

  \caption{Estimated forward vol-adjusted returns as a function of the normalized seasonality features for global soft markets.
  Models are fit using all available training data from January $1950$ through June $2026$. }
\end{figure}

\section{Futures Market Data}
\begin{table}[H]
\centering
\scriptsize
\begin{tabular}{l l l l l}
\hline
\textbf{Symbol} & \textbf{Market} &  \textbf{Tick Size} & \textbf{Trade Slippage Cost} & \textbf{Commission Cost}\\
\hline
\textbf{Bond} & & & & \\
CN & Cdn 10Yr & 0.01 & 1.0 & 4.94\\
FV & US 5Yr & 0.0078125 & 1.0 & 4.34\\
G & Gilts & 0.01 & 1.0 & 4.18\\
JB & JGB & 0.01 & 1.0 & 10.58\\
OE & Bobl & 0.01 & 1.0 & 3.70\\
RX & Bund & 0.01 & 1.0 & 3.70\\
TY & US 10Yr & 0.015625 & 1.0 & 4.54\\
US & US 30Yr & 0.03125 & 1.0 & 4.64\\
\textbf{Currency} & & & & \\
AD & AUD & 0.005 & 1.0 & 6.24\\
BP & GBP & 0.01 & 1.0 & 6.24\\
CD & CAD & 0.005 & 2.0 & 6.24\\
DX & USD & 0.005 & 1.0 & 6.24\\
EC & EUR & 0.005 & 2.0 & 6.24\\
JY & JPY & 0.005 & 1.0 & 6.24\\
SF & CHF & 0.01 & 3.0 & 6.24\\
\textbf{Energy} & & & & \\
CL & Crude & 0.01 & 2.0 & 6.04\\
CO & Brent Crude & 0.01 & 3.0 & 4.64\\
HO & Heating Oil & 0.01 & 3.0 & 6.04\\
NG & Nat Gas & 0.001 & 3.0 & 6.04\\
QS & Gasoil & 0.25 & 2.0 & 4.64\\
XB & RBOB & 0.01 & 7.0 & 6.04\\
\textbf{Equity} & & & & \\
CF & CAC 40 & 0.50 & 5.0 & 3.96\\
ES & S\&P 500 & 0.25 & 1.0 & 5.50\\
GX & DAX & 1.0 & 2.0 & 5.42\\
HI & Hang Seng & 1.0 & 5.0 & 9.14\\
IB & IBEX 35 & 1.0 & 3.0 & 6.26\\
NI & Nikkei & 5.0 & 1.0 & 4.00\\
TP & Topix & 0.50 & 1.0 & 9.32\\
XP & ASX 200 & 1.0 & 1.0 & 3.78\\
Z & FTSE 100 & 0.50 & 2.0 & 4.24\\
\textbf{Grain} & & & & \\
BO & Bean Oil & 0.01 & 3.0 & 6.94\\
C & Corn & 0.25 & 1.0 & 6.94\\
KO & Palm Oil & 1.0 & 1.0 & 14.48\\
KW & KC Wheat & 0.25 & 2.0 & 6.94\\
RS & Canola & 0.10 & 5.0 & 4.94\\
S & Soybean & 0.25 & 3.0 & 6.94\\
SM & Soy Meal & 0.10 & 2.0 & 6.94\\
W & Wheat & 0.25 & 1.0 & 6.94\\
\textbf{Metal} & & & & \\
GC & Gold & 0.10 & 2.0 & 6.14\\
HG & Copper & 0.05 & 4.0 & 6.14\\
PA & Palladium & 0.50 & 5.0 & 6.14\\
PL & Platinum & 0.10 & 9.0 & 6.14\\
SI & Silver & 0.005 & 2.0 & 6.14\\
\textbf{Soft} & & & & \\
CC & Cocoa & 1.0 & 4.0 & 7.24\\
CT & Cotton & 0.01 & 6.0 & 7.24\\
DF & Robusta Coffee & 1.0 & 3.0 & 1.56\\
KC & Coffee & 0.05 & 5.0 & 7.24\\
QC & UK Cocoa & 1.0 & 2.0 & 4.24\\
QW & UK Sugar & 0.10 & 7.0 & 4.24\\
SB & Sugar & 0.01 & 2.0 & 7.24\\
\hline
\end{tabular}
\caption{Futures market universe and trade cost assumptions. Symbols follow Bloomberg market symbology. Data is provided by Commodity Systems Inc (CSI).
Trade slippage cost is expressed in number of ticks per contract traded whereas commission cost is expressed in USD per contract traded.
Costs are estimated from live trading results on a global futures strategy with an  AUM of approximately $\$100$ million USD and an 
annualized volatility of approximately $20\%$.}
\label{table:universe}
\end{table}

\bibliographystyle{plainnat}
\bibliography{Bibliography}

\end{document}